\documentclass[prd,tightenlines,nofootinbib,superscriptaddress]{revtex4}

\usepackage{pre_all}
\usepackage{pre_QK}

\hypersetup{
pdfstartview = {FitH},
}
\hypersetup{
	colorlinks=true,         
	linkcolor=brown,          
	citecolor=red,        
	urlcolor=blue            
}

\begin{document}

\title{Local observables in $\SU_q(2)$ lattice gauge theory}

\author{{\bf Valentin Bonzom}}\email{bonzom@lipn.univ-paris13.fr}
\affiliation{Universit\'e Sorbonne Paris Nord, LIPN, CNRS UMR 7030, F-93430 Villetaneuse, France, EU}

\author{{\bf Ma\"it\'e Dupuis}}\email{mdupuis@perimeterinstitute.ca}
\affiliation{Perimeter Institute, 31 Caroline St North, Waterloo N2L 2Y5, Ontario, Canada}
\affiliation{Department of Applied Mathematics, University of Waterloo, Waterloo, Ontario, Canada}
\author{{\bf Florian Girelli}}\email{fgirelli@uwaterloo.ca}
\affiliation{Department of Applied Mathematics, University of Waterloo, Waterloo, Ontario, Canada}

\author{{\bf Qiaoyin Pan}}\email{qpan@fau.edu}
\affiliation{Department of Physics, Florida Atlantic University, 777 Glades Road, Boca Raton, FL 33431, USA}
\affiliation{Perimeter Institute, 31 Caroline St North, Waterloo N2L 2Y5, Ontario, Canada}
\affiliation{Department of Applied Mathematics, University of Waterloo, Waterloo, Ontario, Canada}

\date{\today}

\begin{abstract}
We consider a deformation of 3D lattice gauge theory in the canonical picture, first classically, based on the Heisenberg double of $\operatorname{SU}(2)$, then at the quantum level. We show that classical spinors can be used to define a fundamental set of local observables. They are invariant quantities which lives on the vertices of the lattice and are labeled by pairs of incident edges. Any function on the classical phase space, $\eg$Wilson loops, can be rewritten in terms of these observables. At the quantum level, we show that spinors become spinor operators. The quantization of the local observables then requires the use of the quantum $\cR$-matrix which we prove to be equivalent to a specific parallel transport around the vertex. We provide the algebra of the local observables, as a Poisson algebra classically, then as a $q$-deformation of $\mathfrak{so}^*(2n)$ at the quantum level. This formalism can be relevant to any theory relying on lattice gauge theory techniques such as topological models, loop quantum gravity or of course lattice gauge theory itself.

\end{abstract}

\maketitle

\tableofcontents 

\section*{Introduction}
The Hamiltonian picture of a lattice gauge theory is specified by the phase space $T^*G$ of a rotator associated to each edge of a lattice \cite{kogut}, for a Lie group $G$. At the vertices, local gauge transformations are generated by the Gauss constraint, which encodes the conservation of the angular momentum of the different rotators meeting at the vertex. This structure, called kinematical, is relevant not only to the discretization of Yang-Mills theory but also for example to loop quantum gravity. The latter aims at describing the quantum nature of space-time using gauge theory techniques \cite{rovelli_2004}, and some class of specific topological models. All those models are based on the same kinematical structure of lattice gauge theory, and they differ in their dynamical aspects. 

Instead of a Lie group $G$, one can generalize the construction by using a Hopf algebra $\cH$ ({\it a.k.a.} a quantum group) \cite{Boulatov:1996bj, Faddeev:1989qx}. At the classical level, this corresponds to replacing the cotangent bundle $T^*G$ with a Heisenberg double \cite{Alekseev:1991tx, Stern:1993rk, Alekseev:1994un}. 
The Hamiltonian picture of Hopf algebra lattice gauge theory is relevant to the construction of topological models which are in particular used to define some quantum computing models \cite{BMCA, CM}, or to define (loop) quantum gravity models \cite{smolin1996, Lewandowski:2008ye, Dupuis:2013haa, Dupuis:2014fya} with a non vanishing cosmological constant.

The symmetry algebra becomes the Drinfeld double $\cH^*\bowtie \cH$ of a given Hopf algebra $\cH$, whose elements decorate the lattice (see also \cite{Girelli:2017lfn} where instead of the Drinfeld double one uses a bicrossproduct Hopf algebra). Recent developments \cite{Bonzom:2014wva, MW} have shown that a clean way to use quantum groups on the lattice is to replace the edges of the lattice with \textit{ribbons}. As a consequence, the local gauge invariance is then expressed in terms of a constraint on elements of $\cH^*$ instead of $T^*G$, which can be interpreted geometrically as a holonomy constrained to be flat. This is therefore a deformed Gauss constraint. 

In any theory, the construction of observables is of course fundamental. While the notion of observables in the gravity case is more subtle than in the Yang-Mills case \cite{Dittrich:2007th, Rovelli:2001bz}, it is customary to call (abusing the terminology) the quantities that are locally gauge invariant, observables (so strictly speaking they could be called more appropriately, kinematical observables). Mathematically these quantities are invariant ($\ie$transforms as scalars) under infinitesimal gauge transformations (which are deformed in the case of quantum groups).

Wilson loops are well-known and natural observables of this type in any gauge theory. They are also extended objects. In the context of loop quantum gravity, it was realized that there are other observables which are more local in nature. Instead of being extended as Wilson lines,  they are associated to the vertices of the lattice \cite{Girelli:2005ii,Livine:2007vk,Freidel:2009ck,  Borja:2010rc, Livine:2011gp}. 

Consider $G=\SU(2)$ as the gauge group (corresponding for instance to both 3D and 4D loop quantum gravity with no cosmological constant). The fundamental degrees of freedom can be taken to be \emph{spinors} ($\ie$living in the fundamental representation of $\SU(2)$; they have \emph{nothing} to do with matter degrees of freedom) living on the ends of the lattice edges. The spinors which meet at an $n$-valent vertex can then be used to define observables labeled by pairs of incident edges, which moreover form a $\u(n)$ algebra. The framework passes on to the quantum level, where spinors become spinor operators ($\ie$tensor operators in the fundamental representation) and give rise to a $\u(n)$ algebra of operators at each $n$-valent vertex.

Later on \cite{Girelli:2017dbk}, the larger algebra $\so^*(2n)$ was identified as the full algebra of observables associated to $n$-valent vertices. These observables are the most fundamental ones since \emph{any} other observable in the holonomy and flux variables, such as Wilson loops, can be rewritten as a function of those fundamental observables \cite{Livine:2013zha}. In other words, they parametrize the invariant subspace of the phase space.

In this paper we work out the generalization to the case of the quantum group $\SU_q(2)$ (with $q$ real). We start with a plain lattice gauge theory based on a ribbon structure, using the classical group $\SL(2,\C)$ but equipped with a non-trivial, deformed, Poisson structure of the Heisenberg double $\mathcal{D}(\SU(2))$. We consider the deformed spinor variables which parametrize this phase space, already introduced in \cite{Dupuis:2014fya}. We show that it is then possible to generalize the construction of the local observables of \cite{Girelli:2017dbk} to the deformed case. We obtain invariants for the deformed action of $\SU(2)$.

We then proceed to the quantization. The quantization of the holonomy-flux algebra was already performed in \cite{Bonzom:2014bua}, which involved tensor operators of spin 1. Here we quantize the spinors directly, which give rise to spinor operators. Those objects have already been developed quite extensively using the full algebraic apparatus of quantum groups \cite{Rittenberg:1991tv, Quesne:1993}, such as the notion of braiding, induced by the quantum $\cR$-matrix. Those algebraic considerations thus provide the guide lines to actually build  local observables directly at the quantum level \cite{Dupuis:2013haa,Dupuis:2013lka}. However, since we are in the world of lattice gauge theory, it is also natural to use the geometric picture to construct the observables in terms of quantum parallel transport. Note that in the non-deformed case, no parallel transport is involved in these local observables. However, in the deformed case, $\AN(2)$ elements play the role of holonomies to transport spinors around the ribbon structure of vertices. It was already noticed in \cite{Bonzom:2014bua} that one can find quantum invariants without using the braiding provided by the $\cR$-matrix. Here, we clarify this aspect and show that these two different approaches, \emph{algebra versus geometry}, actually coincide beautifully. Indeed, \emph{the notion of braided permutation used to construct the tensor operators can be understood as a specific parallel transport along the ribbons}. While this might not come as a surprise to experts in integrable systems, this interpretation in the context of lattice gauge theory is new to the best of our knowledge.

Quantizing the spinors leads to the quantization of the local observables which are build with them. The algebra of those observables around a vertex of valence $n$ is shown to be a $q$-deformation of $\mathfrak{so}^*(2n)$ from \cite{Girelli:2017dbk}, with a $\mathcal{U}_q(\mathfrak{u}(n))$ subalgebra. This is proved by reproducing the Serre-Chevalley relations from our quantized observables.

The setup we have just described corresponds to the kinematical structure of several models. Specifying the \emph{dynamics} then specializes the model. One can $\eg$construct a Hamiltonian to deal with a (deformed) Yang-Mills type theory \cite{Stern:1993rk}, or a Kitaev-like model \cite{MW}. 

In a companion paper \cite{Bonzom:2021ham}, we have considered the dynamics of 3D quantum gravity with a cosmological constant using the present framework. As previously done in the flat case \cite{Bonzom:2011nv}, and in the deformed case using spin 1 operators \cite{Bonzom:2014bua}, we were able to write Hamiltonian constraints in terms of the local observables. Their quantization then leads to quantum Hamiltonian constraints, which in the invariant spin network basis give rise to difference equations. We were then able to show that changes of triangulations under Pachner moves change the coefficients in the spin network basis with the same amplitudes as in the Turaev-Viro model. It therefore derives the path integral approach (the Turaev-Viro model) from the Hamiltonian approach.

The article is organized as follows. In section \ref{sec:holonomy_flux_phase_space}, we recall the phase space structure of a deformed $\SU(2)$ lattice gauge theory. In particular the phase space is defined in terms of fluxes and holonomies. The basic building block is the phase space of a deformed rotator. In section \ref{sec:spinor_phase_space}, we revisit this phase space and parametrize it in terms of spinors. We then proceed to the construction of the local observables associated to the vertices of the lattice.  

In section \ref{sec:phasehopf}, as a preparation for the quantization of the spinors, we recall the quantization of the phase space of the deformed rotator and highlight that the $\cR$-matrix contains information on the quantum fluxes and holonomies. 

In section \ref{sec:quantum_spinors}, we quantize the spinors and obtain explicitly spinor operators. We show that the conjugation by the $\cR-$matrix which is used to build the observables at the quantum level can be interpreted as a parallel transport around the ribbon structure of vertices. Finally, we obtain the quantization of the local observables and prove that they form a deformation of $\mathfrak{so}^*(2n)$ in terms of the Serre-Chevalley relations.


\section{Holonomy-flux phase space}
\label{sec:holonomy_flux_phase_space}

As it is well-known, the phase space of lattice gauge theory is the phase space of a rotator, or spinning top \cite{kogut}, given in terms of the cotangent bundle $T^*G$ where $G$ is the gauge group. In the deformed case, the phase space is deformed, it is not a cotangent bundle anymore. The general notion replacing the cotangent bundle is the Heisenberg double \cite{Semenov1992, Chari1:995guide}. The configuration and momentum variables are typically called holonomies and fluxes, so that we call the usual lattice gauge theory phase space the holonomy-flux phase space. This is in contrast with the spinorial phase space which we will introduce in Section \ref{sec:spinor_phase_space}.

In this section, we review the phase space structure of a lattice gauge theory based on the specific example of the Heisenberg double of $\SU(2)$, $\cD(\SU(2))\cong \SU(2)\bowtie\AN(2)$ which we will work with all along. This can be viewed as the deformed version of an $\SU(2)$ lattice gauge theory. The deformation parameter is $\kappa\in \R^+$.  The standard phase space $T^*\SU(2)$ of an (undeformed) $\SU(2)$ lattice gauge theory is recovered in the limit $\ka\to 0$.

\subsection{Phase space: ribbon and Heisenberg double}
We are interested in  graphs embedded in a 2D canonical surface $\Sigma$. We first consider a single edge for which the  phase space is the Heisenberg double $\cD(\SU(2))$ of $\SU(2)$ with the dual group $\AN(2)$.  $\AN(2)$ is isomorphic to $\SB(2,\bC)$, the special Borel group, which is the group of $2\times 2$ lower triangular matrices with positive real diagonal entries and determinant 1. We parameterize an $\AN(2)$ element $\ell$ as
\be
\ell=\mat{cc}{\lambda & 0 \\ z& \lambda^{-1}}\,,\quad \lambda\in \R^+\,,\,z \in\bC\,.
\label{eq:l_param}
\ee
Note also that $\cD(\SU(2)) \cong \SL(2,\bC)$. We write $\cD(\SU(2))\cong \SU(2)\bowtie \AN(2)$ with $\bowtie$ encoding the mutual action of the two subgroups. This phase space can in fact be derived from a proper discretization \cite{Dupuis:2020ndx} of 3D Euclidean gravity with a negative cosmological constant. A similar derivation can probably be used for other gauge theories. 
\paragraph*{\textbf{Poisson structure.}}
The Poisson structure of the Heisenberg double  is fully determined by the $r$-matrix. Explicitly, the Poisson bracket is given by
\be
\{d_1,d_2\}=-\rT d_1d_2+d_1d_2 r=r d_1d_2-d_1d_2\rT\,,\quad 
\forall d\in \SL(2,\bC)\,,
\label{eq:bivector_SL2C}
\ee
where we used the standard notation $d_1=d\otimes \id$, $d_2=\id \otimes d$ and $r\equiv r_{12}=\sum r_{[1]}\otimes r_{[2]}$, $\rT:=\sum r_{[2]}\otimes r_{[1]}$. 
The last equality is guaranteed by the fact that $(r+\rT)$ is the Casimir of $\cD(\SU(2))$. In the fundamental representation, the $r$-matrix can be written as a $4\times 4$ matrix 
\be
r=\f14 \sum_i \sigma_i \otimes \rho^i = 
\f{i\ka}{4}\mat{cccc}{
1 & 0 & 0 & 0 \\ 0 & -1 & 4 & 0 \\ 0 & 0 & -1 & 0 \\ 0 & 0 & 0 & 1}
\in \su(2)\otimes \an(2)\,,
\label{eq:r_4X4}
\ee
where $\sigma_i, i=1,2,3$ are the Pauli matrices, $\rho^i (i=1,2,3)$ are Lie algebra generators of the Lie algebra $\an(2)$ which can be written in terms of the Pauli matrices as
\be
\rho^j=i\kappa\bigl(\sigma^j-\f12[\sigma^3,\sigma^j]\bigr)=\kappa(i\sigma^j+\epsilon^{3jk}\sigma^k)
\ee
and the Lie algebra of $\an(2)$ is 
\be
\left[ \rho^i,\rho^j \right]=2i\ka (\delta^i_k \delta^j_3 - \delta^i_3 \delta^j_k) \rho^k.
\ee

Note that the two subgroups $\SU(2)$ and $\AN(2)$ can be treated on the same footing. The phase space $\SL(2,\bC)$ can be equivalently described as the Heisenberg double $\cD(\AN(2))$ of $\AN(2)$ with the $r$-matrix $\rt\in\an(2)\otimes \su(2)$ where we simply have that $\rt=\rT,\rt_{21}=r$ since it amounts to exchanging the generators of the two subspaces in \eqref{eq:r_4X4}. The two equivalent descriptions of the phase space $\SL(2,\bC)$ corresponds to the two (and only two possible) {\it Iwasawa decompositions} of a given $\SL(2,\bC)$ element $d$. We denote by $\ell\in \AN(2),u\in\SU(2)$ the elements of the left Iwasawa decomposition $d=\ell u$ 
and by $\lt\in\AN(2),\ut\in\SU(2)$  the elements of the right Iwasawa decomposition $d=\ut\lt$. 
Then \eqref{eq:bivector_SL2C} 
can be decomposed into the Poisson brackets between $\ell$ and $u$:
\be\ba{llll}
\{\ell_1,\ell_2\}=-[\rT,\ell_1\ell_2]\,,&\{\ell_1,u_2\}=-\ell_1\rT u_2\,,& \{u_1,\ell_2\}=\ell_2r u_1\,,& \{u_1,u_2\}=-[r,u_1u_2]\,,
\ea
\label{eq:poisson_left}
\ee
or into the Poisson brackets between $\lt$ and $\ut$:
\be\ba{llll}
\{\lt_1,\lt_2\}=[\rT,\lt_1\lt_2]\,,& \{\lt_1,\ut_2\}=-\ut_2 \rT \lt_1\,,& \{\ut_1,\lt_2\}=\ut_1 r \lt_2\,,& \{\ut_1,\ut_2\}=[r,\ut_1\ut_2]\,.
\ea
\label{eq:poisson_right}
\ee

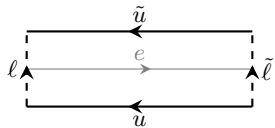
\begin{figure}[h]
	\begin{tikzpicture}[scale=1]

	\coordinate (O) at (0,0);
	\coordinate (A) at (3,0);
	\coordinate (AA) at (0,0.5);
	\coordinate (B) at (3,1);
	\coordinate (CC) at (3,0.5);
	\coordinate (C) at (0,1);

\draw[thin,gray, decoration={markings,mark=at position 0.55 with {\arrow[scale=1.3,>=stealth]{>}}},postaction={decorate}] (AA) --node[midway,above]{$e$}(CC);
	\draw[thick,dashed,decoration={markings,mark=at position 0.55 with {\arrow[scale=1.3,>=stealth]{>}}},postaction={decorate}] (O) -- node[midway, left]{$\ell$}(C); 
	\draw[thick,dashed,decoration={markings,mark=at position 0.55 with {\arrow[scale=1.3,>=stealth]{>}}},postaction={decorate}] (A) -- node[midway, right]{$\lt$}(B);
	\draw[thick,decoration={markings,mark=at position 0.55 with {\arrow[scale=1.3,>=stealth]{>}}},postaction={decorate}] (B) -- node[midway, above]{$\ut$}(C);
	\draw[thick,decoration={markings,mark=at position 0.55 with {\arrow[scale=1.3,>=stealth]{>}}},postaction={decorate}] (A) -- node[midway, below]{$u$}(O);
	
	\end{tikzpicture}
\caption{The ribbon graph associated to the ribbon constraint. The ribbon carries two pairs of variables $(\ell,u)$ and $(\lt,\ut)$. The ribbon constraint is the trivialization of the ribbon loop $\ell u\lt^{-1}\ut^{-1}$.  }
\label{fig:box}
\end{figure}

\paragraph*{\textbf{Ribbon constraint.}}
The equivalence between the left and right Iwasawa decompositions defines a constraint, that we call the \emph{ribbon constraint}
\be
\cC =\ell u\lt^{-1}\ut^{-1}\,.
\label{eq:ribbon_constraint}
\ee
It is easy to check that this is a system of second-class constraints (meaning that they do not close under Poisson brackets). The name ``ribbon" will become clear when we represent graphically these two equal Iwasawa decompositions. Concretely, an edge $e$ is thickened into a ribbon $R(e)$ with 
\begin{itemize}
    \item long sides, parallel to $e$, carrying the $\SU(2)$ elements $u, \tu$ called \emph{holonomies}
    \item short sides carrying the $\AN(2)$ elements $\ell, \tell$ and called \emph{fluxes} \cite{Bonzom:2014wva}.
\end{itemize}
This is represented in Figure~\ref{fig:box} together with a choice of orientations (detailed below). We have fixed the orientation of the long sides decorated with $u$ and $\ut$ to be opposite to that of the edge, which automatically fixes the orientation of the two short sides of a ribbon, so that the ribbon constraint \eqref{eq:ribbon_constraint} is satisfied.

All $\SU(2)$ and $\AN(2)$ subgroup elements are associated to sides of the ribbon and can thus be viewed as holonomies. The ribbon constraint is then interpreted as a trivialization of the path-ordered product of holonomies on the loop surrounding the ribbon.
To emphasize that the phase space we describe here is the deformation of that of the $\Lambda=0$ loop gravity, we use the same terminology and call $\ell,\lt$ fluxes and $u,\ut$ holonomies in the rest of the article. This terminology is consistent with that in \cite{Bonzom:2014wva}.

By solving the ribbon constraint, we obtain the Poisson brackets between $(\lt,\ut)$ and $(\ell,u)$:
\be\ba{llll}
\{\ell_1,\ut_2\}=-\rT\ell_1\ut_2\,,
&\{\lt_1,u_2\}=-\lt_1u_2\rT\,,
&\{u_1,\lt_2\}=\lt_2u_1r,,
&\{\ut_1,\ell_2\}=r\ut_1\ell_2\,,\\
\{\lt_1,\ell_2\}=0\,,
&\{\ut_1,u_2\}=0\,.
\ea
\label{eq:poisson_mix}
\ee
The explicit Poisson brackets between the matrix elements of $\ell,u,\lt,$ and $\ut$ can be found in Appendix \ref{app:Poisson_bracket_SL2C}. The dimension of the phase space for a ribbon is $12-6=6$ upon imposing the ribbon constraint, thus consistent with the dimension of $\SL(2,\bC)$.

\paragraph*{\textbf{$\SU(2)$ transformations.}}
Let us define $X:=\ell\ell^\dagger$ and write  $w=\id +i \vec{\epsilon}\cdot \vec{\sigma}$ an infinitesimal $\SU(2)$ group element. Then, the variation of a phase space function $h$ under a left infinitesimal $\SU(2)$ transformation is given by \cite{Bonzom:2014wva}:
\be
\delta_{\epsilon} h 
=-\lambda^{-2}\ka^{-1}\{ \tr WX,h\}
=-\lambda^{-2}\kappa^{-1} \{2\epsilon_z \lambda^2 +\epsilon_-\lambda z +\epsilon_+\lambda \zb,h \}\,,
\quad\text{wtih }\,W=\mat{cc}{2\epsilon_z & \epsilon_- \\ \epsilon_+ & 0} \,.
\label{eq:SU2_L}
\ee

\paragraph*{\textbf{Change of edge orientations.}}
The way we associate variables to the sides of a ribbon has been described above, as in Figure \ref{fig:box}. Changing the orientation of an edge is an involution $\iota$ which has the following effects on the variables,
\begin{equation} \label{iota}
\begin{aligned}
\iota: &u\mapsto \tu^{-1}\\
&\ell \mapsto \tell^{-1}
\end{aligned}
\end{equation}
and since it is an involution, $\iota(\tu) = u^{-1}$ and $\iota(\tell) = \ell^{-1}$.

\subsection{Ribbon graph and Gauss constraint}
Let $\Gamma$ be a graph embedded in $\Sigma$. We start with the phase space $\prod_e \mathcal{D}(\SU(2))$ where the product is over the edges of $\Gamma$. As we thickened an edge into a ribbon, we now thicken $\Gamma$ into a ribbon graph $\Gamma_{\text{rib}}$ by 
\begin{itemize}
\item thickening every edge into a ribbon in the same way as in Figure \ref{fig:box}, where all ribbons are embedded in $\Sigma$,
\item thickening every $n$-valent vertex of $\Gamma$ into an $n$-gon.
\end{itemize}
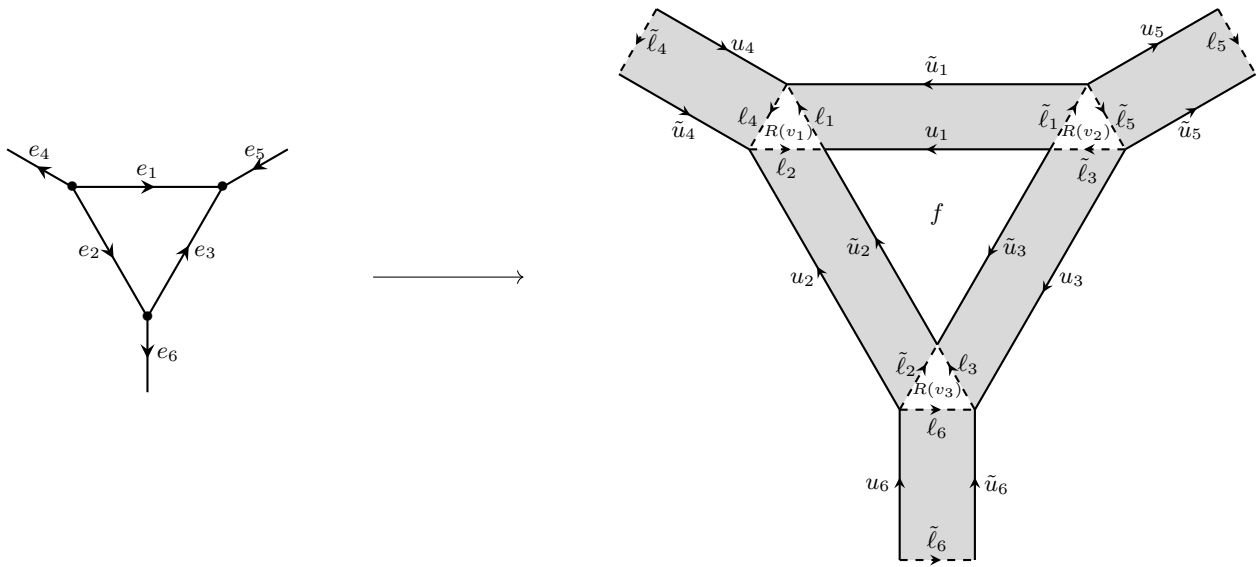
\begin{figure}[h!]
\centering
\begin{tikzpicture}[scale=1]
\coordinate (O) at (0,0);
\coordinate (A) at (2,0);
\coordinate (B) at ([shift=(-60:2cm)]O);
\coordinate (O1) at ([shift=(150:1cm)]O);
\coordinate (A1) at ([shift=(30:1cm)]A);
\coordinate (B1) at ([shift=(-90:1cm)]B);

\draw[thick,decoration={markings,mark=at position 0.55 with {\arrow[scale=1.3,>=stealth]{>}}},postaction={decorate}] (O) -- node[midway,above]{$e_1$}(A); 
\draw[thick,decoration={markings,mark=at position 0.55 with {\arrow[scale=1.3,>=stealth]{>}}},postaction={decorate}] (O) -- node[midway,left]{$e_2$}(B); 
\draw[thick,decoration={markings,mark=at position 0.55 with {\arrow[scale=1.3,>=stealth]{>}}},postaction={decorate}] (B) -- node[midway,right]{$e_3$}(A); 

\draw[thick,decoration={markings,mark=at position 0.55 with {\arrow[scale=1.3,>=stealth]{>}}},postaction={decorate}] (O) -- node[midway,above]{$e_4$}(O1); 
\draw[thick,decoration={markings,mark=at position 0.55 with {\arrow[scale=1.3,>=stealth]{>}}},postaction={decorate}] (A1) -- node[midway,above]{$e_5$}(A); 
\draw[thick,decoration={markings,mark=at position 0.55 with {\arrow[scale=1.3,>=stealth]{>}}},postaction={decorate}] (B) -- node[midway,right]{$e_6$}(B1); 

\draw (O) node {$\bullet$};
\draw (A) node {$\bullet$};
\draw (B) node {$\bullet$};

\coordinate (L) at (4,-1.2);
\coordinate (R) at (6,-1.2);

\draw[->] (L) -- (R);

\coordinate (o) at (10,0.5);
\coordinate (o1) at ([shift=(120:1cm)]o);
\coordinate (o2) at ([shift=(180:1cm)]o);
\coordinate (o3) at ([shift=(150:2cm)]o1);
\coordinate (o4) at ([shift=(150:2cm)]o2);
\coordinate (a) at (13,0.5);
\coordinate (a1) at ([shift=(0:1cm)]a);
\coordinate (a2) at ([shift=(60:1cm)]a);
\coordinate (a3) at ([shift=(30:2cm)]a1);
\coordinate (a4) at ([shift=(30:2cm)]a2);
\coordinate (b) at ([shift=(-60:3cm)]o);
\coordinate (b1) at ([shift=(-60:1cm)]b);
\coordinate (b2) at ([shift=(-120:1cm)]b);
\coordinate (b3) at ([shift=(-90:2cm)]b1);
\coordinate (b4) at ([shift=(-90:2cm)]b2);

\fill[gray!30] (o1) -- (o2) -- (o4) -- (o3) --cycle;
\fill[gray!30] (a1) -- (a2) -- (a4) -- (a3) --cycle;
\fill[gray!30] (b1) -- (b2) -- (b4) -- (b3) --cycle;
\fill[gray!30] (o) -- (o1) -- (a2) -- (a) --cycle;
\fill[gray!30] (o) -- (o2) -- (b2) -- (b) --cycle;
\fill[gray!30] (b) -- (a) -- (a1) -- (b1) --cycle;

\draw[thick,decoration={markings,mark=at position 0.55 with {\arrow[scale=1,>=stealth]{>}}},postaction={decorate}] (a) -- node[midway,above]{$u_1$}(o); 
\draw[thick,decoration={markings,mark=at position 0.55 with {\arrow[scale=1,>=stealth]{>}}},postaction={decorate}] (b) -- node[midway,left]{$\ut_2$}(o); 
\draw[thick,decoration={markings,mark=at position 0.55 with {\arrow[scale=1,>=stealth]{>}}},postaction={decorate}] (a) -- node[midway,right]{$\ut_3$}(b);

\draw[thick,dashed,decoration={markings,mark=at position 0.7 with {\arrow[scale=1,>=stealth]{>}}},postaction={decorate}] (o) -- node[midway,right]{$\ell_1$}(o1);
\draw[thick,dashed,decoration={markings,mark=at position 0.55 with {\arrow[scale=1,>=stealth]{>}}},postaction={decorate}] (o2) -- node[midway,below]{$\ell_2$}(o);
\draw[thick,decoration={markings,mark=at position 0.55 with {\arrow[scale=1,>=stealth]{>}}},postaction={decorate}] (o3) -- node[midway,right]{$u_4$}(o1);
\draw[thick,decoration={markings,mark=at position 0.55 with {\arrow[scale=1,>=stealth]{>}}},postaction={decorate}] (o4) -- node[midway,below]{$\ut_4$}(o2);
\draw[thick,dashed,decoration={markings,mark=at position 0.45 with {\arrow[scale=1,>=stealth]{>}}},postaction={decorate}] (o1) -- node[midway,left]{$\ell_4$}(o2);
\draw[thick,dashed,decoration={markings,mark=at position 0.55 with {\arrow[scale=1,>=stealth]{>}}},postaction={decorate}] (o3) -- node[midway,right]{$\lt_4$}(o4);

\draw[thick,dashed,decoration={markings,mark=at position 0.55 with {\arrow[scale=1,>=stealth]{>}}},postaction={decorate}] (a1) -- node[midway,below]{$\lt_3$}(a);
\draw[thick,dashed,decoration={markings,mark=at position 0.7 with {\arrow[scale=1,>=stealth]{>}}},postaction={decorate}] (a) -- node[midway,left]{$\lt_1$}(a2);
\draw[thick,decoration={markings,mark=at position 0.55 with {\arrow[scale=1,>=stealth]{>}}},postaction={decorate}] (a1) -- node[midway,below]{$\ut_5$}(a3);
\draw[thick,decoration={markings,mark=at position 0.55 with {\arrow[scale=1,>=stealth]{>}}},postaction={decorate}] (a2) -- node[midway,above]{$u_5$}(a4);
\draw[thick,dashed,decoration={markings,mark=at position 0.45 with {\arrow[scale=1,>=stealth]{>}}},postaction={decorate}] (a2) -- node[midway,right]{$\lt_5$}(a1);
\draw[thick,dashed,decoration={markings,mark=at position 0.55 with {\arrow[scale=1,>=stealth]{>}}},postaction={decorate}] (a4) -- node[midway,left]{$\ell_5$}(a3);

\draw[thick,dashed,decoration={markings,mark=at position 0.7 with {\arrow[scale=1,>=stealth]{>}}},postaction={decorate}] (b1) -- node[pos=0.7,right]{$\ell_3$}(b);
\draw[thick,dashed,decoration={markings,mark=at position 0.7 with {\arrow[scale=1,>=stealth]{>}}},postaction={decorate}] (b2) -- node[pos=0.7,left]{$\lt_2$}(b);
\draw[thick,decoration={markings,mark=at position 0.55 with {\arrow[scale=1,>=stealth]{>}}},postaction={decorate}] (b3) -- node[midway,right]{$\ut_6$}(b1);
\draw[thick,decoration={markings,mark=at position 0.55 with {\arrow[scale=1,>=stealth]{>}}},postaction={decorate}] (b4) -- node[midway,left]{$u_6$}(b2);
\draw[thick,dashed,decoration={markings,mark=at position 0.55 with {\arrow[scale=1,>=stealth]{>}}},postaction={decorate}] (b2) -- node[midway,below]{$\ell_6$}(b1);
\draw[thick,dashed,decoration={markings,mark=at position 0.55 with {\arrow[scale=1,>=stealth]{>}}},postaction={decorate}] (b4) -- node[midway,above]{$\lt_6$}(b3);

\draw[thick,decoration={markings,mark=at position 0.55 with {\arrow[scale=1,>=stealth]{>}}},postaction={decorate}] (a2) -- node[midway,above]{$\ut_1$}(o1); 
\draw[thick,decoration={markings,mark=at position 0.55 with {\arrow[scale=1,>=stealth]{>}}},postaction={decorate}] (b2) -- node[midway,left]{$u_2$}(o2); 
\draw[thick,decoration={markings,mark=at position 0.55 with {\arrow[scale=1,>=stealth]{>}}},postaction={decorate}] (a1) -- node[midway,right]{$u_3$}(b1); 

\coordinate (oo) at ([shift=(150:0.55)]o);
\coordinate (aa) at ([shift=(30:0.55)]a);
\coordinate (bb) at ([shift=(-90:0.5)]b);
\coordinate (c) at ([shift=(-30:1.73)]o);

\draw ([shift=(-90:0.05)]oo) node {\small $\scriptstyle R(v_1)$};
\draw ([shift=(-90:0.05)]aa) node {\small $\scriptstyle R(v_2)$};
\draw ([shift=(-90:0.1)]bb) node {\small $\scriptstyle R(v_3)$};
\draw (c) node {$f$};

\end{tikzpicture}
\caption{A piece of a graph $\Gamma$ on the left and the corresponding piece of the ribbon graph $\Gamma_\text{rib}$ on the right.}
\label{fig:ribbon_graph}
\end{figure}
An example is given in Figure~\ref{fig:ribbon_graph}, with three 3-valent vertices and one internal face. 

As such, a ribbon graph contains three types of faces:
\begin{enumerate}[i]
	\item Faces within ribbon edges, for which the ribbon constraint is imposed - these are the faces in grey in Figure~\ref{fig:ribbon_graph};  
	\item Faces surrounded by the short sides of the ribbons. They correspond to the thickened vertices and we call them \emph{ribbon vertices}. In Figure~\ref{fig:ribbon_graph}, these are the three triangular faces $R(v_1), R(v_2)$ and $R(v_3)$. Notice that they are bounded by fluxes only.
	\item Faces surrounded by the long sides of the ribbons - these are the faces of the original graph. They are bounded by $\SU(2)$ holonomies only. 
\end{enumerate}

To finish the combinatorial description of $\Gamma_{\text{rib}}$, notice that the corners of $\Gamma$, $\ie$the portions of $\Sigma$ between pairs of edges incident to a vertex, give rise in $\Gamma_{\text{rib}}$ to vertices (the ends of the long and short sides, and not to be confused with ribbon vertices).

Each ribbon edge thus carries variables $\ell_e, u_e, \tell_e, \tu_e$ which satisfy the ribbon constraint $\mathcal{C}_e = \ell_e u_e \tell_e^{-1} \tu_e^{-1}$. In addition we introduce the Gauss constraints. The Gauss constraint associated to an $n$-valent vertex $v$ imposes that the ordered product of the fluxes around the ribbon vertex $R(v)$ is trivial. 
Explicitly, the Gauss constraint  
reads
\be
\cG_v = \overrightarrow{\prod}_{i=1}^n \ell_{e_i, v}\,,\quad
\ell_{e_i, v} = 
\begin{cases}
\ell_i\quad &\text{if}\quad o_i=1	\\
\lt_i^{-1}\quad &\text{if} \quad o_i=-1
\end{cases}\,,
\label{eq:Gauss_flatness_constraint}
\ee
where   
$o_i=1$ corresponds to an outgoing edge and $o_i=-1$ corresponds to an incoming edge.

The Gauss constraint generates $\SU(2)$ transformations. 
A phase space function $h$ transforms under the infinitesimal rotation parametrized by a infinitesimal vector $\vec{\epsilon}$ as \cite{Bonzom:2014wva}
\begin{equation}\label{eq:generalgaugetrans}
\delta_{\epsilon} h = -\ka^{-1}\prod_{i=1}^n \Uplambda_i^{-2} \{\tr \bigl(W\cG_v\cG_v^\dagger\bigr), h \}\,,\qquad
\text{with $W=\mat{cc}{2\epsilon_z & \epsilon_- \\ \epsilon_+ & 0}$}\,,
\end{equation}
where $\Uplambda_i$ is the first diagonal element of the matrix \eqref{eq:l_param} of the $i$-th flux, $\ell_{e_i,v}$, in $\cG_v$, that is $\lambda_i$ or $\tlambda_i^{-1}$. 

The subspace satisfying the Gauss constraint at every vertex of $\Gamma$ is called the \emph{kinematical} phase space. Its parametrization using observables in terms of spinors and their quantization will be the focus of the present article.

Beyond the kinematical aspects, several choices of dynamics are possible, such as lattice Hamiltonians for Yang-Mills. There is also a topological model called BF, and corresponding to 3D gravity, where the Hamiltonian is a constraint, just like the Gauss constraint. It is called the flatness constraint, and it imposes the holonomies around all faces to be trivial. The classical setup and the quantization of this flatness constraint was initiated in \cite{Bonzom:2014bua} and the extension to spinors has been developed in the companion article \cite{Bonzom:2021ham}.

\subsection{Adjoint ribbon parameterization}
\begin{figure}[h]
	\begin{tikzpicture}[scale=1]

	\coordinate (O) at (0,0);
	\coordinate (A) at (3,0);
	\coordinate (AA) at (0,0.5);
	\coordinate (B) at (3,1);
	\coordinate (CC) at (3,0.5);
	\coordinate (C) at (0,1);

\draw[thin,gray, decoration={markings,mark=at position 0.55 with {\arrow[scale=1.3,>=stealth]{>}}},postaction={decorate}] (AA) --node[midway,above]{$e$}(CC);
	\draw[thick,dashed,decoration={markings,mark=at position 0.55 with {\arrow[scale=1.3,>=stealth]{>}}},postaction={decorate}] (O) -- node[midway, left]{$\ell^{\dagger -1}$}(C); 
	\draw[thick,dashed,decoration={markings,mark=at position 0.55 with {\arrow[scale=1.3,>=stealth]{>}}},postaction={decorate}] (A) -- node[midway, right]{$\lt^{\dagger -1}$}(B);
	\draw[thick,decoration={markings,mark=at position 0.55 with {\arrow[scale=1.3,>=stealth]{>}}},postaction={decorate}] (B) -- node[midway, above]{$\ut$}(C);
	\draw[thick,decoration={markings,mark=at position 0.55 with {\arrow[scale=1.3,>=stealth]{>}}},postaction={decorate}] (A) -- node[midway, below]{$u$}(O);
	
	\end{tikzpicture}
\caption{The conjugate ribbon structure defined in terms of upper triangular matrices.}
\label{fig:otherparam}
\end{figure}
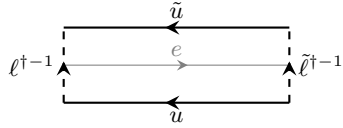

We have been working with the ribbon constraint \eqref{eq:ribbon_constraint}, but there is another version available. Indeed, we have worked with $\AN(2)$ in terms of lower triangular matrices. But instead, we could use upper triangular matrices. The equivalence between the two formulations can be seen by using the adjoint on $\mathcal{C}$. It is also convenient to take the inverse, so that $\SU(2)$ elements are left invariant. This gives the following adjoint ribbon constraint
\be
\cC =\ell u\lt^{-1}\ut^{-1} \qquad\rightarrow\qquad \cC^{-1\dagger}= \ell^{-1\dagger} u\lt^{\dagger}\ut^{-1}. 
\label{eq:ribbon_constraintbis}
\ee
It amounts to replacing $\ell$ and $\tell$ with respectively $\ell^{-1\dagger}$ and  $\tell^{-1\dagger}$ (and similarly with $u, \tu$ but obviously $u^{-1\dagger}=u$ for any $u\in\SU(2)$). Therefore, only the short side structure is changed, as in Figure~\ref{fig:otherparam}.
The associated transformation preserving the Lie algebra $\an(2)$ is given by $\rho^i\rightarrow -\rho^{i\dagger}$. As a consequence, one  switches the $r$-matrix by  $r\rightarrow r^\dagger = -\rT$. All Poisson brackets are given in Appendix \ref{app:Poisson_bracket_SL2C}. 

Under this parametrization, the Gauss constraint is transformed accordingly as $\cG_v \rightarrow \cG_v ^{\dagger -1}$, which transforms $\cG_v\cG_v^\dagger \to (\cG_v\cG_v^\dagger)^{-1}$. We thus have the same action on phase space functions as with the previous generators of gauge transformations if we consider
\begin{equation}\label{eq:generalgaugetransbis}
\delta_{\epsilon} h = -\frac{1}{\ka}\prod_{i=1}^n \Uplambda_i^{-2} \{\tr \tilde W\left(\cG_v\cG_v^\dagger\right)^{-1}, h \}\,,\qquad
\text{with } \tilde W=\mat{cc}{0 & - \epsilon_- \\ - \epsilon_+ & 2\epsilon_z}\,,
\end{equation}
where $\Uplambda_i$ is still $\lambda_i$ or $\tlambda_i^{-1}$ according to the orientation of the edge $e_i$.

In fact, this adjoint parametrization is not only an alternative one but a necessary piece to construct the complete kinematical phase space because $\ell$ and $\lt$ only contain respectively $z$ and $\tz$ in their matrix elements, but neither $\bar{z}$ nor $\bar{\tz}$. We will see in Section \ref{sec:phasehopf} that both $z$ and $\zb$ (as well as $\tz$ and $\bar{\tz}$) are needed to construct the $\UQ$ generators upon quantization. 

\section{Spinorial phase space for a deformed lattice gauge theory}
\label{sec:spinor_phase_space}

We have just described above the kinematical phase space using the holonomy-flux variables. We now describe the same space in terms of \emph{spinors}. They live on the half-edges of the lattice and will make it easier to construct \emph{local}, gauge invariant quantities, $\ie$observables.
Indeed, invariant functions of fluxes, for example, do not Poisson close \cite{Girelli:2005ii}. The right variables to build a (Poisson) closed algebra of observables are the spinors.

To avoid confusion, we emphasize that they do not encode matter degrees of freedom, they are just a different parametrization of the phase-space. They were initially introduced in the loop quantum gravity formalism as a parametrization of the $T^*\SU(2)$ phase space \cite{Girelli:2005ii, Freidel:2009ck,Freidel:2010tt,Livine:2011zz,Bonzom:2011nv, Dupuis:2010iq,Dupuis:2011fz}. We intend here to construct the deformed spinors which provide an alternative parametrization of the deformed holonomy-flux phase space, which will allow us to construct the (deformed) notion of observables for this set up.

\medskip

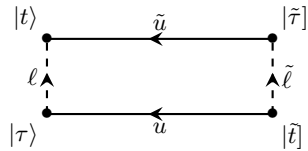
\begin{figure}[h!]
\centering
\begin{tikzpicture}
	\coordinate (O) at (0,0);
	\coordinate (A) at (3,0);
	\coordinate (B) at (3,1);
	\coordinate (C) at (0,1);
	
	\draw[thick,dashed,decoration={markings,mark=at position 0.55 with {\arrow[scale=1.3,>=stealth]{>}}},postaction={decorate}] (O) -- node[midway, left]{$\ell$ }(C); 
	\draw[thick,dashed,decoration={markings,mark=at position 0.55 with {\arrow[scale=1.3,>=stealth]{>}}},postaction={decorate}] (A) -- node[midway, right]{$\lt$ }(B);
	\draw[thick,decoration={markings,mark=at position 0.55 with {\arrow[scale=1.3,>=stealth]{>}}},postaction={decorate}] (B) -- node[midway, above]{$\ut$}(C);
	\draw[thick,decoration={markings,mark=at position 0.55 with {\arrow[scale=1.3,>=stealth]{>}}},postaction={decorate}] (A) -- node[midway, below]{$u$}(O);
	
	\draw (O) node{$\bullet$} node[below left]{$|\tau\ra$};
	\draw (A) node{$\bullet$} node[below right]{$|\tt]$};
	\draw (B) node{$\bullet$} node[above right]{$|\ttau]$};
	\draw (C) node{$\bullet$} node[above left]{$|t\ra$};

\end{tikzpicture}
\caption{$\SU(2)$-covariant spinors in the ribbon picture. Note that we can replace $\ell$ and $\tell$ respectively by  $\ell^{-1\dagger}$ and  $\lt^{-1\dagger}$. }
\label{fig:kappa_spinors}
\end{figure}

We start with some deformed spinors that allow us to parametrize the $\AN(2)$ elements. We will then use them to define $\SU(2)$-covariant spinors which are the key objects of this section. In Section \ref{sec:quantum_spinors}, they will be quantized as spinor operators, which are spin-$1/2$ tensor operators for $\UQ$ or $\UQI$ \cite{Rittenberg:1991tv}. 
Graphically, they can be naturally associated to the four corners of the ribbon, see Figure \ref{fig:kappa_spinors}, which will be clear by the end of Section \ref{sec:tilde_covariant_spinors}.

\subsection{Basic variables}
Our building blocks are two independent spinors  $|\zeta \ra, |\tilde{\zeta}\ra \in \bC^2$ and their conjugate $\la \zeta| \in \bC^2$ and $\la \tilde{\zeta}| \in \bC^2$,
\be
|\zeta \ra=\left( \tabl{c}{ \zeta_0\\ \zeta_1}\right)\,,\quad 
\la \zeta|=(\bzeta_0, \, \bzeta_1)\,,\quad|\tilde{\zeta} \ra=\left( \tabl{c}{ \tilde{\zeta}_0\\ \tilde{\zeta}_1}\right)\,,\quad 
\la \tilde{\zeta}|=(\bar{\tilde{\zeta}}_0, \, \bar{\tilde{\zeta}}_1)\,,\quad
\label{eq:undeformed_spinor}
\ee
such that $$
\{\zeta_A, \, \bzeta_B\} =-i \delta_{AB}\,, \quad \{\zeta_A,\tzeta_B\}=\{\bzeta_A,\btzeta_B\}=\{\zeta_A,\btzeta_B\}=\{\bzeta_A,\tzeta_B\}=0\,,\forall A,B=0,1.$$
We also introduce the dual spinor
\be
|\zeta]=\mat{c}{-\bzeta_{1} \\  \bzeta_{0}}
=\mat{cc}{0 & -1 \\ 1 & 0 } |\bzeta \ra,\qquad
[\zeta|=\mat{cc}{-\zeta_{1} &  \zeta_{0}}\,.
\label{eq:undeformed_dual_spinor}
\ee
which is orthogonal to $|\zeta\rangle$, $\langle \zeta|\zeta]=0$. Similarly, one defines the dual of the tilde spinor $|\tzeta]$. We denote $N_{A}=\zeta_{A}\bzeta_{A}$, the modulus of the spinor components for $A=0,1$ and $N=N_{0}+N_{1}$ for their norm. A spinor and its dual have the same norm $\la \zeta|\zeta\ra =[\zeta|\zeta] = N$. The modulus generates dilation on the complex variables:
\be
\{ N_A, \zeta_B \}=i \delta_{AB}\zeta_A\,,
\qquad
\{N_A, \bar{\zeta}_B\}=-i\delta_{AB}\bzeta_B\,.
\label{eq:Poisson_undeformed_spinor}
\ee

Let us  now  define the deformed variables $|\zeta^\kappa\ra$ from $|\zeta\ra$, with its dual $\la \zeta^\kappa |$ and norm $\la \zeta^\kappa |\zeta^\kappa\ra$ as in \cite{Dupuis:2014fya}:
\bes 
\zeta^\kappa_A &\equiv & \zeta_A\sqrt{\f{\sinh(\f{\kappa}{2}N_A)}{\f{\kappa}{2}N_A}}
\,,
\quad 
\bar{\zeta}^\kappa_A
\equiv \overline{\zeta^\kappa_A} \,,
\label{eq:kappa_spinor}\\
 \la \zeta^\kappa |\zeta^\kappa \ra
&=& \sum_A   \bar{\zeta}_A^\kappa \zeta_A^\kappa 
\,=\, \sum_A 
\f{2}{\kappa}\sinh(\f{\kappa N_A}{2})
\,=\,
\f{1}{\kappa}\sum_A  \left(e^{\f{\kappa N_{A}}2}-e^{\f{-\kappa N_{A}}2}\right)\geq 0 \,, \textrm{ with }  N_A= \bzeta_{A} \zeta_{A}\,.
\label{eq:norm_kappa_spinor}
\ees
They satisfy the following Poisson brackets 
\be
\{\zeta_A^\kappa, \bar{\zeta}^\kappa_B\}= -i \,\delta_{AB}\,  \cosh\Bigl(\f{\kappa N_A}{2}\Bigr),
\quad
\{N_{A}, \zeta_B^\kappa\}=i \,\delta_{AB} \,\zeta^\kappa_A,
\quad
\{ N_{A},\bar{\zeta}_B^\kappa\}=-i \,\delta_{AB}\, \bar{\zeta}^\kappa_A\,.
\label{eq:Poisson_nontilde}
\ee
It is easy to check that we recover the undeformed Poisson brackets \eqref{eq:Poisson_undeformed_spinor} when $\ka\dr 0$. The deformed variable $|\tzeta^\ka\ra$ is defined from $|\tzeta\ra$ by \eqref{eq:kappa_spinor} and \eqref{eq:norm_kappa_spinor} where all $\zeta_A$ are replaced by $\tzeta_A$. 

\paragraph*{\textbf{Change of edge orientations.}}
Since there are no differences between $|\zeta^\kappa\rangle$ and $|\tilde{\zeta}^\kappa\rangle$, and since changing the orientation of an edge exchanges the two sectors, it is natural to lift the involution $\iota$ to the spinor space as follows,
\begin{equation} \label{iotazeta}
\iota(\zeta^\kappa_{A}) = \tilde{\zeta}^\kappa_{A},\qquad \iota(\tilde{\zeta}^\kappa_{A}) = \zeta^\kappa_{A}\,,\qquad\text{for }\, A=0,1\,.
\end{equation}

\paragraph*{\textbf{Recontructing the fluxes.}} We will use $\zeta^\kappa_{0,1}$ to reconstruct $\ell$ and $\tilde{\zeta}^\kappa_{0,1}$ to reconstruct $\tell$. Since the spinors in the tilde and non-tilde sectors are identical whereas $\iota(\ell)\neq \tell$, $\ell$ can not be the same function of $\zeta^\kappa_{0,1}$ as $\tell$ is of $\tilde{\zeta}^\kappa_{0,1}$. We use
\be\ba{ll}
\lambda \equiv  \exp(\f{\ka}{4}(N_1-N_0))\,, &
z \equiv-\ka\bzeta_0^\ka \zeta_1^\ka\,,\\ [0.15cm]
\tlambda \equiv \exp(\f{\ka}{4}(\Nt_0-\Nt_1))\,, &
\tz \equiv  \ka \btzeta^\ka_0 \tzeta^\ka_1 \,.
\ea
\label{eq:defi_lambda_z}
\ee
By applying $\iota$ to \eqref{eq:defi_lambda_z}, we recover as expected that $\iota(\lambda) = \tilde{\lambda}^{-1}$ and $\iota(z) = -\tz$. 
The $\AN(2)$ matrices $\ell$ and $\tell$ become functions of the spinors,
\be
\ell(\zeta^\ka_{0,1},\bzeta^\ka_{0,1})=
\lb\tabl{cc}{
\lambda & 0\\
z & \lambda^{-1}
} \rb\,,\quad
\lt (\tzeta^\ka_{0,1},\btzeta^\ka_{0,1})=
 \mat{cc}{\tlambda & 0 \\ \tz & \tlambda^{-1}}\,,
\label{eq:l_lt_param}
\ee
and $\iota(\ell) = \lt^{-1}$.  
It is easy to check that these $\AN(2)$ matrix elements do satisfy the expected Poisson brackets \eqref{eq:poisson_matrix_elements}. Let us point out that $z$, $\zb$, $\lambda$ all commute with $N=N_{0}+N_{1}$, $\{N,\bar{z}\}=\{N,\lambda\}=\{N,z\}=0$.

While the deformed variables $|\zeta^\ka\ra$ and $|\tzeta^\ka\ra$ are important in parametrizing the $\AN(2)$ elements and generating the (infinitesimal) rotation transformations \cite{Bonzom:2014wva}, they are not yet the spinors we will use to reconstruct the holonomy-flux phase space, because they do \emph{not} transform covariantly under the $\SU(2)$ action.  

\subsection{Covariant Spinors}
\label{sec:covariant_spinors}
Let us now define the variables which transform covariantly as spin 1/2 under $\SU(2)$, $\ie$either \eqref{eq:generalgaugetrans} or \eqref{eq:generalgaugetransbis}, depending on if we consider the ribbon variable $\ell$ or $\ell^{-1\dagger}$. We consider the first case, where we deal with $\ell$. 
We recall  that $X=\ell\ell^\dagger$ with $\ell$ an $\AN(2)$ element now parametrized as in \eqref{eq:l_lt_param} whose entries are defined in terms of the spinor variables given in  \eqref{eq:defi_lambda_z}.

\paragraph*{\textbf{Covariant spinor.}} 
An $\SU(2)$-covariant spinor (henceforth spinor) $|T\ra$ is defined by the transformation law
\be
\delta_\epsilon |T\ra \equiv (w-\id)|T\ra=i\mat{cc}{\epsilon_z & \epsilon_- \\ \epsilon_+ & -\epsilon_z}|T\ra\,,
\label{eq:spinor_transformation}
\ee
where we recall that $w=\id+ i\vec{\epsilon}\cdot \vec{\sigma}$ is an infinitesimal $\SU(2)$ group element. As shown in \cite{Dupuis:2014fya}, the only two independent solutions (up to normalization) to equate the RHS of \eqref{eq:spinor_transformation} with the RHS of \eqref{eq:SU2_L} are $|t\ra$ and its dual $|t]$ defined as 
\be 
|t\ra = \mat{c}{t_{-}\\ t_{+}} =\mat{c}{e^{\f{\ka N_1}{4}}\zeta^\ka_0 \\ e^{-\f{\ka N_0}{4}}\zeta^\ka_1}\,,
\qquad
|t]=\mat{cc}{0&-1\\1&0} |\overline{t}\ra = \mat{c}{-\tb_{+}\\  \tb_{-}}
=\mat{c}{-e^{-\f{\ka N_0}{4}}\bzeta^\ka_1 \\ e^{\f{\ka N_1}{4}}\bzeta^\ka_0}\,.
\label{eq:def_t}
\ee
The norm is a function of  the non-deformed norm $N$,
\be
\la t|t \ra =[t|t] = \f{2}{\kappa}\sinh\left( \f{\kappa}{2}N \right) \,.
\nn\ee
The Poisson brackets of the components are 
\be\ba{lll}
\ba{l}
\{t_-,t_+\}=\f{i\kappa}{2}\,t_-\,t_+
\,,\\[0.15cm]
\{\tb_-,\tb_+\}=-\f{i\kappa}{2}\,\tb_-\,\tb_+\,,
\ea
&
\ba{l}
\{t_-,\tb_-\}=\f{i\ka}{2}\left(t_-\,\tb_- -\f{2}{\ka} e^{\f{\kappa}{2}N} \right)
\,,\\[0.15cm]
\{t_+,\tb_+\}=-\f{i\ka}{2}\left( t_+\,\tb_+ +\f{2}{\ka} e^{-\f{\kappa}{2}N} \right)\,,
\ea
&
\{t_-,\tb_+\}=\{t_+,\tb_-\}=0\,.
\ea
\label{eq:poisson_t}
\ee

\paragraph*{\textbf{Braided covariant spinor.}} 
The spinor $|t\rangle$ can be ``parallelly transported'' by $\ell^{-1}$, which produces another spinor, whose transformation law under $\SU(2)$ is called \emph{braided}. Explicitly, using \eqref{eq:defi_lambda_z}, we have 
\be 
\ell^{-1}\,|t\ra
=
\mat{cc}{\lambda^{-1} & 0\\ -z & \lambda}\,
\mat{c}{e^{\f{\ka N_1}{4}}\zeta^\ka_0 \\ e^{-\f{\ka N_0}{4}}\zeta^\ka_1}
=
e^{\f{\ka N}{4}}\mat{c}{e^{-\f{\ka N_1}{4}}\zeta_0^\ka \\e^{\f{\ka N_0}{4}}\zeta^\ka_1} \,,
\ee
which prompts the definition of the following spinor\footnote{It differs from the spinor $|\tau\rangle$ of \cite{Dupuis:2014fya} by its normalization.},
\be
|\tau\ra \equiv 
e^{-\f{\ka N}{4}} \ell^{-1}\,|t\ra.
\label{eq:braided-cov}
\ee
The dual of $|\tau\ra$ is 
\be
|\tau]
=\mat{cc}{0&-1\\1&0} |\btau \ra = \mat{c}{-\btau_{+}\\  \btau_{-}}
=\mat{c}{-e^{\f{\ka N_0}{4}}\bzeta^\ka_1 \\  e^{-\f{\ka N_1}{4}}\bzeta_0^\ka}
=e^{\f{\kappa}{4}N}\ell^{-1}|t]\,.
\nn\ee

On the other hand, as the ribbon structure can be equivalently represented by either $\ell$ or $\ell^{-1\,\dagger}$ as shown in \eqref{eq:ribbon_constraintbis}, one expects that 
$|\tau\ra$ can also be defined 
by parallelly transporting $|t\ra$ with $\ell^\dagger$.
This is indeed the case, 
\be\label{eq:otherspinor}
|\tau\ra = e^{\f{\ka N}{4}} \ell^\dagger |t\ra\,, \quad |\tau] = e^{-\f{\ka N}{4}}\ell^\dagger |t]\,.
\ee 
Hence whether we use $\ell$ or $\ell^{-1\dagger}$ we get essentially the same object.

\medskip

The Poisson brackets of the components of $|\tau\ra$ are the same as those of $|t\ra$ and $|t]$ with $\tau_A$ replacing $t_{-A}$ and $\btau_A$ replacing $\tb_{-A}$, $\ie$
\be
\ba{l}
\{\tau_-,\tau_+\}= -\f{i\ka}{2}\tau_-\tau_+\,, \\
\{\btau_-,\btau_+\}= \f{i\ka}{2}\btau_-\btau_+\,,
\ea\quad
\ba{l}
\{\tau_-,\btau_-\}= -\f{i\ka}{2}(\tau_-\btau_- +\f{2}{\ka}e^{-\f{\ka}{2} N}) \,, \\
\{\tau_+,\btau_+\}= \f{i\ka}{2}(\tau_+\btau_+ -\f{2}{\ka}e^{\f{\ka}{2} N} )\,,
\ea\quad
\{\tau_-,\btau_+\}= \{\btau_-,\tau_+\} =0\,.
\label{eq:poisson_tau}
\ee
It will also be useful to compute the Poisson brackets between $\{t_A,\bt_A\}$ and $\{\tau_A,\btau_A\}$. They give
\be\ba{llll}
\{t_-, \btau_-\} = -i\cosh\f{\ka N_0}{2}\,, &
\{t_-, \btau_+\} = -\f{i\ka}{2}t_-\btau_+\,,&
\{t_+, \btau_-\} = \f{i\ka}{2}t_+\btau_-\,,&
\{t_+, \btau_+\} = -i\cosh\f{\ka N_1}{2}\,, \\[0.15cm]
\{\tb_-, \tau_-\} = i\cosh\f{\ka N_0}{2}\,,&
\{\tb_-, \tau_+\} = \f{i\ka}{2}\tb_-\tau_+\,,&
\{\tb_+, \tau_-\}=-\f{i\ka}{2}\tb_+\tau_-\,,&
\{\tb_+, \tau_+\}=i\cosh\f{\ka N_1}{2}\,,\\[0.15cm]
\{t_A,\tau_B\}=0\,,&
\{\tb_A,\btau_B\}=0\,.
\ea
\label{eq:poisson_t_tau}
\ee

$|\tau\rangle$ defines what we call a \emph{braided spinor}. Indeed, it transforms as a spinor under the $\SU(2)$ transformations generated by \eqref{eq:SU2_L}, but with a group element $w'$ related to $w$ through $\ell$.  
Since  triangular matrices are not stable under conjugation by $\SU(2)$ group elements, we need to introduce another $\SU(2)$ group element to stabilize the transformation. Let $^{(w)}\ell\in \AN(2)$ and $w'\in \SU(2)$ be defined by the Iwasawa decomposition
\begin{equation}
w\ell =^{(w)}\!\ell w'\,.
\end{equation}
Then we say that $\ell$ transforms as
\begin{equation}
\ell\quad \xrightarrow{w\in \SU(2)} \quad
 ^{(w)}\!\ell 
 = w\ell w'^{-1}\in\AN(2)\,.
\end{equation}
Going at the infinitesimal level \cite{Bonzom:2014wva},
\be
 w\sim \id +i \vec{\epsilon}\cdot \vec{\sigma} 
 =\id +i\mat{cc}{\epsilon_z & \epsilon_- \\\epsilon_+ &-\epsilon_z}\,,\quad
 w'\sim \id +i \vec{\epsilon'}\cdot \vec{\sigma} 
 =\id +i\mat{cc}{\epsilon'_z & \epsilon'_- \\\epsilon'_+ &-\epsilon'_z}\,,
\ee
the relation between $\vec{\epsilon}$ and $\vec{\epsilon'}$ is given by
\be\ba{lll}
\epsilon'_\pm  &=& \lambda^{-2} \epsilon_\pm\,, \\[0.2cm]
\epsilon'_z &=& \epsilon_z +\f12 (\lambda^{-1}z\epsilon_- +\lambda^{-1}\zb \epsilon_+ )\,.
\ea\ee
One can then check that, remarkably, the transformation generated by \eqref{eq:SU2_L} is a rotation by (the infinitesimal version of) $w'$
\be
\delta_{\epsilon}|\tau\ra = -\lambda^{-2} \kappa^{-1} \{\tr WX, |\tau\ra \} 
=-\lambda^{-2}\ka^{-1} \{2\epsilon_z \lambda^{2} + \epsilon_-\lambda z +\epsilon_+ \lambda \zb ,|\tau\ra \} 
= i\mat{c}{\epsilon'_z \tau_- +\epsilon'_- \tau_+ \\ \epsilon'_+ \tau_- - \epsilon'_z \tau_+}\sim(w'-\id)|\tau\ra\,,
\label{eq:SU2_tau}
\ee
$|\tau]$ is also a braided covariant spinor. The transformation \eqref{eq:SU2_tau} can also be written as a non-braided one, but generated with $X^{\text{op}}:= \ell^{\dagger}\ell$ instead of $X$, 
\be
\delta_{\epsilon}|\tau\ra
= \lambda^2 \ka^{-1} \{ \tr W' (X^{\text{op}})^{-1}, |\tau\ra  \} 
= \lambda^2 \ka^{-1} \{ 2\epsilon'_z \lambda^{-2} -\epsilon'_-\lambda^{-1} z-\epsilon'_+\lambda^{-1}\zb,|\tau\ra \}
=(w'-\id)|\tau\ra\,,
\label{eq:SU2_tau_2}
\ee
with $W'=\epsilon'_z (\id +\sigma_z)+ \epsilon'_-\sigma_+ +\epsilon'_+\sigma_-$. 
 $\{|t\ra\,,|t]\}$ and $\{|\tau\ra\,,|\tau]\}$ can viewed as two orthogonal complete basis of the space $\C^2\otimes \C^2$. We have seen the orthogonality $[t|t\ra=[\tau|\tau\ra=0$ above. Their completeness is guaranteed by the fact that
\be
|t\ra\la t| +|t][t| =\la t|t \ra \mat{cc}{1&0\\0&1} \equiv 
\la \tau|\tau \ra \mat{cc}{1&0\\0&1} =|\tau\ra\la\tau|+|\tau][\tau|\,.
\label{eq:complete}
\ee

\subsection{The tilde spinors}
\label{sec:tilde_covariant_spinors}

Covariant spinors and braided covariant spinors for the tilde sector, the ``tilde covariant spinors", are defined in a similar way as the non-tilde ones. We have 
\be\ba{ll}
|\tt\ra= \iota(|t\rangle) =
\mat{c}{\tt_-\\ \tt_+ } =
\mat{c}{e^{\f{\ka \Nt_1}{4}}\tzeta^\ka_0 \\ e^{-\f{\ka \Nt_0}{4}}\tzeta^\ka_1}
\,,\quad & 
|\tt]=\mat{cc}{0&-1\\1&0}|\btt\ra
=\mat{c}{-\btt_+\\\btt_- }
=\mat{c}{-e^{-\f{\ka \Nt_0}{4}}\btzeta^\ka_1 \\ e^{\f{\ka \Nt_1}{4}}\btzeta^\ka_0} \,,\\
|\ttau\ra= \iota(|\tau\rangle) = 
\mat{c}{\ttau_- \\ \ttau_+}=
\mat{c}{e^{-\f{\ka \Nt_1}{4}}\tzeta^\ka_0 \\ e^{\f{\ka \Nt_0}{4}}\tzeta^\ka_1}
\,,\quad &
|\ttau]=\mat{cc}{0&-1\\1&0}|\bttau\ra 
=\mat{c}{-\bttau_+ \\ \bttau_- }
=\mat{c}{ -e^{\f{\ka \Nt_0}{4}}\btzeta^\ka_1 \\ e^{-\f{\ka \Nt_1}{4}}\btzeta^\ka_0 }\,,
\ea
\label{eq:tt_ttau}
\ee
whose norms are given by
\be
\la\tt|\tt\ra=[\tt|\tt]=
\la\ttau|\ttau\ra =[\ttau|\ttau]=\f{2}{\ka}\sinh \lb \f{\ka}{2}\Nt \rb\,.
\ee
They are independent of the non-tilde spinors, $\ie$all the components Poisson commute with those of the non-tilde spinors.  
The Poisson brackets of the tilde spinor components are the same as the non-tilde ones:
\begin{align}
&\ba{l}
\{\tt_-,\tt_+\}= \f{i\ka}{2}\tt_-\tt_+\,, \\[0.15cm]
\{\btt_-,\btt_+\}= - \f{i\ka}{2}\btt_-\btt_+\,, 
\ea &&
\ba{l}
\{\tt_-,\btt_-\}= \f{i\ka}{2}\left(\tt_-\,\btt_- -\f{2}{\ka} e^{\f{\kappa}{2}N} \right)
 \,,\\[0.15cm]
\{\tt_+,\btt_+\}= -\f{i\ka}{2}\left(\tt_+\,\btt_+ +\f{2}{\ka} e^{-\f{\kappa}{2}N} \right)
\,,
\ea &&
\{\tt_-,\btt_+\}=
\{\btt_-,\tt_+\}=0 \,,
\label{eq:poisson_tt}
\\[0.2cm]
&\ba{l}
\{\ttau_-,\ttau_+\}= -\f{i\kappa}{2}\,\ttau_-\,\ttau_+\,, \\[0.15cm]
\{\bttau_-,\bttau_+\}= \f{i\ka}{2}\bttau_-\bttau_+\,,
\ea &&
\ba{l}
\{\ttau_-,\bttau_-\}= -\f{i\ka}{2}\lb \ttau_-\bttau_- +\f{2}{\ka}e^{-\f{\ka}{2}\Nt} \rb\,, \\[0.15cm]
\{\ttau_+,\bttau_+\}=  \f{i\ka}{2}\lb \ttau_+\bttau_+ -\f{2}{\ka}e^{\f{\ka}{2} \Nt} \rb\,, 
\ea &&
\ba{l}
\{\ttau_-,\bttau_+\}= \{\bttau_-,\ttau_+\}= 0\,.
\ea 
\label{eq:poisson_ttau}
\end{align}

Note however that $\lt$ is not the same function of $\tilde{\zeta}^{\kappa}_0, \tilde{\zeta}^{\kappa}_1$ as $\ell$ is of $\zeta^{\kappa}_0, \zeta^{\kappa}_1$, see \eqref{eq:defi_lambda_z}, \eqref{eq:l_lt_param}. In fact , we have $\iota(\ell) = \tell^{-1}$ where we recall that $\iota$ defined in \eqref{iotazeta} is an operator which adds tildes to $\zeta^\kappa_{0,1}$ and their complex conjugates. As a consequence, the relation between $|\tt\rangle$ and $|\ttau\rangle$ is not obtained by adding tildes to $|\tau\rangle = e^{-\frac{\kappa N}{4}}\ell^{-1}|t\rangle = e^{\frac{\kappa N}{4}}\ell^{\dagger}|t\rangle$. Instead we act with $\iota$ to get
\begin{equation} \label{eq:tilde_braided_covariant}
|\ttau\rangle = e^{-\frac{\kappa \tilde{N}}{4}} \tell |\tt\rangle = e^{\frac{\kappa \tilde{N}}{4}} \tell^{-1\dagger} |\tt\rangle.
\end{equation}

Since the Poisson brackets of the tilde spinors are the same (with tildes) as the non-tilde ones, the generator of $\SU(2)$ transformations for the tilde spinors is given by 
\begin{equation}
\iota\bigl( -\lambda^{-2} \kappa^{-1} \tr (WX)\bigr) = -\tilde{\lambda}^2 \kappa^{-1} \tr W (\Xt^{\text{op}})^{-1}\,,
\end{equation}
where $\Xt^{\text{op}} = \tell^\dagger \tell$. 
This is consistent with the Gauss constraint \eqref{eq:Gauss_flatness_constraint}, which is a product of $\ell$ and $\lt^{-1}$ depending on the orientations of the ribbons.  Explicitly, we can expand $\tr W(\Xt^{\text{op}})^{-1}$ as
\be
\tr W(\Xt^{\text{op}})^{-1} = 2\epsilon_z \tlambda^{-2} - \epsilon_-\tlambda^{-1}\tz - \epsilon_+\tlambda^{-1}\btz\,.
\ee
It is straightforward then using the Poisson brackets from Appendix \ref{app:Poisson_bracket_SL2C} to show that the tilde spinors \eqref{eq:tt_ttau} satisfy the following equations 
\begin{align}
&\delta_{\epsilon} |\tt\ra 
= -\tlambda^{2} \ka^{-1} \{\tr W(\Xt^{\text{op}})^{-1}, |\tt\ra \} 
=(w-\id)|\tt\ra\,, \quad
&&\delta_{\epsilon} |\tt] 
= -\tlambda^{2} \ka^{-1} \{\tr W(\Xt^{\text{op}})^{-1}, |\tt] \} 
=(w-\id)|\tt]\,,
\label{eq:SU2_tt} \\
&\delta_{\epsilon} |\ttau\ra 
=-\tlambda^{2} \ka^{-1} \{\tr W(\Xt^{\text{op}})^{-1}, |\ttau\ra \}
=(w''-\id)|\ttau\ra \,, \quad
&&\delta_{\epsilon} |\ttau] 
=-\tlambda^{2} \ka^{-1} \{\tr W(\Xt^{\text{op}})^{-1}, |\ttau] \}
=(w''-\id)|\ttau] \,,
\label{eq:SU2_ttau}
\end{align}
where the infinitesimal $\SU(2)$ elements $w=\id +i \vec{\epsilon}\cdot \vec{\sigma}$ and $w''=\id +i \vec{\epsilon''}\cdot \vec{\sigma}$ are related by the right $\SU(2)$ transformation of $\lt$, $\ie$
\be
\lt\quad \xrightarrow{w\in \SU(2)} \quad
 \lt^{(w)} = w''^{-1}\lt w\in\AN(2)\,, \quad
 w= \id +i \vec{\epsilon}\cdot \vec{\sigma} 
 =\id +i\mat{cc}{\epsilon_z & \epsilon_- \\\epsilon_+ &-\epsilon_z}\,,\quad
 w''= \id +i \vec{\epsilon''}\cdot \vec{\sigma} 
 =\id +i\mat{cc}{\epsilon''_z & \epsilon''_- \\\epsilon''_+ &-\epsilon''_z}\,.
\ee
Thus the two infinitesimal parameters $\vec{\epsilon}$ and $\vec{\epsilon''}$ are related by
\be\ba{lll}
\epsilon''_\pm  &=& \tlambda^{2} \epsilon_\pm\,, \\[0.2cm]
\epsilon''_z &=& \epsilon_z -1/2 (\tlambda z\epsilon_- +\lambda \btz \epsilon_+ )\,.
\ea\ee

Just like there are two ways to write the transformations of $|\tau\rangle$ and $|\tau]$, there are also two for $|\ttau\rangle$ and $|\ttau]$. While we have seen above the equivalent of \eqref{eq:SU2_tau}, the equivalent of \eqref{eq:SU2_tau_2} is
\be
\delta_\epsilon|\ttau\ra = \iota\bigl(\lambda^{2}\kappa^{-1}) \{ \tr W'' \iota(X^{\text{op}-1}), |\ttau\rangle\} = \tlambda^{-2} \ka^{-1} \{\tr W''\Xt, |\ttau\ra \}\,, \quad
\delta_\epsilon|\ttau] =\tlambda^{-2} \ka^{-1} \{\tr W''\Xt, |\ttau] \}\,,
\ee
and it is clear that $|\ttau\ra$ and $|\ttau]$ are braided spinors in the same sense as $|\tau\rangle, |\tau]$.

There is a nice geometric interpretation of the relations \eqref{eq:braided-cov} and \eqref{eq:tilde_braided_covariant} which define the braided covariant spinors. If we consider $|t\rangle$ to sit at a vertex of $\Gamma_{\text{rib}}$ which is the target of the short side carrying $\ell$, then $|\tau\rangle = e^{-\f{\ka N}{4}} \ell^{-1}\,|t\ra$ sits on the vertex of $\Gamma_{\text{rib}}$ at the source of the short side carrying $\ell$. In other words, $|\tau\rangle$ results from parallelly transporting $|t\rangle$ by $\ell^{-1}$. Similarly $|\tilde{\tau}\rangle$ is the result of parallelly transporting $|\tilde{t}\rangle$ by $\tell$. This is represented in Figure \ref{fig:kappa_spinors}.

\subsection{Recovering the holonomy-flux variables from the spinors}
\label{sec:Recovering}

We assign the four spinors $|t\ra$, $|\tau\ra$, $|\ttau]$, $|\tt]$ to the corners of the ribbon edge as in  Figure \ref{fig:kappa_spinors}. We assume the norm matching condition $N=\Nt$ so that the tilde spinors and their corresponding non-tilde spinors have the same norm:
\be
\la t|t \ra = [ \ttau| \ttau ] 
= \f{2}{\ka}\sinh\f{\ka N}{2}\,.
\label{eq:norm_condition}
\ee
The holonomies $u,\ut\in \SU(2)$ can be parametrized in terms of these spinors
\be
u=
\f{|\tau \ra [ \tt|-|\tau]\la \tt|}{\sqrt{\la \tau|\tau \ra \la \tt|\tt\ra}}\,,\quad
\ut=
\f{|t\ra [\ttau|-|t] \la \ttau|}{\sqrt{\la t|t\ra\la\ttau|\ttau\ra}}\,,\quad
\quad
\textrm{ with } N=\tilde N\,,
\label{eq:u_from_spinor}
\ee
so that the following parallel transport relations along the long sides of the ribbon are satisfied,
\be\ba{llll}
u |\tt ] = |\tau\ra\,, &
u|\tt\ra=-|\tau]\,, & 
u^{-1}|\tau\ra=|\tt]\,, &
u^{-1}|\tau]=-|\tt\ra\,, \\[0.15cm]
\ut |\ttau\ra =- |t]\,, & 
\ut |\ttau] = |t\ra\,, &
\ut^{-1} |t] = -|\ttau\ra\,, &
\ut^{-1} |t\ra = |\ttau]\,.
\label{eq:u_acton_spinors}
\ea\ee
 On the other hand, the fluxes $\ell\,,\lt\in \AN(2)$ can also be reconstructed by the deformed spinors as
\be
\ell=\f{e^{-\f{\ka N}{4}}|t\ra\la \tau| + e^{\f{\ka N}{4}}|t][\tau|}{\sqrt{\la t|t\ra \la \tau|\tau\ra}}\,,\quad
\lt=\f{e^{\f{\ka \Nt}{4}}|\ttau\ra\la \tt| + e^{-\f{\ka \Nt}{4}}|\ttau][\tt|}{\sqrt{\la \tt|\tt\ra \la \ttau|\ttau\ra}}\,.
\label{eq:flux_from_spinor}
\ee
Their inverses
\be
\ell^{-1}=\f{e^{\f{\ka N}{4}}|\tau\ra\la t| + e^{-\f{\ka N}{4}}|\tau][t|}{\sqrt{\la t|t\ra \la \tau|\tau\ra}}\,,\quad
\lt^{-1}=\f{e^{-\f{\ka \Nt}{4}}|\tt\ra\la \ttau| + e^{\f{\ka \Nt}{4}}|\tt][\ttau|}{\sqrt{\la \tt|\tt\ra \la \ttau|\ttau\ra}}
\label{eq:inverse_flux_from_spinor}
\ee
can be checked by the orthogonality and completeness \eqref{eq:complete} of the two bases $\{|t\ra\,,|t]\}$ and $\{|\tau\ra\,,|\tau]\}$. Likewise for the tilde sectors.  
The parallel transport relations between the spinors and braided spinors by the fluxes can be perfectly reflected by \eqref{eq:flux_from_spinor} and \eqref{eq:inverse_flux_from_spinor}.

Therefore, the spinor assignment of Figure~\ref{fig:kappa_spinors} fully illustrates the parallel transport relations of the four spinors. Finally, they solve the ribbon constraint:
\be
\left\{\ba{l}
\ell u|\tt ] = \ell|\tau\ra = e^{-\frac{\kappa N}{4}}|t\ra \\
\ut\lt |\tt] = e^{-\frac{\kappa N}{4}} \ut | \ttau ] = e^{-\frac{\kappa N}{4}}|t\ra
\ea\right.
\quad\Longrightarrow \quad
\cC\equiv \ell u \lt^{-1}\ut^{-1}=\id \,,
\ee
and the same can be done with the equivalent ribbon constraint $\ell^{-1\dagger} u \lt^{\dagger}\ut^{-1}=\id$. These spinors thus live on the constraint surface generated by the ribbon constraint $\cC$. The matrix components defined in \eqref{eq:u_from_spinor} also satisfy the desired Poisson brackets (see \eqref{eq:poisson_matrix_elements}).

As shown in \cite{Dupuis:2014fya}, the phase space $\SL(2,\bC)$ with the Poisson structure \eqref{eq:bivector_SL2C} for one ribbon is equivalent to $\cS_\ka \times \tilde{\cS}_\ka//\cM$, with $\cS_\ka=\{|t\ra\in\bC^2\backslash\{ \la t|t\ra=0\}\}$ the spinor phase space with the Poisson structure \eqref{eq:poisson_t}, $\tilde{\cS}_\ka=\{|\tt\ra\in\bC^2\backslash\{ \la \tt|\tt\ra=0\}\}$ the phase space with the Poisson structure \eqref{eq:poisson_tt}, and $\cM:=N-\Nt$ the norm matching constraint. It is a simple check that the dimension of such a phase space is $8-2=6$, matching that of the holonomy-flux phase space constructed in Section \ref{sec:holonomy_flux_phase_space}.

\subsection{\texorpdfstring{$\ka\rightarrow 0$}{kappa to 0} limit: the non-deformed phase space}
We furthermore write the fluxes in terms of the spinors. Consider the Hermitian matrices $\ell\ell^\dagger =X \equiv \ka X_0\id-\ka \vec{X}\cdot \vec{\sigma}$  and $\ell^\dagger\ell\equiv X^{\text{op}}= \ka X^{\text{op}}_0\id -\ka \vec{X}^{\text{op}}\cdot\vec{\sigma}$. Their components can be  represented in term of  the spinors $|t\ra$ and $|\tau\ra$, 
\be
 X_0=\f{1}{\ka}\sqrt{1+\f{\ka^2}{4} \la t|t \ra^2}= X^{\op}_0 = 
\f{1}{\ka}\sqrt{1+\f{\ka^2}{4} \la \tau|\tau\ra^2}, \quad  \vec{X}= \f12\la t | \vec{\sigma}|t\ra,  \quad \vec{X}^{\op}=\f12 \la \tau | \vec{\sigma}|\tau \ra.
\label{eq:def_X_Xop}
\ee
Similarly, $\lt\lt^\dagger\equiv\Xt = \ka\Xt_0\id-\ka\vec{\Xt}\cdot \vec{\sigma}$ and $\lt^\dagger\lt\equiv\Xt^{\op}=\ka\Xt^{\op}_0\id-\ka\vec{\Xt}^{\op}\cdot \vec{\sigma}$ can be written with the tilde spinors. Explicitly,
\be
\Xt_0=\f{1}{\ka}\sqrt{1+\f{\ka^2}{4}[\ttau|\ttau]^2}=\Xt_0^\op=\f{1}{\ka}\sqrt{1+\f{\ka^2}{4}[\tt|\tt]^2}\,,\quad
\vec{\Xt}=\f12[\ttau|\vec{\sigma}|\ttau]\,,\quad
\vec{\Xt}^{\op}=\f12[\tt|\vec{\sigma}|\tt]\,.
\label{eq:def_Xt_Xtop_spinor}
\ee
These objects transform as vectors under the $\SU(2)$ transformation as $\Xt=\ut^{-1} X\ut$ and $u^{-1}X^{\text{op}}u=\Xt^{\text{op}}$, consistently with \eqref{eq:u_acton_spinors}, and as such can be seen as the deformation of the flat flux vectors. They capture the hyperbolic geometry of the discretization of $\Sigma$ \cite{Bonzom:2014wva}. In particular, the Gauss constraint for a three-valent node encodes the closure of a hyperbolic triangle, whose side lengths and angles can be fully characterized in terms of the vectors $X$s or $X^{\text{op}}$s associated to the corresponding sides (see \cite{Bonzom:2014wva}).

When $\ka\rightarrow 0$, $|t\ra,|\tt]$ is identical to $|\tau\ra,|\ttau]$ respectively, as it can be directly seen from their definition \eqref{eq:def_t}, \eqref{eq:braided-cov} and \eqref{eq:tt_ttau}. We recover then the flat case where there is only one pair of spinors associated to each edge. The flux vectors $\vec{X}$ and $\vec{\Xt}$ become the standard flat flux vectors which we denote $\vx$ and $\vtx$ respectively. As a consistency check, one can take the $\ka\rightarrow 0$ limit for $X$ \eqref{eq:def_X_Xop} and $\Xt$ \eqref{eq:def_Xt_Xtop_spinor} defined in terms of the spinors, or more explicitly in terms of the $\ka$-deformed spinor variables as in \eqref{eq:defi_lambda_z}. Let us rewrite
\be\ba{lll}
X=\mat{cc}{\lambda^2 & \lambda \zb\\ \lambda z & \lambda^{-2}+|z|^2}
&=&
\mat{cc}{e^{\f{\ka (N_1-N_0)}{2}} & -\ka e^{\f{\ka(N_1-N_0)}{4}}\zeta^{\ka}_0\bzeta^{\ka}_1 \\ 
-\ka e^{\f{\ka(N_1-N_0)}{4}}\zeta^{\ka}_1\bzeta^{\ka}_0 & e^{\f{\ka(N_0-N_1)}{2}}+4\sinh\f{\ka N_1}{2}\sinh\f{\ka N_0}{2} }
\\[0.5cm]
&\xrightarrow{\ka\rightarrow 0}&
\mat{cc}{1+ \f{\ka (N_1-N_0)}{2} & -\ka \zeta_0\bzeta_1 \\
-\ka \zeta_1\bzeta_0 & 1+ \f{\ka (N_0-N_1)}{2}}
=(1+\f{\ka N}{2})\id -\ka \vx\cdot \vec{\sigma}\,,\\[0.7cm]
\Xt=\mat{cc}{\tlambda^2 & \tlambda \btz \\ \tlambda \tz & \tlambda^{-2}+|\tz|^2}
&=&
\mat{cc}{e^{\f{\ka (\Nt_0-\Nt_1)}{2}} & \ka e^{\f{\ka(\Nt_0-\Nt_1)}{4}}\tzeta^{\ka}_0\btzeta^{\ka}_1 \\ 
\ka e^{\f{\ka(\Nt_0-\Nt_1)}{4}}\tzeta^{\ka}_1\btzeta^{\ka}_0 & e^{\f{\ka(\Nt_1-\Nt_0)}{2}}+4\sinh\f{\ka \Nt_1}{2}\sinh\f{\ka \Nt_0}{2} }\\[0.5cm]
&\xrightarrow{\ka\rightarrow 0}&
\mat{cc}{1+ \f{\ka (\Nt_0-\Nt_1)}{2} & \ka \tzeta_0\btzeta_1 \\
\ka \tzeta_1\btzeta_0 & 1+ \f{\ka (\Nt_1-\Nt_0)}{2}}
=(1+\f{\ka \Nt}{2})\id -\ka \vtx\cdot \vec{\sigma}\,.
\ea\ee 
where $\vx:=\f12\la \zeta|\vec{\sigma}|\zeta\ra$ and $\vtx:=\f12[\tzeta|\vec{\sigma}|\tzeta]$. 
On the other hand, $u\stackrel{N=\Nt}{\simeq}\ut\equiv g$. 
The flat limit of the holonomy and the flux vector components can be checked to satisfy the Poisson brackets \cite{Dupuis:2019yds}
\be\ba{lll}
\{x^i,g\}=\frac{i}{2} \sigma^ig \,,&
\{x^i,x^j\}=\epsilon^{ijk}x^k\,,&
\{g,g\}\stackrel{N=\Nt}{\simeq}0\,,\\[0.15cm]
\{\tx^i,g\}=\frac{i}{2}g\sigma^i \,,&
\{\tx^i,\tx^j\}=-\epsilon^{ijk}\tx^k\,,&
\{x^i,\tx^j\}=0\,.
\ea\ee
Therefore, the $\ka\rightarrow 0$ limit of the flux vectors $\vec{X}$ and $\vec{\Xt}$ \eqref{eq:def_X_Xop} recover the flat fluxes 
\be
\vec{X}=-\f{1}{2\ka}\tr(X\vec{\sigma})\xrightarrow{\ka\rightarrow 0} \vx\,,\quad
\vec{\Xt}=-\f{1}{2\ka}\tr(\Xt\vec{\sigma})\xrightarrow{\ka\rightarrow 0} \vtx\,.
\ee
The same limit can be achieved for $\vec{X}^{\op}$ and $\vec{\Xt}^{\op}$ as $|t\ra\xleftrightarrow{\ka\rightarrow 0}|\tau\ra$ and $|\tt]\xleftrightarrow{\ka\rightarrow 0}|\ttau]$.

\medskip

\section{Spinorial observables}
\label{sec:scalar_product}

\subsection{The spinorial phase space}
For a given graph $\Gamma$, we take the Cartesian product of the spinor phase spaces over all edges of $\Gamma$. An edge $e$ carries the spinors $|t_e\rangle, |\tau_e\rangle, |\tt_e\rangle, |\ttau_e\rangle$ and their duals. We have already seen in Section \ref{sec:Recovering} that those variables reconstruct the holonomy-flux variables in a way which automatically solves the ribbon constraint in each ribbon. We are thus left with imposing the Gauss constraint at each vertex of $\Gamma$. 

Let us consider an $n$-valent vertex $v$ of $\Gamma$. We then pick an arbitrary edge incident to it, which we denote by $e_1$, and then going counterclockwise starting from $e_1$, we label the other incident edges by   $e_2,\cdots,e_n$ and identify $e_{n+1}\equiv e_1$. In the ribbon graph $\Gamma_{\text{rib}}$, $v$ gives rise to an $n$-gon $R(v)$ and each edge $e_i$ to a ribbon edge $R(e_i)$. Each of them shares a vertex with its two neighbor ribbons, one clockwise and one counter-clockwise.

It is convenient to unify the notation for spinors as follows,
\be
\ba{llll}
t^-=|t\ra\, & 
t^+=|t]\, &
\tau^-=|\tau\ra \,&
\tau^+=|\tau] \\
\tt^-=|\tt\ra \,& 
\tt^+=|\tt] \,&
\ttau^-=|\ttau\ra\, &
\ttau^+=|\ttau]
\ea
\ee
or component-wise
\be
\ba{llll}
t^-_A=t_A\,, \quad &
t^+_A=(-1)^{\f12-A}\tb_{-A}\,,\quad &
\tau^-_A=\tau_A\,,\quad &
\tau^+_A=(-1)^{\f12-A}\btau_{-A}\,,\quad \\
\tt^-_A=\tt_A\,, \quad &
\tt^+_A=(-1)^{\f12-A}\btt_{-A}\,,\quad &
\ttau^-_A=\ttau_A\,,\quad &
\ttau^+_A=(-1)^{\f12-A}\bttau_{-A}\,,
A=\pm \f12\,.
\ea\ee
We use the same notation as in \eqref{eq:Gauss_flatness_constraint} to denote the fluxes on the boundary edges of $R(v)$ as $\ell_{e_i,v}$. 
Denote the spinor sitting at the source vertex of $\ell_{e_i,v}$ to be $t^\epsilon_{e_iv}$ and that sitting at its target is $\tau^\epsilon_{e_iv}$. Referring to Figure \ref{fig:kappa_spinors}, they are explicitly
\be \label{SpinorsNotationsAtVertex}
\left|\ba{l} 
t^\epsilon_{e_iv} = t^\epsilon_i\\
\tau^\epsilon_{e_iv} = \tau^\epsilon_i\\
\ea\right.
\text{if }
o_i=+\,,\quad
\left|\ba{l}
t^\epsilon_{e_iv} = \tt^\epsilon_i\\
\tau^\epsilon_{e_iv} = \ttau^\epsilon_i\\
\ea\right.
\text{if }
o_i=-\,.
\ee
Indeed, each vertex in $\Gamma_{\text{rib}}$ is assigned two spinors from two different ribbons. For instance, the spinors $\tau^{\epsilon}_{e_{i+1}v}$ and $t^{\epsilon}_{e_iv}$ sit at the vertex where $\ell_{e_i,v}$ and $\ell_{e_{i+1},v}$ intersect. 
We now show in the following proposition that these two spinors, except $t^\epsilon_{e_n}$ sitting at the base vertex, are all braided-covariant under the $\SU(2)$ transformation generated by the Gauss constraint. 
\begin{prop} \label{prop:SU(2)LatticeGaugeGenerators}
The spinors $\tau^{\epsilon_{i+1}}_{e_{i+1},v}$ and $t^{\epsilon_i}_{e_i,v}$ $(i=1,\cdots,n)$ which sit on the same vertex of $\Gamma_{\text{rib}}$ are braided-covariant under the $\SU(2)$ transformation defined in \eqref{eq:generalgaugetrans} by the braided infinitesimal $\SU(2)$ parameter denoted by $w^{(i)}=\id +i\mat{cc}{\epsilon_z^{(i)} & \epsilon_-^{(i)} \\ \epsilon_+^{(i)} & \epsilon_z^{(i)}}$. If we parameterize $\ell_{e_i,v}=\mat{cc}{\Uplambda_i & 0 \\ \fz_i & \Uplambda_i^{-1}}$, then the transformation reads
\begin{subequations}
\begin{align}
\delta_\epsilon t^\epsilon_{e_iv} &= 
-\ka^{-1}\lb\prod_{k=1}^n\Uplambda_k^{-2}\rb \{\tr W\cG\cG^\dagger , t^\epsilon_{e_iv} \} = (w^{(i+1)}-\id)t^\epsilon_{e_iv}\,,\\
\delta_\epsilon \tau^\epsilon_{e_iv} &= 
-\ka^{-1}\lb\prod_{k=1}^n\Uplambda_k^{-2}\rb \{\tr W\cG\cG^\dagger , \tau^\epsilon_{e_iv} \} = (w^{(i)}-\id)\tau^\epsilon_{e_iv}\,,
\end{align}
\label{eq:SU2_transformation_anySpinor}
\end{subequations}
where parameters in $w^{(i)}$ are defined by induction as 
\be
\left|\ba{l}
\epsilon_\pm^{(i)} = \Uplambda_{i}^{-2} \epsilon^{(i+1)}_{\pm}\\[0.15cm]
\epsilon_z^{(i)} = \epsilon_z^{(i+1)} 
+ \f12 \lb \Uplambda_{i}^{-1}\fz_{i} \epsilon_-^{(i+1)} + \Uplambda_{i}^{-1}\fbz_{i} \epsilon_+^{(i+1)} \rb
\ea\right.\,,
\quad\text{and }\quad
\left|\ba{l}
\epsilon_\pm^{(n+1)} \equiv \epsilon_\pm\\[0.15cm]
\epsilon_z^{(n+1)} \equiv \epsilon_z
\ea\right.\,, \,i=1,\cdots,n\,.
\label{eq:braided_epsilon}
\ee
\end{prop}
\begin{proof}
We prove this proposition using the following induction result of the $\SU(2)$ transformation for any function $f$ from \cite{Bonzom:2014wva}\footnote{We use different conventions from \cite{Bonzom:2014wva} hence why the expressions look different.},
\be
-\ka^{-1}\lb\prod_{k=1}^n\Uplambda_k^{-2}\rb\{\tr W\cG\cG^\dagger,f\} \equiv 
-\ka^{-1}\sum_{k=1}^n \Uplambda_k^{-2}\tr W^{(k+1)}\{\ell_{e_k,v}\ell_{e_k,v}^\dagger,f\}\,,\quad
\text{with}\quad
\left|\ba{l}
W^{(k)} = -\Uplambda_k^{-2}\ell_{e_k,v}^\dagger W^{(k+1)} \ell_{e_k,v} \\
W^{(n+1)}\equiv W =\mat{cc}{2\epsilon_z & \epsilon_- \\ \epsilon_+ & 0}
\ea\right.\,,
\label{eq:SU2_transform_any_function_2}
\ee
and the Poisson brackets $\{\Uplambda_i^2,t_{e_iv,A}^\epsilon \}=(-1)^{\f12-A}\f{i\ka}{2}\Uplambda_i^2t_{e_iv,A}^\epsilon$, $\{\Uplambda_i^2,\tau_{e_iv,A}^\epsilon \}=(-1)^{\f12-A}\f{i\ka}{2}\Uplambda_i^2\tau_{e_iv,A}^\epsilon$ and
\be\ba{lll}
&\left|\ba{l}
\{\Uplambda_i \fz_i,t_{e_iv,-}^\epsilon \}=-i\ka\Uplambda_i^2 t_{e_iv,+}^\epsilon\,, \\
\{\Uplambda_i \fz_i,t_{e_iv,+}^\epsilon\}=0\,,
\ea\right.\quad
&\left|\ba{l}
\{\Uplambda_i \fbz_i ,t_{e_iv,-}^\epsilon \}=0 \\
\{\Uplambda_i \fbz_i ,t_{e_iv,+}^\epsilon \} =-i\ka\Uplambda_i^2t_{e_iv,-}^\epsilon \,,
\ea\right.\quad
\\[0.3cm]
&\left|\ba{l}
\{\Uplambda_i \fz_i,\tau_{e_iv,-}^\epsilon \}=-\f{i\ka}{2}\Uplambda_i \fz_i\tau_{e_iv,-}^\epsilon -i\ka\tau_+^\epsilon\,,\\
\{\Uplambda_i \fz_i,\tau_{e_iv,+}^\epsilon\}=\f{i\ka}{2}\Uplambda_i\fz_i \tau_{e_iv,+}^\epsilon\,,
\ea\right.\quad
&\left|\ba{l}
\{\Uplambda_i \fbz_i,\tau_{e_iv,-}^\epsilon\} =-\f{i\ka}{2}\Uplambda_i\fbz_i\tau_{e_iv,-}^\epsilon\,,\\
\{\Uplambda_i\fbz_i,\tau_{e_iv,+}^\epsilon \}=\f{i\ka}{2}\Uplambda_i \fbz_i\tau_{e_iv,+}^\epsilon -i\ka \tau_{e_iv,-}^\epsilon\,.
\ea\right. 
\ea
\label{eq:poisson_ti_taui}
\ee
The braided matrix $W^{(k)}$ reads explicitly 
\be
W^{(k)}=\mat{cc}{2\epsilon_z^{(k)} & \epsilon_-^{(k)} \\ \epsilon_+^{(k)} & 0} 
\ee
where the vector components of $\vec{\epsilon}^{(k)}$ are defined inductively in \eqref{eq:braided_epsilon} or explicitly
\be
\epsilon^{(k)}_\pm = \lb \prod_{i=k}^n \Uplambda_i^{-2} \rb\epsilon_\pm\,,\quad
\epsilon^{(k)}_z = \epsilon_z +\f12\sum_{i=k}^n\lb \prod_{j=i}^n \Uplambda_j^{-2}\rb \lb\epsilon_-\Uplambda_i\fz_i + \epsilon_+\Uplambda_i\fbz_i \rb\,.
\ee
Expanding the right-hand side of \eqref{eq:SU2_transform_any_function_2}, the $\SU(2)$ transformation for $t^\epsilon_{e_iv,A}$ is
\begin{align}
\delta_\epsilon t^\epsilon_{e_iv} 
&= -\ka^{-1} \Uplambda_i^{-2}\lb 2\epsilon_z^{(i-1)}\{\Uplambda_i^2,t^\epsilon_{e_iv}\} 
+\epsilon_-^{(i+1)}\{\Uplambda_i\fz_i,t^\epsilon_{e_iv} \}
+\epsilon_+^{(i+1)}\{\Uplambda_i\fbz_i,t^\epsilon_{e_iv} \} \rb
=\mat{cc}{\epsilon_z^{(i+1)}& \epsilon_-^{(i+1)}\\
 \epsilon_+^{(i+1)} &-\epsilon_z^{(i+1)} }t^\epsilon_{e_iv}\,,\\[0.2cm]
\delta_\epsilon \tau^\epsilon_{e_iv} 
&= -\ka^{-1} \Uplambda_i^{-2}\lb 2\epsilon_z^{(i+1)}\{\Uplambda_i^2,\tau^\epsilon_{e_iv}\} 
+\epsilon_-^{(i+1)}\{\Uplambda_i\fz_i,\tau^\epsilon_{e_iv} \}
+\epsilon_+^{(i+1)}\{\Uplambda_i\fbz_i,\tau^\epsilon_{e_iv} \} \rb
=\mat{cc}{\epsilon_z^{(i)}& \epsilon_-^{(i)}\\
 \epsilon_+^{(i)} &-\epsilon_z^{(i)} }\tau^\epsilon_{e_iv}\,,
 \end{align}
where the right-hand sides of both equations above are calculated via \eqref{eq:braided_epsilon} and \eqref{eq:poisson_ti_taui}. This proves \eqref{eq:SU2_transformation_anySpinor}.
\end{proof}

We will build local invariant quantities by taking scalar products between spinors from different edges which meet at the same vertex of $\Gamma$. Due to the ribbon structure, they might meet at the same vertex of $\Gamma_{\text{rib}}$ or at different vertices of $\Gamma_{\text{rib}}$. In the latter case, parallel transport around the ribbon vertex is required to evaluate the scalar product at a common vertex in $\Gamma_{\text{rib}}$. An example of the situation is given for a 3-valent vertex in Figure \ref{fig:three_valent_node}. 
One can form (quadratic) scalar products of spinors from two adjacent links $e_i$ and $e_{i+1}$. The symmetry transformation is induced at the vertex where the ribbons meet and if they sit at the same vertex, this ensures that the scalar product is invariant. One can also define observables for spinors not sitting at the same vertex. But in this case, it is necessary to parallel transport one spinor to the other in order to ensure invariance.

\medskip

\subsection{Invariants from spinors sitting at the same vertex in \texorpdfstring{$\Gamma_{\text{rib}}$}{the ribbon graph}}

The spinors $t^{\epsilon_i}_{e_iv}$ and $\tau^{\epsilon_{i+1}}_{e_{i+1}v}$ sit at the same vertex in $\Gamma_{\text{rib}}$. One can build directly quadratic observables denoted $E^{\epsilon_i,\epsilon_{i+1}}_{i,i+1}$ with these two spinors by forming their scalar products
\be
E^{\epsilon_i,\epsilon_{i+1}}_{i,i+1}=
\epsilon_i\sum_{A=\pm 1/2} (-1)^{\f12+A}t^{\epsilon_i}_{e_iv,-A}\tau^{\epsilon_{i+1}}_{e_{i+1}v,A}
=\begin{cases}
	\epsilon_{i}\sum_{A=\pm 1/2}(-1)^{\f12+A} t^{\epsilon_i}_{i,-A} \tau^{\epsilon_{i+1}}_{i+1,A}\,,\quad
	\text{ for }\quad o_i=o_{i+1}=1\\[0.15cm]
	\epsilon_{i}\sum_{A=\pm 1/2}(-1)^{\f12+A} t^{\epsilon_i}_{i,-A} \ttau^{\epsilon_{i+1}}_{i+1,A}\,,\quad
	\text{ for }\quad o_i=-o_{i+1}=1\\[0.15cm]
	\epsilon_{i}\sum_{A=\pm 1/2} (-1)^{\f12+A}\tt^{\epsilon_i}_{i,-A}\tau^{\epsilon_{i+1}}_{i+1,A}\,,\quad
	\text{ for }\quad -o_i=o_{i+1}=1\\[0.15cm]
	\epsilon_{i}\sum_{A=\pm 1/2}(-1)^{\f12+A} \tt^{\epsilon_i}_{i,-A}\ttau^{\epsilon_{i+1}}_{i+1,A}\,,\quad
	\text{ for }\quad o_i=o_{i+1}=-1
\end{cases}
\,.
\label{eq:E12_classical}
\ee 
Consider for instance $o_i=o_{i+1}=1$, $E^{\epsilon_i,\epsilon_{i+1}}_{i,i+1}$ encodes four possible options of scalar products depending on the signs of $\epsilon_i=\pm$ and $\epsilon_{i+1}=\pm$.
\be
\epsilon_{i}\sum_{A=\pm 1/2}(-1)^{\f12+A} t^{\epsilon_i}_{i,-A} \tau^{\epsilon_{i+1}}_{i+1,A}
=\begin{cases}
	[t^{\epsilon_i}|\tau^{\epsilon_{i+1}} \ra \quad&\text{ for} \quad \epsilon_i=\epsilon_{i+1}=- \\
	[t^{\epsilon_i}|\tau^{\epsilon_{i+1}}] \quad&\text{ for} \quad \epsilon_i=-,\epsilon_{i+1}=+ \\
	\la t^{\epsilon_i}|\tau^{\epsilon_{i+1}}\ra \quad&\text{ for} \quad \epsilon_i=+,\epsilon_{i+1}=-\\
	\la t^{\epsilon_i}|\tau^{\epsilon_{i+1}}] \quad&\text{ for} \quad \epsilon_i=\epsilon_{i+1}=+ 
\end{cases}\,.
\label{eq:epsilon_cases}
\ee
They are by definition invariant under the $\SU(2)$ transformation acting on the vertex of $\Gamma_{\text{rib}}$ where the two spinors meet. Indeed, under an $\SU(2)$ transformation with $g\in \SU(2)$,  $[t^{\epsilon_i}|\rightarrow [t^{\epsilon_i}|g^{-1}\,, \la t^{\epsilon_i}|\rightarrow \la t^{\epsilon_i}|g^{-1}\,, |\tau^{\epsilon_{i+1}}\ra \rightarrow g|\tau^{\epsilon_{i+1}}\ra\,, |\tau^{\epsilon_{i+1}}]\rightarrow g|\tau^{\epsilon_{i+1}}]$ and so clearly all $E^{\epsilon_i,\epsilon_{i+1}}_{i,i+1}$ defined in  \eqref{eq:epsilon_cases} are invariant under $\SU(2)$ transformations. Since those transformations are generated by the Gauss constraint as shown in Proposition \ref{prop:SU(2)LatticeGaugeGenerators}, we find directly the following corollary.
\begin{corollary}
The scalar product  $E^{\epsilon_i,\epsilon_{i+1}}_{i,i+1}$ defined in \eqref{eq:E12_classical} is invariant under the infinitesimal gauge transformation $\delta_\epsilon$ generated by the Gauss constraint defined in \eqref{eq:generalgaugetrans}, $\ie$
\be
\delta_\epsilon E^{\epsilon_i,\epsilon_{i+1}}_{i,i+1}=0\,.
\ee
\end{corollary}
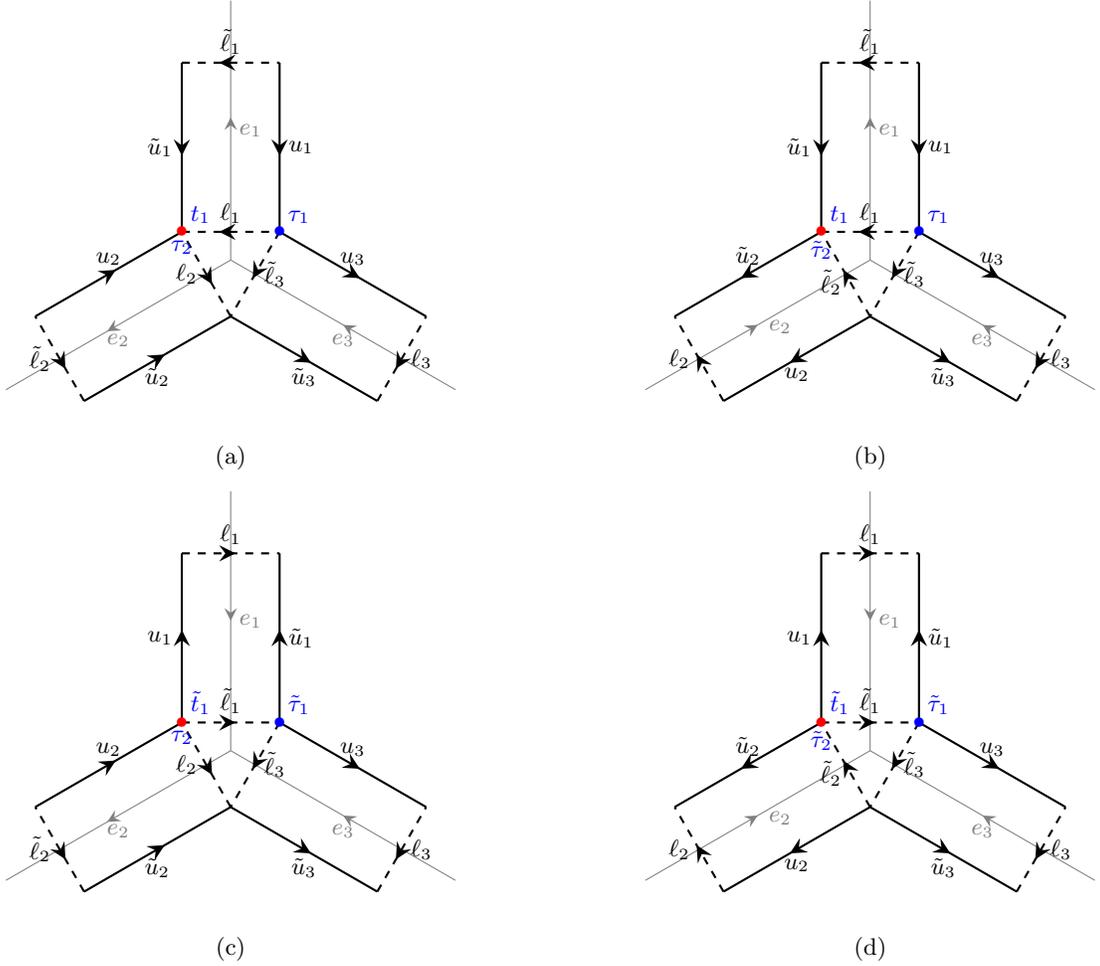
\begin{figure}[h!]
	\centering
	\begin{minipage}{0.45\textwidth}
\centering
	\begin{tikzpicture}[scale=1.5]
\coordinate (o) at (0,0);
\coordinate (a1) at ([shift=(150:0.5cm)]o);
\coordinate (a2) at ([shift=(30:0.5cm)]o);
\coordinate (b1) at ([shift=(90:1.5cm)]a1);
\coordinate (b2) at ([shift=(90:1.5cm)]a2);
\coordinate (a3) at ([shift=(-90:0.5cm)]o);
\coordinate (c1) at ([shift=(-150:1.5cm)]a3);
\coordinate (c2) at ([shift=(-150:1.5cm)]a1);
\coordinate (d1) at ([shift=(-30:1.5cm)]a3);
\coordinate (d2) at ([shift=(-30:1.5cm)]a2);

\coordinate (A) at ([shift=(90:2.3cm)]o);
\coordinate (C) at ([shift=(-30:2.3cm)]o);
\coordinate (B) at ([shift=(210:2.3cm)]o);

\draw[gray,decoration={markings,mark=at position 0.55 with {\arrow[scale=1.5,>=stealth]{>}}},postaction={decorate}] (o) -- node[right,pos=.5]{$e_1$}(A);
\draw[gray,decoration={markings,mark=at position 0.55 with {\arrow[scale=1.5,>=stealth]{>}}},postaction={decorate}] (o) -- node[below,pos=.5]{$e_2$}(B);
\draw[gray,decoration={markings,mark=at position 0.55 with {\arrow[scale=1.5,>=stealth]{<}}},postaction={decorate}] (o) -- node[below,pos=.5]{$e_3$}(C);

\draw[thick,dashed,decoration={markings,mark=at position 0.55 with {\arrow[scale=1.5,>=stealth]{<}}},postaction={decorate}] (a1) -- node[above,pos=.5]{$\ell_1$}(a2);
\draw[thick,dashed,decoration={markings,mark=at position 0.55 with {\arrow[scale=1.5,>=stealth]{>}}},postaction={decorate}] (a2) -- node[right,pos=.5]{$\lt_3$}(a3);
\draw[thick,dashed,decoration={markings,mark=at position 0.55 with {\arrow[scale=1.5,>=stealth]{<}}},postaction={decorate}] (a3) -- node[left,pos=.5]{$\ell_2$}(a1);

\draw[thick,dashed,decoration={markings,mark=at position 0.55 with {\arrow[scale=1.5,>=stealth]{<}}},postaction={decorate}] (b1) -- node[above,pos=.5]{$\lt_1$}(b2);
\draw[thick,dashed,decoration={markings,mark=at position 0.55 with {\arrow[scale=1.5,>=stealth]{<}}},postaction={decorate}] (c1) -- node[left,pos=.5]{$\lt_2$}(c2);
\draw[thick,dashed,decoration={markings,mark=at position 0.55 with {\arrow[scale=1.5,>=stealth]{<}}},postaction={decorate}] (d1) -- node[right,pos=.5]{$\ell_3$}(d2);

\draw[thick,decoration={markings,mark=at position 0.55 with {\arrow[scale=1.5,>=stealth]{>}}},postaction={decorate}] (b1) -- node[left,pos=.5]{$\ut_1$}(a1);
\draw[thick,decoration={markings,mark=at position 0.55 with {\arrow[scale=1.5,>=stealth]{>}}},postaction={decorate}] (b2) -- node[right,pos=.5]{$u_1$}(a2);

\draw[thick,decoration={markings,mark=at position 0.55 with {\arrow[scale=1.5,>=stealth]{>}}},postaction={decorate}] (c1) -- node[below,pos=.5]{$\ut_2$}(a3);
\draw[thick,decoration={markings,mark=at position 0.55 with {\arrow[scale=1.5,>=stealth]{>}}},postaction={decorate}] (c2) -- node[above,pos=.5]{$u_2$}(a1);

\draw[thick,decoration={markings,mark=at position 0.55 with {\arrow[scale=1.5,>=stealth]{<}}},postaction={decorate}] (d1) -- node[below,pos=.5]{$\ut_3$}(a3);
\draw[thick,decoration={markings,mark=at position 0.55 with {\arrow[scale=1.5,>=stealth]{<}}},postaction={decorate}] (d2) -- node[above,pos=.5]{$u_3$}(a2);

\draw[red] (a1) node{$\bullet$};
\draw[blue] (a1) node[above right]{$t_1$};
\draw[blue] (a1) node[below]{$\tau_2$};		
\draw[blue] (a2) node{$\bullet$};
\draw[blue] (a2) node[above right]{$\tau_1$};
	\end{tikzpicture}
\subcaption{}
\label{fig:three_valent_node_oo}
\end{minipage} 
\quad
	\begin{minipage}{0.45\textwidth}
\centering
	\begin{tikzpicture}[scale=1.5]
\coordinate (o) at (0,0);
\coordinate (a1) at ([shift=(150:0.5cm)]o);
\coordinate (a2) at ([shift=(30:0.5cm)]o);
\coordinate (b1) at ([shift=(90:1.5cm)]a1);
\coordinate (b2) at ([shift=(90:1.5cm)]a2);
\coordinate (a3) at ([shift=(-90:0.5cm)]o);
\coordinate (c1) at ([shift=(-150:1.5cm)]a3);
\coordinate (c2) at ([shift=(-150:1.5cm)]a1);
\coordinate (d1) at ([shift=(-30:1.5cm)]a3);
\coordinate (d2) at ([shift=(-30:1.5cm)]a2);

\coordinate (A) at ([shift=(90:2.3cm)]o);
\coordinate (C) at ([shift=(-30:2.3cm)]o);
\coordinate (B) at ([shift=(210:2.3cm)]o);

\draw[gray,decoration={markings,mark=at position 0.55 with {\arrow[scale=1.5,>=stealth]{>}}},postaction={decorate}] (o) -- node[right,pos=.5]{$e_1$}(A);
\draw[gray,decoration={markings,mark=at position 0.55 with {\arrow[scale=1.5,>=stealth]{<}}},postaction={decorate}] (o) -- node[below,pos=.4]{$e_2$}(B);
\draw[gray,decoration={markings,mark=at position 0.55 with {\arrow[scale=1.5,>=stealth]{<}}},postaction={decorate}] (o) -- node[below,pos=.5]{$e_3$}(C);

\draw[thick,dashed,decoration={markings,mark=at position 0.55 with {\arrow[scale=1.5,>=stealth]{<}}},postaction={decorate}] (a1) -- node[above,pos=.5]{$\ell_1$}(a2);
\draw[thick,dashed,decoration={markings,mark=at position 0.55 with {\arrow[scale=1.5,>=stealth]{>}}},postaction={decorate}] (a2) -- node[right,pos=.5]{$\lt_3$}(a3);
\draw[thick,dashed,decoration={markings,mark=at position 0.55 with {\arrow[scale=1.5,>=stealth]{>}}},postaction={decorate}] (a3) -- node[left,pos=.4]{$\lt_2$}(a1);

\draw[thick,dashed,decoration={markings,mark=at position 0.55 with {\arrow[scale=1.5,>=stealth]{<}}},postaction={decorate}] (b1) -- node[above,pos=.5]{$\lt_1$}(b2);
\draw[thick,dashed,decoration={markings,mark=at position 0.55 with {\arrow[scale=1.5,>=stealth]{>}}},postaction={decorate}] (c1) -- node[left,pos=.5]{$\ell_2$}(c2);
\draw[thick,dashed,decoration={markings,mark=at position 0.55 with {\arrow[scale=1.5,>=stealth]{<}}},postaction={decorate}] (d1) -- node[right,pos=.5]{$\ell_3$}(d2);

\draw[thick,decoration={markings,mark=at position 0.55 with {\arrow[scale=1.5,>=stealth]{>}}},postaction={decorate}] (b1) -- node[left,pos=.5]{$\ut_1$}(a1);
\draw[thick,decoration={markings,mark=at position 0.55 with {\arrow[scale=1.5,>=stealth]{>}}},postaction={decorate}] (b2) -- node[right,pos=.5]{$u_1$}(a2);

\draw[thick,decoration={markings,mark=at position 0.55 with {\arrow[scale=1.5,>=stealth]{<}}},postaction={decorate}] (c1) -- node[below,pos=.5]{$u_2$}(a3);
\draw[thick,decoration={markings,mark=at position 0.55 with {\arrow[scale=1.5,>=stealth]{<}}},postaction={decorate}] (c2) -- node[above,pos=.5]{$\ut_2$}(a1);

\draw[thick,decoration={markings,mark=at position 0.55 with {\arrow[scale=1.5,>=stealth]{<}}},postaction={decorate}] (d1) -- node[below,pos=.5]{$\ut_3$}(a3);
\draw[thick,decoration={markings,mark=at position 0.55 with {\arrow[scale=1.5,>=stealth]{<}}},postaction={decorate}] (d2) -- node[above,pos=.5]{$u_3$}(a2);

\draw[red] (a1) node{$\bullet$};
\draw[blue] (a1) node[above right]{$t_1$};
\draw[blue] (a1) node[below]{$\ttau_2$};
\draw[blue] (a2) node{$\bullet$};
\draw[blue] (a2) node[above right]{$\tau_1$};
	\end{tikzpicture}
\subcaption{}
\label{fig:three_valent_node_oi}
\end{minipage}  \\
	\begin{minipage}{0.45\textwidth}
\centering
	\begin{tikzpicture}[scale=1.5]
\coordinate (o) at (0,0);
\coordinate (a1) at ([shift=(150:0.5cm)]o);
\coordinate (a2) at ([shift=(30:0.5cm)]o);
\coordinate (b1) at ([shift=(90:1.5cm)]a1);
\coordinate (b2) at ([shift=(90:1.5cm)]a2);
\coordinate (a3) at ([shift=(-90:0.5cm)]o);
\coordinate (c1) at ([shift=(-150:1.5cm)]a3);
\coordinate (c2) at ([shift=(-150:1.5cm)]a1);
\coordinate (d1) at ([shift=(-30:1.5cm)]a3);
\coordinate (d2) at ([shift=(-30:1.5cm)]a2);

\coordinate (A) at ([shift=(90:2.3cm)]o);
\coordinate (C) at ([shift=(-30:2.3cm)]o);
\coordinate (B) at ([shift=(210:2.3cm)]o);

\draw[gray,decoration={markings,mark=at position 0.55 with {\arrow[scale=1.5,>=stealth]{<}}},postaction={decorate}] (o) -- node[right,pos=.5]{$e_1$}(A);
\draw[gray,decoration={markings,mark=at position 0.55 with {\arrow[scale=1.5,>=stealth]{>}}},postaction={decorate}] (o) -- node[below,pos=.5]{$e_2$}(B);
\draw[gray,decoration={markings,mark=at position 0.55 with {\arrow[scale=1.5,>=stealth]{<}}},postaction={decorate}] (o) -- node[below,pos=.5]{$e_3$}(C);

\draw[thick,dashed,decoration={markings,mark=at position 0.55 with {\arrow[scale=1.5,>=stealth]{>}}},postaction={decorate}] (a1) -- node[above,pos=.5]{$\lt_1$}(a2);
\draw[thick,dashed,decoration={markings,mark=at position 0.55 with {\arrow[scale=1.5,>=stealth]{>}}},postaction={decorate}] (a2) -- node[right,pos=.5]{$\lt_3$}(a3);
\draw[thick,dashed,decoration={markings,mark=at position 0.55 with {\arrow[scale=1.5,>=stealth]{<}}},postaction={decorate}] (a3) -- node[left,pos=.5]{$\ell_2$}(a1);

\draw[thick,dashed,decoration={markings,mark=at position 0.55 with {\arrow[scale=1.5,>=stealth]{>}}},postaction={decorate}] (b1) -- node[above,pos=.5]{$\ell_1$}(b2);
\draw[thick,dashed,decoration={markings,mark=at position 0.55 with {\arrow[scale=1.5,>=stealth]{<}}},postaction={decorate}] (c1) -- node[left,pos=.5]{$\lt_2$}(c2);
\draw[thick,dashed,decoration={markings,mark=at position 0.55 with {\arrow[scale=1.5,>=stealth]{<}}},postaction={decorate}] (d1) -- node[right,pos=.5]{$\ell_3$}(d2);

\draw[thick,decoration={markings,mark=at position 0.55 with {\arrow[scale=1.5,>=stealth]{<}}},postaction={decorate}] (b1) -- node[left,pos=.5]{$u_1$}(a1);
\draw[thick,decoration={markings,mark=at position 0.55 with {\arrow[scale=1.5,>=stealth]{<}}},postaction={decorate}] (b2) -- node[right,pos=.5]{$\ut_1$}(a2);

\draw[thick,decoration={markings,mark=at position 0.55 with {\arrow[scale=1.5,>=stealth]{>}}},postaction={decorate}] (c1) -- node[below,pos=.5]{$\ut_2$}(a3);
\draw[thick,decoration={markings,mark=at position 0.55 with {\arrow[scale=1.5,>=stealth]{>}}},postaction={decorate}] (c2) -- node[above,pos=.5]{$u_2$}(a1);

\draw[thick,decoration={markings,mark=at position 0.55 with {\arrow[scale=1.5,>=stealth]{<}}},postaction={decorate}] (d1) -- node[below,pos=.5]{$\ut_3$}(a3);
\draw[thick,decoration={markings,mark=at position 0.55 with {\arrow[scale=1.5,>=stealth]{<}}},postaction={decorate}] (d2) -- node[above,pos=.5]{$u_3$}(a2);

\draw[red] (a1) node{$\bullet$};
\draw[blue] (a1) node[above right]{$\tt_1$};
\draw[blue] (a1) node[below]{$\tau_2$};		
\draw[blue] (a2) node{$\bullet$};
\draw[blue] (a2) node[above right]{$\ttau_1$};
	\end{tikzpicture}
\subcaption{}
\label{fig:three_valent_node_io}
\end{minipage}
\quad
\begin{minipage}{0.45\textwidth}
\centering
	\begin{tikzpicture}[scale=1.5]
\coordinate (o) at (0,0);
\coordinate (a1) at ([shift=(150:0.5cm)]o);
\coordinate (a2) at ([shift=(30:0.5cm)]o);
\coordinate (b1) at ([shift=(90:1.5cm)]a1);
\coordinate (b2) at ([shift=(90:1.5cm)]a2);
\coordinate (a3) at ([shift=(-90:0.5cm)]o);
\coordinate (c1) at ([shift=(-150:1.5cm)]a3);
\coordinate (c2) at ([shift=(-150:1.5cm)]a1);
\coordinate (d1) at ([shift=(-30:1.5cm)]a3);
\coordinate (d2) at ([shift=(-30:1.5cm)]a2);

\coordinate (A) at ([shift=(90:2.3cm)]o);
\coordinate (C) at ([shift=(-30:2.3cm)]o);
\coordinate (B) at ([shift=(210:2.3cm)]o);

\draw[gray,decoration={markings,mark=at position 0.55 with {\arrow[scale=1.5,>=stealth]{<}}},postaction={decorate}] (o) -- node[right,pos=.5]{$e_1$}(A);
\draw[gray,decoration={markings,mark=at position 0.55 with {\arrow[scale=1.5,>=stealth]{<}}},postaction={decorate}] (o) -- node[below,pos=.4]{$e_2$}(B);
\draw[gray,decoration={markings,mark=at position 0.55 with {\arrow[scale=1.5,>=stealth]{<}}},postaction={decorate}] (o) -- node[below,pos=.5]{$e_3$}(C);

\draw[thick,dashed,decoration={markings,mark=at position 0.55 with {\arrow[scale=1.5,>=stealth]{>}}},postaction={decorate}] (a1) -- node[above,pos=.5]{$\lt_1$}(a2);
\draw[thick,dashed,decoration={markings,mark=at position 0.55 with {\arrow[scale=1.5,>=stealth]{>}}},postaction={decorate}] (a2) -- node[right,pos=.5]{$\lt_3$}(a3);
\draw[thick,dashed,decoration={markings,mark=at position 0.55 with {\arrow[scale=1.5,>=stealth]{>}}},postaction={decorate}] (a3) -- node[left,pos=.4]{$\lt_2$}(a1);

\draw[thick,dashed,decoration={markings,mark=at position 0.55 with {\arrow[scale=1.5,>=stealth]{>}}},postaction={decorate}] (b1) -- node[above,pos=.5]{$\ell_1$}(b2);
\draw[thick,dashed,decoration={markings,mark=at position 0.55 with {\arrow[scale=1.5,>=stealth]{>}}},postaction={decorate}] (c1) -- node[left,pos=.5]{$\ell_2$}(c2);
\draw[thick,dashed,decoration={markings,mark=at position 0.55 with {\arrow[scale=1.5,>=stealth]{<}}},postaction={decorate}] (d1) -- node[right,pos=.5]{$\ell_3$}(d2);

\draw[thick,decoration={markings,mark=at position 0.55 with {\arrow[scale=1.5,>=stealth]{<}}},postaction={decorate}] (b1) -- node[left,pos=.5]{$u_1$}(a1);
\draw[thick,decoration={markings,mark=at position 0.55 with {\arrow[scale=1.5,>=stealth]{<}}},postaction={decorate}] (b2) -- node[right,pos=.5]{$\ut_1$}(a2);

\draw[thick,decoration={markings,mark=at position 0.55 with {\arrow[scale=1.5,>=stealth]{<}}},postaction={decorate}] (c1) -- node[below,pos=.5]{$u_2$}(a3);
\draw[thick,decoration={markings,mark=at position 0.55 with {\arrow[scale=1.5,>=stealth]{<}}},postaction={decorate}] (c2) -- node[above,pos=.5]{$\ut_2$}(a1);

\draw[thick,decoration={markings,mark=at position 0.55 with {\arrow[scale=1.5,>=stealth]{<}}},postaction={decorate}] (d1) -- node[below,pos=.5]{$\ut_3$}(a3);
\draw[thick,decoration={markings,mark=at position 0.55 with {\arrow[scale=1.5,>=stealth]{<}}},postaction={decorate}] (d2) -- node[above,pos=.5]{$u_3$}(a2);

\draw[red] (a1) node{$\bullet$};
\draw[blue] (a1) node[above right]{$\tt_1$};
\draw[blue] (a1) node[below]{$\ttau_2$};
\draw[blue] (a2) node{$\bullet$};
\draw[blue] (a2) node[above right]{$\ttau_1$};
	\end{tikzpicture}
\subcaption{}
\label{fig:three_valent_node_ii}
\end{minipage}
\caption{A node with three incident edges $e_1,e_2,e_3$ ({\it in gray}) and the correspondent ribbon graph. The four possible orientations for $e_1$ and $e_2$ with a fix orientation $o_3=-1$ for $e_3$ are illustrated separately. The spinors defining the scalar product $E^{\epsilon_1,\epsilon_2}_{12}$ can be read at the common vertex ({\it in red}) of the ribbons associated to $e_1$ and $e_2$.}
\label{fig:three_valent_node}
\end{figure}

\subsection{Invariants from spinors sitting at different vertices in \texorpdfstring{$\Gamma_{\text{rib}}$}{the ribbon graph}}
We now explain how to build invariants for an arbitrary pair of edges $i,j=1, \dotsc, n$ incident to an $n$-valent vertex. As before, we can work with the ribbon decorated with $\ell, \tell$ or $\ell^{-1\dagger},\tell^{-1\dagger}$. We choose to explicit the case where we use $\ell, \tell$, the other case is obtained in  a similar way.

Consider first $j=i+1$ so that the edges share a vertex in $\Gamma_{\text{rib}}$. Then we know of the invariant $E_{i,i+1}^{\epsilon_i, \epsilon_{i+1}}$. We can also try to define an observable in terms of $\tau_i$ and $\tau_{i+1}$. We have showed that the scalar product of $t_i$ and $\tau_{i+1}$ is an observable. On the other hand, we know that $t_i$ is the result of transporting $\tau_i$ by $\ell_{e_iv}$, see \eqref{eq:braided-cov}, \eqref{eq:tilde_braided_covariant} ($\ell_{e_iv}$ is defined in \eqref{eq:Gauss_flatness_constraint}). Therefore we can in fact transport $\tau_{i+1}$ by $\ell_{e_i v}$ so that it sits at the same vertex as $\tau_i$ in $\Gamma_{\text{rib}}$. Obviously one gets the same invariant as in \eqref{eq:E12_classical}.

\begin{prop}
Up to coefficients $e^{\pm\f{\ka N_i}{4}}$, we have that 
\be
E^{\epsilon_i,\epsilon_{i+1}}_{i,i+1}
\propto\begin{cases}
\epsilon_i \sum_{A=\pm 1/2} (-1)^{\f12+A}\tau^{\epsilon_i}_{i,-A} \left(\ell_{i}\mone \tau^{\epsilon_{i+1}}_{i+1}\right)_{A}
\sim\epsilon_i \sum_{A=\pm 1/2} (-1)^{\f12+A}\tau^{\epsilon_i}_{i,-A} \left(\ell^\dagger_{i} \tau^{\epsilon_{i+1}}_{i+1}\right)_{A}
\,,&
	\text{ for } o_i=o_{i+1}=1\\[0.15cm]
\epsilon_i \sum_{A=\pm 1/2} (-1)^{\f12+A}\tau^{\epsilon_i}_{i,-A} \left(\ell_i^{-1} \ttau^{\epsilon_{i+1}}_{i+1}\right)_{A}
\sim\epsilon_i \sum_{A=\pm 1/2} (-1)^{\f12+A}\tau^{\epsilon_i}_{i,-A} \left(\ell_i^\dagger \ttau^{\epsilon_{i+1}}_{i+1}\right)_{A}
\,,&
	\text{ for } o_i=-o_{i+1}=1\\[0.15cm]
\epsilon_i	\sum_{A=\pm 1/2}(-1)^{\f12+A} \ttau^{\epsilon_i}_{i,-A}\left(\tell_i \tau^{\epsilon_{i+1}}_{i+1}\right)_{A}
\sim\epsilon_i	\sum_{A=\pm 1/2}(-1)^{\f12+A} \ttau^{\epsilon_i}_{i,-A}\left(\tell_i^{-1\,\dagger} \tau^{\epsilon_{i+1}}_{i+1}\right)_{A}
\,,&
	\text{ for }-o_i=o_{i+1}=1\\[0.15cm]
\epsilon_i	\sum_{A=\pm 1/2}(-1)^{\f12+A} \ttau^{\epsilon_i}_{i,-A}\left(\tell_i \ttau^{\epsilon_{i+1}}_{i+1}\right)_{A}
\sim\epsilon_i	\sum_{A=\pm 1/2}(-1)^{\f12+A} \ttau^{\epsilon_i}_{i,-A}\left(\tell_i^{-1\,\dagger} \ttau^{\epsilon_{i+1}}_{i+1}\right)_{A}
\,,&
	\text{ for } o_i=o_{i+1}=-1\,.
\end{cases}
\label{eq:E12_classical-bis}
\ee
\end{prop}
\begin{proof}
Consider the definition \eqref{eq:E12_classical} and focus on the first case, with $o_i=o_{i+1}=1$. Then, we apply \eqref{eq:braided-cov}, $\lb \ell_i \tau^{\epsilon_i}_i\rb_A\propto t^{\epsilon_i}_{i,A}$ and that $\lb \ell_i^{-1\,\dagger} \tau^{\epsilon_i}_i\rb_A\propto t^{\epsilon_i}_{1,A}$ up to coefficients $e^{\pm\f{\ka N_i}{4}}$. We further have 
\be
\left(\ell\mone\right)_{AB}= (-1)^{B-A}\ell_{-B-A}\,,\quad
\lb \ell^{\dagger} \rb_{AB} = (-1)^{B-A} \lb\ell^{-1\,\dagger}\rb_{-B-A}\,.
\label{eq:ellandinverse}
\ee
Putting these equalities together, we get the proposition. 
\end{proof}

In the quantization scheme, since we need to order the Hilbert spaces, and build the spinor operators using some braided permutation to the following Hilbert space we will need to set up a reference point. This is called the \emph{cilium}. We will see that the notion of braided permutation is nothing else than the quantum version of the parallel transport we are discussing. As a consequence, the notion of quantum observable based on the braiding will be associated to the formulation \eqref{eq:E12_classical-bis} instead of \eqref{eq:E12_classical}.

\medskip

We generalize this construction to edges $e_i, e_j$ incident to the same vertex in $\Gamma$ but with $j\neq i+1$. 
To simplify the notations of  \eqref{SpinorsNotationsAtVertex}, we denote $\tau^{\epsilon_i}_i \equiv \tau^{\epsilon_i}_{e_iv}$ and similarly for the other spinors.
Up to parallel transport by $\ell_{e_i v}, \ell_{e_j v}$, we can always build our observables from the spinors $\tau_{i}^{\epsilon_i}, \tau_{j}^{\epsilon_j}$. The recipe is to parallel transport $\tau_{j}^{\epsilon_j}$ around the ribbon vertex to meet $\tau_{i}^{\epsilon_i}$ at the same vertex in $\Gamma_{\text{rib}}$. This is done by introducing $\cL_{ij}$ (resp. $\cL_{ij}^{-1\dagger}$), the $\AN(2)$ holonomy consisting of the product of $\ell^{-1}$ and $\tell$ (resp. $\ell^{\dagger}$ and $\tell^{-1\dagger}$) clockwise around $R(v)$ from $j$ to $i$,
\begin{prop}
The quantity 
\be
E^{\epsilon_i,\epsilon_j}_{ij}
=\begin{cases}
	\epsilon_i\sum_{A=\pm 1/2}(-1)^{\f12+A} \tau^{\epsilon_i}_{i,-A} \left(\cL_{ij} \tau^{\epsilon_j}_{j}\right)_{A}
	\sim\epsilon_i\sum_{A=\pm 1/2}(-1)^{\f12+A} \tau^{\epsilon_i}_{i,-A} \left(\cL_{ij}^{-1\,\dagger} \tau^{\epsilon_j}_{j}\right)_{A}
	\,,\quad
	\text{ for }\quad o_i=o_j=1\\[0.15cm]
	\epsilon_i\sum_{A=\pm 1/2} (-1)^{\f12+A}\tau^{\epsilon_i}_{i,-A} \left(\cL_{ij} \ttau^{\epsilon_j}_{j}\right)_{A}
	\sim\epsilon_i\sum_{A=\pm 1/2} (-1)^{\f12+A}\tau^{\epsilon_i}_{i,-A} \left(\cL_{ij}^{-1\,\dagger} \ttau^{\epsilon_j}_{j}\right)_{A}
	\,,\quad
	\text{ for }\quad o_i=-o_j=1\\[0.15cm]
	\epsilon_i\sum_{A=\pm 1/2}(-1)^{\f12+A} \ttau^{\epsilon_i}_{i,-A}\left(\cL_{ij} \tau^{\epsilon_j}_{j}\right)_{A}
	\sim\epsilon_i\sum_{A=\pm 1/2}(-1)^{\f12+A} \ttau^{\epsilon_i}_{i,-A}\left(\cL_{ij}^{-1\,\dagger} \tau^{\epsilon_j}_{j}\right)_{A}
	\,,\quad
	\text{ for }\quad -o_i=o_j=1\\[0.15cm]
	\epsilon_i\sum_{A=\pm 1/2}(-1)^{\f12+A} \ttau^{\epsilon_i}_{i,-A}\left(\cL_{ij} \ttau^{\epsilon_j}_{j}\right)_{A}
	\sim\epsilon_i\sum_{A=\pm 1/2}(-1)^{\f12+A} \ttau^{\epsilon_i}_{i,-A}\left(\cL_{ij}^{-1\,\dagger} \ttau^{\epsilon_j}_{j}\right)_{A}
	\,,\quad
	\text{ for }\quad o_i=o_j=-1
\end{cases}
\label{eq:Eij_classical-bis}
\ee
is an observable, $\ie$ $\delta_\epsilon  E^{\epsilon_i,\epsilon_j}_{e_ie_j} =0$.
\end{prop}
Different expressions can be obtained if one uses $t^{\epsilon_i}_{e_iv}$ or $t_{e_j v}^{\epsilon_j}$ instead.

\subsection{Poisson algebra of observables} 

Let us now compute the observable Poisson algebra formed by the quadratic invariant $E_{ij}^{\epsilon_i,\epsilon_j}$. When $q=1$, it is well-known that they form a $\so^*(2n)$ Poisson algebra \cite{Girelli:2017dbk} with a $\u(n)$ subalgebra, where $n$ is the valency of the vertex. When $q\neq 1$, this algebra is deformed as we now describe.

To distinguish different kinds of observables, we define 
\be
\fe_i \equiv \fe_{i,i}= E_{i,i}^{+,-} \equiv E_{i,i}^{-,+}\,,\quad
\fe_{i,i+1}=E_{i,i+1}^{+,-}\,,\quad 
\fe_{i+1,i}=E_{i,i+1}^{-,+}\,,\quad
\ff_{i,i+1}=E_{i,i+1}^{-,-}\,,\quad 
\fft_{i,i+1}=-E_{i,i+1}^{+,+}\,.
\label{eq:qSO*2n_generators_ii+1}
\ee
$\fe_{i,i+1}$ and $\fe_{i+1,i}$ are related by complex conjugation, and likewise for $\ff_{i,i+1}$ and $\fft_{i,i+1}$. That is,
\be
\fe_{i+1,i} = \overline{\fe_{i,i+1}}\,,\quad
\fft_{i,i+1} = \overline{\ff_{i,i+1}}\,.
\ee
With no loss of generality, we can take the orientation $o_i=o_{i+1}=-1$ and write these generators explicitly,
\begin{subequations}
\begin{align}
\fe_i=N_i\,,\quad
\fe_{i,i+1}=\la \tt_i | \ttau_{i+1} \ra 
\equiv \btt_{i,-}\ttau_{i+1,-} + \btt_{i,+}\ttau_{i+1,+}
= e^{\f{\ka}{4}(N_{i,1}-N_{i+1,1})}\btzeta_{i,0}^\ka \tzeta_{i+1,0}^\ka 
+ e^{-\f{\ka}{4}(N_{i,0}-N_{i+1,0})}\btzeta_{i,1}^\ka \tzeta_{i+1,1}^\ka\,,
\label{eq:qUn_generators_classical_1}\\[0.15cm]
\fe_{i+1,i} = [ \tt_i | \ttau_{i+1} ] 
\equiv \tt_{i,-}\bttau_{i+1,-} + \tt_{i,+}\bttau_{i+1,+}
= e^{\f{\ka}{4}(N_{i,1}-N_{i+1,1})}\tzeta_{i,0}^\ka \btzeta_{i+1,0}^\ka 
+ e^{-\f{\ka}{4}(N_{i,0}-N_{i+1,0})}\tzeta_{i,1}^\ka \btzeta_{i+1,1}^\ka\,,
\label{eq:qUn_generators_classical_2}\\[0.3cm]
\ff_{i,i+1} = [ \tt_i | \ttau_{i+1} \ra 
\equiv \tt_{i,-}\ttau_{i+1,+} - \tt_{i,+}\ttau_{i+1,-} 
= e^{-\f{\ka}{4}(N_{i,0}+N_{i+1,1})}\tzeta_{i,1}^\ka \tzeta_{i+1,0}^\ka 
- e^{\f{\ka}{4}(N_{i,1}+N_{i+1,0})}\tzeta_{i,0}^\ka \tzeta_{i+1,1}^\ka\,,
\label{eq:qSO*2n_generators_classical_1}\\[0.15cm]
\fft_{i,i+1} =-\la \tt_i | \ttau_{i+1} ] 
\equiv \btt_{i,-}\bttau_{i+1,+} - \btt_{i,+}\bttau_{i+1,-} 
= e^{-\f{\ka}{4}(N_{i,1}+N_{i+1,0})}\btzeta_{i,0}^\ka \btzeta_{i+1,1}^\ka
-e^{\f{\ka}{4}(N_{i,0}+N_{i+1,1})}\btzeta_{i,1}^\ka \btzeta_{i+1,0}^\ka  \,.
\label{eq:qSO*2n_generators_classical_2}
\end{align}
\label{eq:qSO*2n_generators_classical}
\end{subequations}
Indeed, $\fe_{i,i+1}$ is holomorphic in spinor variables at the $i$-th site and anti-holomorphic at the $(i+1)$-th site while $\fe_{i+1,i}$ is in the opposite way. On the other hand, $\ff_{i,i+1}$ (\resp $\fft_{i,i+1}$) is holomorphic (\resp anti-holomorphic) at both sites. 
The holomorphic functions in spinor variables will be quantized to annihilation operators while the anti-holomorphic ones will be quantized to creation operators which we will see in Section \ref{sec:quantum_spinors}.

Other generators $\fe_{ij}\,,\fe_{ji}\,,\ff_{ij}$ and $\fft_{ij}$ with $j>i+1$ can be defined recursively as follows. 
\begin{subequations}
\begin{align}
\fe_{ij}&=\f{1}{\f{2}{\ka}\sinh\f{\ka \fe_{j-1}}{2}}\lb\fe_{i,j-1}\fe_{j-1,j}+e^{\f{\ka \fe_{j-1}}{2}}\fft_{i,j-1}\ff_{j-1,j} \rb
\equiv \f{1}{\f{2}{\ka}\sinh\f{\ka \fe_{i+1}}{2}}\lb\fe_{i,i+1}\fe_{i+1,j}+e^{\f{\ka \fe_{i+1}}{2}}\fft_{i,i+1}\ff_{i+1,j} \rb\,,
\label{eq:qSO*_generators_ij_1}\\
\fe_{ji}&=\f{1}{\f{2}{\ka}\sinh\f{\ka \fe_{j-1}}{2}}\lb\fe_{j-1,i}\fe_{j,j-1}+e^{-\f{\ka \fe_{j-1}}{2}}\ff_{i,j-1}\fft_{j-1,j} \rb
\equiv \f{1}{\f{2}{\ka}\sinh\f{\ka \fe_{i+1}}{2}}\lb\fe_{i+1,i}\fe_{j,i+1}+e^{-\f{\ka \fe_{i+1}}{2}}\ff_{i,i+1}\fft_{i+1,j} \rb\,,
\label{eq:qSO*_generators_ij_2}\\
\ff_{ij}&=\f{1}{\f{2}{\ka}\sinh\f{\ka \fe_{j-1}}{2}}\lb e^{-\f{\ka\fe_{j-1}}{2}}\ff_{i,j-1}\fe_{j-1,j}+\fe_{j-1,i}\ff_{j-1,j} \rb
\equiv \f{1}{\f{2}{\ka}\sinh\f{\ka \fe_{i+1}}{2}}\lb e^{-\f{\ka\fe_{i+1}}{2}}\ff_{i,i+1}\fe_{i+1,j}+\fe_{i+1,i}\ff_{i+1,j} \rb\,,
\label{eq:qSO*_generators_ij_3}\\
\fft_{ij}&=\f{1}{\f{2}{\ka}\sinh\f{\ka \fe_{j-1}}{2}}\lb e^{\f{\ka\fe_{j-1}}{2}} \fft_{i,j-1}\fe_{j,j-1}+\fe_{i,j-1}\fft_{j-1,j}\rb
\equiv \f{1}{\f{2}{\ka}\sinh\f{\ka \fe_{i+1}}{2}}\lb e^{\f{\ka\fe_{i+1}}{2}}\fft_{i,i+1}\fe_{j,i+1}+\fe_{i,i+1}\fft_{i+1,j} \rb\,.
\label{eq:qSO*_generators_ij_4}
\end{align}
\label{eq:qSO*_generators_ij}
\end{subequations}
Remarkably, the generators \eqref{eq:qSO*2n_generators_classical} and \eqref{eq:qSO*_generators_ij} can be recovered geometrically. 
To do this, without loss of generality, we will use the definition \eqref{eq:Eij_classical-bis} of $E^{\epsilon_i,\epsilon_j}_{ij}$ and take $o_i=o_{i+1}=\cdots =o_j=-1$ for convenience. Then the generators given in \eqref{eq:qSO*_generators_ij} can be equivalently given by\footnote{Indeed, the parallel transport can also be done by using $\cL_{ij}$ instead of $\cL_{ij}^{-1\,\dagger}$. We have chosen the latter one so that $\fe_{ij}$ and $\fe_{ji}$ will be naturally quantized to the standard generators of $\UUQn$ as we will see in \eqref{eq:Chevalley_to_CartanWeyl} using \eqref{eq:Eij_in_bosons}. }
\be
\ba{ll}
\fe_{ij}=\lb  \prod_{k=i}^{j-1}e^{\f{\ka \fe_k}{4}} \rb \la \ttau_i | \cL_{ij}^{-1\,\dagger} |\ttau_j\ra 
\,,\quad &
\fe_{ji}=\lb  \prod_{k=i}^{j-1}e^{-\f{\ka \fe_k}{4}} \rb [ \ttau_i | \cL_{ij}^{-1\,\dagger} |\ttau_j ]\quad\\[0.15cm]
\ff_{ij}=\lb  \prod_{k=i}^{j-1}e^{-\f{\ka \fe_k}{4}} \rb [ \ttau_i | \cL_{ij}^{-1\,\dagger} |\ttau_j \ra\,,\quad &
\fft_{ij}=-\lb  \prod_{k=i}^{j-1}e^{\f{\ka \fe_k}{4}} \rb \la \ttau_i | \cL_{ij}^{-1\,\dagger} |\ttau_j ]
\ea\quad
\text{ with }\quad
\cL_{ij}^{-1\,\dagger}=\lt_i^{-1\,\dagger}\lt_{i+1}^{-1\,\dagger}\cdots\lt_{j-1}^{-1\,\dagger}\,.
\label{eq:qSO*_generators_ij_flux}
\ee
As a consistency check, when $j=i+1$, \eqref{eq:qSO*2n_generators_classical} can be written using \eqref{eq:qSO*_generators_ij_flux} with only the braided spinors $|\ttau\ra, |\ttau]$ and one flux $\lt_i$ as
\be\ba{ll}	
\fe_{i,i+1}=e^{\f{\ka \Nt_i}{4}}\la \ttau_i|\lt_i^{-1\,\dagger}|\ttau_{i+1}\ra \equiv \la \tt_i|\ttau_{i+1}\ra\,,\quad &
\fe_{i+1,i}=e^{-\f{\ka \Nt_i}{4}}[ \ttau_i|\lt_i^{-1\,\dagger}|\ttau_{i+1}] \equiv [ \tt_i|\ttau_{i+1}]\,,
\\[0.15cm]
\ff_{i,i+1}=e^{-\f{\ka \Nt_i}{4}}[ \ttau_i|\lt_i^{-1\,\dagger}|\ttau_{i+1}\ra \equiv [ \tt_i|\ttau_{i+1}\ra\,,\quad &
\fft_{i,i+1}=-e^{\f{\ka \Nt_i}{4}}\la \ttau_i|\lt_i^{-1\,\dagger}|\ttau_{i+1}] \equiv - \la \tt_i|\ttau_{i+1}]\,.
\ea\ee 
We can also switch the indices for the generators $\ff_{i,i+p} \,(p\in\N^+)$, $\fft_{i,i+p}$ and define 
\be
\ff_{i+p,i}:=\lb  \prod_{k=0}^{p-1}e^{-\f{\ka \fe_{i+k}}{4}}\rb[ \ttau_{i+p}|\cL_{i,i+p}^{\dagger}|\tt_i \ra \equiv -\ff_{i,i+p}\,\quad
\fft_{i+p,i}:=-\lb  \prod_{k=0}^{p-1}e^{\f{\ka \fe_{i+k}}{4}} \rb\la \ttau_{i+p}|\cL_{i,i+p}^{\dagger}|\tt_i] = -\fft_{i,i+p}\,.
\ee
The Poisson algebra formed by the generators defined in \eqref{eq:qSO*2n_generators_classical} is given in the following two propositions.
\begin{prop}
	$\fe_i,$ $\fe_{i,i+1}$ and $\fe_{i+1,i}$ defined in \eqref{eq:qUn_generators_classical_1} and \eqref{eq:qUn_generators_classical_2} form a $\ka$-deformed $\u(n)$ Poisson algebra. They satisfy the following Poisson brackets
\be\begin{split}
\{\fe_i,\fe_j\}=0\,,\quad
\{\fe_i,\fe_{j,j+1}\} = i(\delta_{i,j+1}-\delta_{ij})\fe_{j,j+1}\,,\quad
\{\fe_i,\fe_{j+1,j}\} = i(\delta_{ij}-\delta_{i,j+1})\fe_{j+1,j}\,,\\
\{\fe_{i,i+1},\fe_{j+1,j}\} =\delta_{ij}\f{2i}{\ka}\sinh\f{\ka(\fe_{i+1}-\fe_{i})}{2} 
\,.
\end{split}
\label{eq:qUn_Poisson}
\ee
\end{prop}
\begin{prop}
	$\fe_i,$ $\fe_{i,i+1},$ $\fe_{i+1,i},$ $\ff_{i,i+1}$ and $\fft_{i,i+1}$ defined in  \eqref{eq:qSO*2n_generators_classical} form a $\ka$-deformed $\so^*(2n)$ Poisson algebra. They satisfy \eqref{eq:qUn_Poisson} and the following Poisson brackets
\be\begin{split}
&\{\fe_{i,i+1},\ff_{j,j+1}\} = i\delta_{i,j+1}\lb \ff_{i-1,i+1} +\f{\ka}{2}\fe_{i,i+1}\ff_{i-1,i+1}\rb\,,\quad
\{\fe_{i+1,i},\ff_{j,j+1}\} = i\delta_{i,j-1}\lb e^{\f{\ka \fe_i}{2}}  \ff_{i,i+2} -\f{\ka}{2}\fe_{i+1,i}\ff_{i+1,i+2}\rb\,,\\
&\{\fe_{i,i+1},\fft_{j,j+1}\} = -i\delta_{i,j-1}\lb e^{-\f{\ka \fe_i}{2}} \fft_{i,i+2} + \f{\ka}{2}\fe_{i,i+1}\fft_{i+1,i+2}\rb \,,\quad
\{\fe_{i+1,i},\fft_{j,j+1}\} = -i\delta_{i,j+1}\lb \fft_{i-1,i+1}-\f{\ka}{2}\fe_{i+1,i}\fft_{i-1,i+1} \rb\,,\\
&\{\ff_{i,i+1},\fft_{j,j+1}\}=-i\delta_{ij}\f{2}{\ka}\sinh\f{\ka(\fe_i+\fe_{i+1})}{2} 
+i\delta_{i,j-1}\lb \fe_{i+2,i} -\f{\ka}{2}\ff_{i,i+1}\fft_{i+1,i+2} \rb
+i\delta_{i,j+1}\lb \fe_{i-1,i+1} +\f{\ka}{2}\ff_{i,i+1}\fft_{i-1,i} \rb\,,\\
&\{\fe_i,\ff_{j,j+1}\}=i(\delta_{ij}+\delta_{ij+1})\ff_{j,j+1}\,,\quad
\{\fe_i,\fft_{j+1,j}\}=-i(\delta_{ij}+\delta_{ij+1})\fft_{j,j+1}\,,\quad
\{\ff_{i,i+1},\ff_{j,j+1}\}=\{\fft_{i,i+1},\fft_{j,j+1}\}=0\,.
\end{split}
\label{eq:qSO*2n_Poisson_algebra}
\ee
\end{prop}
\begin{proof}
The Poisson algebra \eqref{eq:qSO*2n_Poisson_algebra} can be directly calculated with (the tilde version of) the Poisson brackets \eqref{eq:poisson_t}, \eqref{eq:poisson_tau} and \eqref{eq:poisson_t_tau}. 
To get the first three lines of \eqref{eq:qSO*2n_Poisson_algebra}, it is also useful to use the following Poisson brackets.
\be\ba{ll}
 e^{\f{3\ka\Nt}{4}}\{\btt_-,e^{-\f{\ka \Nt}{2}}\ttau_- \} = ie^{\f{\ka}{4}(\Nt_1-\tN_0)}\equiv i \tlambda^{-1}\,, &
 e^{\f{3\ka \Nt}{4}}\{\btt_-,e^{-\f{\ka \Nt}{2}}\ttau_+ \} = 0 \\[0.15cm]
 e^{\f{3\ka \Nt}{4}}\{\btt_+,e^{-\f{\ka \Nt}{2}}\ttau_- \} = -i \ka \btzeta^\ka_1\bzeta^\ka_0 \equiv -i \btz \,, &
 e^{\f{3\ka \Nt}{4}}\{\btt_+,e^{-\f{\ka \Nt}{2}}\ttau_+ \} =ie^{\f{\ka}{4}(\Nt_0-\tN_1)}\equiv i\tlambda\,,\\[0.3cm]
 e^{-\f{3\ka \Nt}{4}}\{\bttau_- ,e^{\f{\ka \Nt}{2}}\tt_-\} = i e^{\f{\ka}{4}(\Nt_0-\Nt_1)}\equiv i\tlambda\,, &
 e^{-\f{3\ka \Nt}{4}}\{\bttau_- ,e^{\f{\ka \Nt}{2}}\tt_+\} = 0\,,\\[0.15cm]
 e^{-\f{3\ka \Nt}{4}}\{\bttau_+ ,e^{\f{\ka \Nt}{2}}\tt_-\} = i\ka\btzeta^\ka_1 \tzeta^\ka_0 \equiv i\btz\,,&
 e^{-\f{3\ka \Nt}{4}}\{\bttau_+ ,e^{\f{\ka \Nt}{2}}\tt_+\} = ie^{\f{\ka}{4}(\Nt_1-\Nt_0)}\equiv i \tlambda^{-1}\,.
\ea\ee
We use these result to show $\eg$$\{\fe_{i,i+1},\ff_{i-1,i}\}$. We first write that
\be\begin{split}
\{\fe_{i,i+1},e^{-\f{\ka \Nt_i}{2}} \ff_{i-1,i}\} 
&= \{\btt_{i,-}\ttau_{i+1,-} + \btt_{i,+}\ttau_{i+1,+}, \tt_{i-1,-}(e^{-\f{\ka \Nt_i}{2}}  \ttau_{i,+}) -\tt_{i-1,+}(e^{-\f{\ka \Nt_i}{2}} \ttau_{i,-})\}\\
&=ie^{-\f{3\ka \Nt_i}{4}} \lb \tt_{i-1,-}\tlambda_i \ttau_{i+1,+} + \tt_{i-1,+} \btz_{i}\ttau_{i+1,+} - \tt_{i-1,+}\tlambda_i^{-1}\ttau_{i+1,-} \rb\\
&\equiv ie^{-\f{3\ka \Nt_i}{4}}[\tt_{i-1}|\lt_{i}^{-1\,\dagger}|\ttau_{i+1}\ra 
\equiv  ie^{-\f{\ka \Nt_i}{2}} \ff_{i-1,i+1}\,,
\end{split}\ee
where the left-hand side can also be separated into 
$$\{\fe_{i,i+1},e^{-\f{\ka \Nt_i}{2}} \ff_{i-1,i}\} 
= e^{-\f{\ka \Nt_i}{2}}\{\fe_{i,i+1},\ff_{i-1,i}\}
- \f{i\ka}{2}e^{-\f{\ka \Nt_i}{2}}\fe_{i,i+1}\ff_{i-1,i} \,.$$
We then conclude that 
\be
\{\fe_{i,i+1},\ff_{j,j+1}\} = i\delta_{i,j+1}\lb \ff_{i-1,i+1} +\f{\ka}{2}\fe_{i,i+1}\ff_{i-1,i+1}\rb
\nn\ee
hence the first Poisson bracket in \eqref{eq:qSO*2n_Poisson_algebra}. The first three lines of \eqref{eq:qSO*2n_Poisson_algebra} can computed in the similar way.
\end{proof}

\medskip

Let us now discuss the quantization of the model. 


\section{From phase space to Hopf algebras}
\label{sec:phasehopf}

The relevant structures for this quantization are the  Hopf algebras $\UQ, $ $\UQI,$ and $\SU_q(2), \, \SUQI(2)$ with $q$ real. The necessity to have the Hopf algebras $\UQI,$ and $\SUQI(2)$ was perhaps not fully appreciated in the previous work \cite{Dupuis:2014fya}, though it appeared already in \cite{Stern:1993rk}.  

We are interested in  the quantization of the Poisson brackets \eqref{eq:poisson_left} and \eqref{eq:poisson_right} for a single ribbon. To this aim, we construct the operators associated to the classical variables (the holonomy-flux algebra) and  introduce the Hilbert space structure on which we represent these operators.

\subsection{Poisson bracket quantization }
As a first step, we introduce the deformation parameter, $q=e^{\hbar \kappa}$. Then  the classical  $r$-matrix is quantized as $r\rightarrow R$ with  
\be R=\mat{cccc}{q^{\f14} & 0 & 0 & 0 \\
 0 & q^{-\f14} &  q^{-\f14}(q^{\f12}-q^{-\f12}) & 0 \\ 
0 &0 & q^{-\f14} & 0 \\
 0 & 0 & 0 & q^{\f14}}\,
\approx \id\otimes \id +i\hbar r+O(\hbar^2)\,.
\label{eq:R_and_r} 
\ee
Note  that one obtains the inverse matrix $R^{-1}$ if one replaces $q$ by $q^{-1}$.  

\medskip 

We quantize the holonomies and fluxes to be matrices of operators $\ell\rightarrow\lh\,,u\rightarrow \uh\,,\lt\rightarrow\lth\,,\ut\rightarrow\uth$. 
The  quantization of the Poisson brackets \eqref{eq:poisson_left} and \eqref{eq:poisson_right} gives the following commutation relations for the matrices of operators \cite{Stern:1993rk,Alekseev:1994un}
\begin{align}
\ba{llll}
\RT\uh_1\uh_2 = \uh_2\uh_1\RT\,, &
R \lh_1\lh_2=\lh_2\lh_1 R\,, &
\lh_1\RT^{-1}\uh_2 =\uh_2\lh_1\,,&
\lh_2R^{-1}\uh_1 = \uh_1\lh_2\,, \\[0.15cm]
\RT^{-1} \uth_1\uth_2 =\uth_2\uth_1 \RT^{-1} \,, &
R^{-1} \lth_1\lth_2 = \lth_2 \lth_1 R^{-1} \,, &
\uth_2 \RT \lth_1 =\lth_1 \uth_2\,,&
\uth_1 R \lth_2 =\lth_2 \uth_1\,,\\[0.15cm]
\lth_1 \uh_2 \RT^{-1} = \uh_2 \lth_1 \,, &
R \uth_1 \lh_2 = \lh_2 \uth_1\,, &
\RT^{-1} \lh_1 \uth_2 = \uth_2 \lh_1\,, &
\uh_1\lth_2 R = \lth_2 \uh_1 \,.
\label{eq:RTT_all}
\ea
\end{align}
The Poisson brackets \eqref{eq:poisson_left}, \eqref{eq:poisson_right} and \eqref{eq:poisson_mix} are recovered at the first order through the map $[\hat{A},\hat{B}]=i\hbar\widehat{ \{A,B\}}$. 
 Note that $R^{-1}$ appears because of the minus sign difference  between the classical Poisson structures respectively defined in \eqref{eq:poisson_left} and in \eqref{eq:poisson_right}.  

\medskip 

The classical Casimir $r+\rT$ can be quantized as $\RT R$ and requesting this operator  to be a Casimir implies that 
\be
[\RT R, \lh_1\lh_2]\,=\,[\RT R, \uh_2\uh_1]\,=\,[\RT R,\lth_2\lth_1]\,=\, [\RT R,\uth_1\uth_2] =0\,.
\label{eq:Casimir_RRT}
\ee
Using this in \eqref{eq:RTT_all} leads to the following equivalent commutation relations
\be\ba{ccc}
\RT\uh_1\uh_2 = \uh_2\uh_1\RT 
&\Longleftrightarrow &
R^{-1}\uh_1\uh_2 = \uh_2\uh_1 R^{-1}\,, \\[0.2cm]
\RT^{-1} \uth_1\uth_2 =\uth_2\uth_1 \RT^{-1}
&\Longleftrightarrow &
R \uth_1\uth_2 = \uth_2\uth_1 R \,,
\ea
\label{eq:U_Ut_commutation_2}
\ee
which are more amenable to identify the relevant structure. 

The relations \eqref{eq:RTT_all} and \eqref{eq:U_Ut_commutation_2} define the algebra structure of the Hopf algebras $\UQI, \, \UQ$\footnote{Strictly speaking, these are the matrix elements of $L$ which belong to $\Fun_{q^{-1}}(\AN(2))$.}  and $\SU_{q}(2), \, \SUQI(2)$
\be\left|\ba{lll}
\lt\in \AN(2) &\rightarrow & \lth \in \AN_{q}
(2)\equiv \Fun_{q}(\AN(2)) \cong \UQ\\
\ut\in \SU(2) &\rightarrow & \uth \in \SU_{q}(2) \equiv \Fun_q(\SU(2))\\
\ell \in \AN(2) &\rightarrow & \lh \in \Fun_{q^{-1}}(\AN(2))\cong \UQI\\
u\in \SU(2) & \rightarrow & \uh\in \SUQI(2) 
\ea\right.\,,
\ee
We have in particular 
\be
\lh=\mat{cc}{K^{-1} & 0 \\ -q^{\f14}(q^{\f12}-q^{-\f12})J_+ & K}\,,\quad
\lth=\mat{cc}{\Kt & 0 \\ q^{-\f14}(q^{\f12}-q^{-\f12})\Jt_+ & \Kt^{-1}}\,,
\label{eq:quantum_fluxes}
\ee
where $(J_\pm,K=q^{\f{J_z}{2}})$ and $(\Jt_\pm,\Kt=q^{\f{\Jt_z}{2}})$ are two commuting copies of the $\UQ$ generators (see Appendix \ref{sec:UQ_and_SUq2}). The antipodes $\bS(\lh)$ and $S(\lth)$ (see \eqref{eq:cop_S_counit} and \eqref{eq:com_Sb_counit}) are given by acting the correspondent antipodes on all the matrix elements. That is
\begin{align}
\bS(\lh)&=\mat{cc}{\bS(K^{-1}) & 0 \\ -q^{\f14}(q^{\f12}-q^{-\f12})\bS(J_+) & \bS(K)}
=\mat{cc}{K & 0 \\ q^{-\f14}(q^{\f12}-q^{-\f12})J_+ & K^{-1}}
\,,\\
S(\lth)&=\mat{cc}{S(\Kt) & 0 \\ q^{-\f14}(q^{\f12}-q^{-\f12})S(\Jt_+) & S(\Kt^{-1})}
=\mat{cc}{\Kt^{-1} & 0 \\ -q^{\f14}(q^{\f12}-q^{-\f12})\Jt_+ & \Kt}\,.
\end{align}
The definitions of those Hopf algebras are given in Appendix \ref{sec:UQ_and_SUq2}. 
We note that the left Iwasawa decomposition leads to elements in the Hopf algebras, $\UQI$ and $\SUQI(2)$ while the right decomposition leads to 
elements in the Hopf algebras, $\UQ$ and $\SU_{q}(2)$. At the classical level, this is reflected in the presence of the minus sign difference between \eqref{eq:poisson_left}, \eqref{eq:poisson_right}, the  Poisson structures respectively for the elements $u$, $\ell$ of the left Iwasawa decomposition and for the elements $\ut$, $\lt$ of the right Iwasawa decomposition.

\subsection{The $\cR$-matrix contains the information about the flux and the holonomy}
Let us add some additional comments on the defining relations
\be
\lh_1\lh_2R^{-1}=R^{-1}\lh_2\lh_1\,,\quad
R^{-1}\uh_1\uh_2 =\uh_2\uh_1R^{-1}\,,\quad
\lth_1\lth_2R=R\lth_2\lth_1\,,\quad
R\uth_1\uth_2=\uth_2\uth_1R\,.
\label{eq:L_U_Lt_Ut_def}
\ee 
It is well-known \cite{saleur} that they can be obtained from the {\it quantum Yang-Baxter equation} (QYBE) 
\be
\cR_{12}\cR_{13}\cR_{23} = \cR_{23}\cR_{13}\cR_{12}\,,
\label{eq:QYBE}
\ee 
where we have used the standard notation $\cR_{12}=\sum \cR_{(1)}\otimes \cR_{(2)}\otimes\id\,,\cR_{23}=\id\otimes \cR_{(1)}\otimes \cR_{(2)}\,,\cR_{13}=\cR_{(1)}\otimes \id \otimes \cR_{(2)}$. 

The solution relevant to us   is specifically 
\be
\cR = q^{J_z \otimes J_z} \sum_{n=0}^{\infty} 
\f{(1-q^{-1})^n}{[n]!} q^{\f{n(n-1)}{4}} \lb q^{\f{J_z}{2}}J_+ \rb^n \otimes \lb q^{-\f{J_z}{2}}J_- \rb^n\,.
\label{eq:cR0}
\ee
where $[n]:=\f{q^{n/2}-q^{-n/2}}{q^{1/2}-q^{-1/2}}$ is called a $q$-number.

In the above quantization scheme, we have used this solution in the $\frac12\otimes \frac12$ representation, with the  generators  represented as $2\times 2$ matrices
\be
\rho(J_-)=\mat{cc}{0 & 0 \\ 1 & 0}\,,\quad
\rho(J_+)=\mat{cc}{0 & 1 \\ 0 & 0}\,,\quad
\rho(K)=\mat{cc}{q^{\f14} & 0 \\ 0 & q^{-\f14}}\quad  \rightarrow  \quad R=\rho(\cR)
\label{eq:UQ_fundamental_representation}
\ee

All the relations \eqref{eq:L_U_Lt_Ut_def} can be seen as different realizations of the QYBE \eqref{eq:QYBE} written in a specific representation. Indeed, in terms of the components of the $\cR$-matrix, the Yang-Baxter equation is written as
\be
\sum_{k_1,k_2,k_3}\cR^{i_1\phantom{k_1}i_2}_{\phantom{i_1}k_1\phantom{i_2}k_2}\cR'^{k_1\phantom{j_1}i_3}_{\phantom{k_1\,}j_1\phantom{i_3}k_3}\cR''^{k_2\phantom{j_2}k_3}_{\phantom{k_2\,\,\,}j_2\phantom{k_3}j_3}
=\sum_{k_1,k_2,k_3}\cR''^{i_2\phantom{k_2}i_3}_{\phantom{i_2\,\,\,}k_2\phantom{k_2}k_3}
\cR'^{i_1\phantom{k_1}k_3}_{\phantom{i_1\,}k_1\phantom{k_3}j_3}
\cR^{k_1\phantom{j_1}k_2}_{\phantom{k_1}j_1\phantom{k_2}j_2}\,,
\label{eq:QYBE_comp}
\ee
where $\cR,\cR',\cR''$ are different copies of the $\cR$-matrix. 
The first two indices ($i,j$) of $\cR^{i\phantom{j}k}_{\phantom{i}j\phantom{k}l}$ are the indices for $\cR_{(1)}$ and the last two indices ($k,l$) are the indices for $\cR_{(2)}$ given in the decomposition $\cR=\sum \cR_{(1)}\otimes \cR_{(2)}$.

\medskip 
 
Let us fix the representation of $\cR_{(2)},\cR'_{(2)}$ and $\cR''$ to be the fundamental representation of $\UQ$, then the indices $(i_{2},i_{3}),(j_{2},j_{3}),(k_{2},k_{3})\in\{-\f12,\f12\}$ in \eqref{eq:QYBE_comp}. 
In this representation, we then have \cite{saleur}
\begin{equation}
(\lth^k_{\phantom{k}l})^\alpha_{\phantom{\alpha}\beta}=\cR^{\alpha\phantom{\beta}k}_{\phantom{\alpha}\beta\phantom{k}l}
\end{equation}
where the indices $k,l=\pm \frac12$ are the indices labelling the matrix elements  of $\lth$, while $\alpha, \beta$ are the indices of the $\UQ$ generators in any representation. The QYBE \eqref{eq:QYBE_comp} thus reduces to $\lth_1\lth_2R=R\lth_2\lth_1$. 

On the other hand, fixing the representation of $\cR,\cR'_{(1)}$ and $\cR''_{(1)}$ to be the fundamental representation and using that 
\begin{equation}
(\uth^i_{\phantom{i}j})^\alpha_{\phantom{\alpha}\beta}=\cR^{i\phantom{j}\alpha}_{\phantom{i}j\phantom{\alpha}\beta}
\end{equation}
when $i,j \in \{-\frac12, \frac12\}$, the QYBE reduces to $R\uth_1\uth_2=\uth_2\uth_1R$.

In the same spirit, the first two equations in \eqref{eq:L_U_Lt_Ut_def} are the QYBE for the $\cR$-matrix of $\UQI$ in a given representation. 
Note that the $\cR$-matrix for $\UQI$ is simply the inverse of the $\cR$-matrix for $\UQ$. 
Therefore, the $\cR$-matrix captures the quantum holonomy and quantum flux information in its two sub-spaces. This gives a more geometrical interpretation to the $\cR$-matrix in terms of quantum ``holonomies'' either in some deformation of $\AN(2)$ or $\SU(2)$\footnote{Although we stick to the terminology that $\ell$ and $\lt$ are called fluxes, they are $\AN(2)$ holonomies in the ribbon picture as each is assigned to a side of the ribbon.}.

The construction of tensor operators (such as spinor and vector operators) usually requires some braiding defined in terms of the $\cR$-matrix to transform appropriately \cite{Rittenberg:1991tv, Quesne:1993}. We will show how this braiding  can be re-interpreted in a more geometrical setting, $\ie$in terms of parallel transport.

\section{Quantum spinorial representation of deformed lattice gauge theory}
\label{sec:quantum_spinors}

This section contains some of the key-results of the paper. In particular, after quantizing the deformed spinors, we will show how the definition of spinor operators on different Hilbert spaces, usually performed via the $\cR$-matrix, can be done using some parallel transport. This leads to a new geometrical interpretation of the $\cR$-matrix. We will also provide the quantization of the observables \eqref{eq:E12_classical} and show that they form a deformation of $\mathfrak{so}^*(2n)$. 

\subsection{Quantizing the spinors}

The quantization of the deformed variables $\zk_A, \bzeta_A^\ka , N_A$  will give rise to the $q$-deformation of the  Jordan map for $\su(2)$.  
Indeed these variables can be quantized as $q$-boson operators:
 the variables  $\zk_A$ are quantized as $q$-boson annihilation operators, the variables $\bzeta^\ka_A$ as  $q$-boson creation operators and the variables $N_A$ as number operators. 
 Explicitly, 
 \be \ba{lll}
 (\zeta^\kappa_0, \zeta^\kappa_1)\dr (a,b)\,, \quad &
 (\bzeta^\kappa_0, \bzeta^\kappa_1)\dr   (a^\dagger, b^\dagger)\,, \quad & 
 (N_0, N_1) \dr (N_a, N_b)\,,\\
 (\tzeta^\ka_0,\tzeta^\ka_1 )\dr (\at,\bt)\,,\quad &
 (\btzeta^\ka_0,\btzeta^\ka_1 )\dr (\at^\dagger ,\bt^\dagger)\,,\quad &
 (\Nt_0,\Nt_1) \dr (\Nt_a,\Nt_b)\,.
 \ea
 \label{eq:quantization_spinors}
 \ee
 These $q$-harmonic oscillators obey the following commutation rules
 \be \label{eq:a_adagger}
 a a^\dagger - q^{\mp\demi}a^\dagger a = q^{\pm \f{N_a}{2}}\,,\quad
 a^{\dagger}a-q^{\pm \f12}aa^{\dagger} =-q^{\pm\f{N_a+1}{2}}\,,\quad
 [N_a,a^\dagger]=a^\dagger\,,\quad
 [N_a,a]=-a\,,
 \ee
 from which one can deduce
 \be\label{eq:def_a}
 q^{N_{a}/2}a^\dagger= q^{1/2}a^\dagger q^{N_{a}/2}, \quad q^{N_{a}/2} a = q^{-1/2} a q^{N_{a}/2}\,,\quad
a^\dagger a = [N_a]\equiv\f{q^{N_a/2}-q^{-N_a/2}}{q^\demi - q^{-\demi}} , \quad aa^\dagger= [N_{a}+1] \,. 
 \ee
Similar relations hold for the operators ($b$, $b^\dagger$, $N_b$) and the tilde variables. The different sets of $q$-boson operators ($a, a^\dagger, N_a$)  ($b, b^\dagger, N_b$),  ($\at,\adt,\Nt_a$) and ($\bt,\bdt,\Nt_b$) all commute with each other. 

States can be labeled by their occupation numbers, $|n_a\rangle = a^{\dagger n_a}|0\rangle/\sqrt{[n_a]}$ and $|n_b\rangle = b^{\dagger n_b}|0\rangle/\sqrt{[n_b]}$, and
\begin{equation}
|n_a, n_b\rangle_{\text{HO}} = |n_a\rangle \otimes |n_b\rangle.
\end{equation}

The $q$-deformed Jordan map is \cite{Biedenharn:1996vv},
\be \label{qJordanMap}
J_+=a^\dagger b\,,\quad
J_-=ab^\dagger\,,\quad
K=q^{\f{J_z}{2}}= q^{\f{N_a-N_b}{4}}\,,\quad
\Jt_+=\at^\dagger \bt\,,\quad
\Jt_-=\at\bt^\dagger\,,\quad
\Kt=q^{\f{\Jt_z}{2}}= q^{\f{\Nt_a-\Nt_b}{4}}\,.
\ee
 Indeed, with the quantization map \eqref{eq:quantization_spinors}, we recover the classical generators $z, \bz, \lambda$ and $\tz,\btz,\tlambda$ at the linear $\hbar$-order of the quantum fluxes \eqref{eq:quantum_fluxes} by taking $q=e^{\ka\hbar}=1+\ka\hbar +O(\hbar^2)$,
 \be
\left|\ba{rll}
-q^{\f14}(q^{\f12}-q^{-\f12})J_+  &\dr & z = -\ka{\bzk}_0 \zk_1  \\[0.15cm]
K\mone  & \dr &  \lambda= \exp(\f{\kappa}{4}(N_1-N_0)) 
\ea\right.,\quad
\left|\ba{rll}
q^{\f14}(q^{\f12}-q^{-\f12})\Jt_+ &\dr & \tz=\ka\btzeta^\ka_0\tzeta_1^\ka \\[0.15cm]
\Kt &\dr & \tlambda =\exp (\f{\ka}{4}(\Nt_0-\Nt_1))
\ea\right.. 
\ee

\medskip 

We define the right adjoint action\footnotemark{}, denoted as $\coact$ (\resp $\bcoact$), of $\UQ$ (\resp $\UQI$)  on some operator $\cO$:
\bes
&& J_\pm \coact \cO =  S(J_\pm) \cO K+ S(K\mone) \cO J_\pm
=-q^{\pm\f12}J_\pm \cO K+ K \cO J_\pm\,, \quad 
K  \coact  \cO = S(K)  \cO K
=K^{-1}\cO K\,, 
\\[0.15cm]
&& J_\pm \bcoact \cO =  \bS(J_\pm) \cO K^{-1}+ \bS(K) \cO J_\pm
=-q^{\mp\f12}J_\pm \cO K^{-1}+ K^{-1} \cO J_\pm\,, \quad 
K  \bcoact  \cO = \bS(K)  \cO K
=K^{-1}\cO K\,.
\ees
\footnotetext{
Given a generator $x$ of a Hopf algebra $H$ with coproduct $\cop(x)=\sum x_{(1)}\otimes x_{(2)}$, there are two kinds of adjoint actions on operators $\cO$'s of $H$ namely the left adjoint action $x \act \cO :=\sum x_{(1)}\cO S(x_{(2)})$ and the right adjoint action $x\coact \cO :=\sum S(x_{(1)}) \cO x_{(2)}$, where $S$ is the antipode of $H$.
}
Let $\cV^j$ be the irreducible representation of $\UQ$ of dimension $2j+1$. The basis state $\jm\in\cV^j$ of fixed magnetic number $m$ is the Fock state $|n_a,n_b\ra_{\text{HO}}$,
\be
\jm=|j+m, j-m\rangle_{\text{HO}}
\ee
$\ie$ $j=\f12 (n_a+n_b)$ and $m=\f12(n_a-n_b)$. The $q$-bosons act on those states as
\be\ba{ll}
a^{\dagger} \jm =\sqrt{[j+m+1]}\,|j+\f12,m+\f12\ra\,, &
a \jm =\sqrt{[j+m]}\,|j-\f12,m-\f12\ra\,, \\[0.15cm]
b^{\dagger} \jm = \sqrt{[j-m+1]}\, |j+\f12,m-\f12\ra\,, &
b \jm = \sqrt{[j-m]}\, |j-\f12,m+\f12\ra\,, \\[0.15cm]
N_a \jm = (j+m)\,\jm\,, &
N_b \jm =(j-m) \,\jm\,.
\ea\ee

\begin{figure}[h!]
	\centering
	\begin{minipage}{0.45\textwidth}
\centering
	\begin{tikzpicture}[scale=1.5]
\coordinate (o) at (0,0);
\coordinate (a1) at ([shift=(150:0.5cm)]o);
\coordinate (a2) at ([shift=(30:0.5cm)]o);
\coordinate (b1) at ([shift=(90:1.5cm)]a1);
\coordinate (b2) at ([shift=(90:1.5cm)]a2);
\coordinate (a3) at ([shift=(-90:0.5cm)]o);
\coordinate (c1) at ([shift=(-150:1.5cm)]a3);
\coordinate (c2) at ([shift=(-150:1.5cm)]a1);
\coordinate (d1) at ([shift=(-30:1.5cm)]a3);
\coordinate (d2) at ([shift=(-30:1.5cm)]a2);

\coordinate (A) at ([shift=(90:2.3cm)]o);
\coordinate (C) at ([shift=(-30:2.3cm)]o);
\coordinate (B) at ([shift=(210:2.3cm)]o);

\draw[gray,decoration={markings,mark=at position 0.55 with {\arrow[scale=1.5,>=stealth]{<}}},postaction={decorate}] (o) -- node[right,pos=.5]{$e_1$}(A);

\draw[thick,dashed,decoration={markings,mark=at position 0.55 with {\arrow[scale=1.5,>=stealth]{>}}},postaction={decorate}] (a1) -- node[above,pos=.5]{$\lth$}(a2);

\draw[thick,dashed,decoration={markings,mark=at position 0.55 with {\arrow[scale=1.5,>=stealth]{>}}},postaction={decorate}] (b1) -- node[above,pos=.5]{$\lh$}(b2);

\draw[thick,decoration={markings,mark=at position 0.55 with {\arrow[scale=1.5,>=stealth]{<}}},postaction={decorate}] (b1) -- node[left,pos=.5]{$$}(a1);
\draw[thick,decoration={markings,mark=at position 0.55 with {\arrow[scale=1.5,>=stealth]{<}}},postaction={decorate}] (b2) -- node[right,pos=.5]{$$}(a2);

\draw[red] (a2) node{$\bullet$};
\draw (a1) node{$\bullet$};
\draw (a2) node{$\bullet$};
\draw[blue] (a1) node[below left]{$\bftt^\epsilon$};
\draw[blue] (a2) node[below right]{$\mtt^\epsilon$};	
\draw[blue] (b1) node[above left]{$\mt^\epsilon$};
\draw[blue] (b2) node[above right]{$\bft^\epsilon$};	
	\end{tikzpicture}
\end{minipage} 
\caption{The reference ribbon. The spinor operators $\bft^\epsilon$ and $\bftt^\epsilon$ are   $\UQ$ quantum spinors, while $\mt^\epsilon$ and $\mtt^\epsilon$ are $\UQI$ quantum spinors.}
\label{fig:reference}
\end{figure}
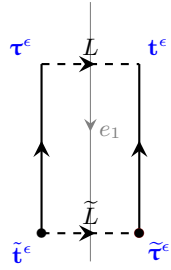

With the quantization map given above, we are now ready to define the $\UQ$ and $\UQI$ quantum spinors, 
which decorate the ribbon as in figure \ref{fig:reference}. 
A $\UQ$ (\resp $\UQI$) quantum spinor, denoted as ${\bf T}=\mat{c}{{\bf T}_-\\{\bf T}_+}$, by definition should transform under the $\UQ$ (\resp $\UQI$) adjoint action as a spinor, $\ie$
\be
J_\pm \bullet {\bf T}_\pm=0\,,\quad
J_\pm\bullet {\bf T}_\mp={\bf T}_\mp\,,\quad
K\bullet {\bf T}_\pm =q^{\mp\f14}{\bf T}_\pm\,,
\label{eq:adjoint_action_for_spinors}
\ee
where $\bullet$ is the right adjoint action (which can be either $\coact$ or $\bcoact$).
\begin{remark}
According to Biedenharn's terminology \cite{Biedenharn:1996vv}, the relations \eqref{eq:adjoint_action_for_spinors} define what he calls {\it conjugate spinors}. This is what we will call the \textit{right adjoint quantum spinors} in this article. A left adjoint quantum spinor, or a quantum spinor according to  Biedenharn's terminology, is defined by the $\UQ$ or $\UQI$ left adjoint action. Denote uniformly the $\UQ$ or $\UQI$ left adjoint action by $\circ$, then the left adjoint action of the  generators on a left adjoint quantum spinor, say ${\bf T}'$, is
\be
J_\pm \circ {\bf T}'_\pm=0\,,\quad
J_\pm \circ {\bf T}'_\mp={\bf T}'_\mp\,,\quad
K \circ {\bf T}'_\pm =q^{\pm\f14}{\bf T}'_\pm\,.
\nn\ee
Note the different behavior under the action of $K$ compared to \eqref{eq:adjoint_action_for_spinors}. A $\UQ$ right adjoint quantum spinor $_q{\bf T}$ can be obtained via a $\UQI$ left adjoint quantum spinor $_{q^{-1}}{\bf T}'$ with the relation $_q{\bf T}_A = (-1)^{\f12-A}q^{\f{A}{2}}{}_{q^{-1}}{\bf T}'_A$, while a $\UQI$ right adjoint quantum spinor $_{q^{-1}}{\bf T}$ can be obtained via an $\UQ$ left adjoint quantum spinor $_q{\bf T}'$ with the relation $_{q^{-1}}{\bf T}_A = (-1)^{\f12-A}q^{-\f{A}{2}}{}_{q}{\bf T}'_A$.
\end{remark}

A spinor operator is a special example of a tensor operator $\T^{j=\demi}$. A tensor operator $\T^j$ associated with the representation $j$ transforms under the adjoint action as an element of the representation $j$. The Wigner-Eckart theorem provides the matrix elements of any tensor operator $\T^j$.

\begin{theorem}[Wigner-Eckart Theorem for $\UQ$ \cite{Biedenharn:1996vv} ]
The matrix element of a tensor operator $\T^j$ of rank $j$ with $j$  an irreducible representation of $\UQ$  is proportional to the $q$-WCG coefficient:
\be
\jml \T^j_m \jmr = N^j_{j_1j_2} 
\,_{q}C^{\,j_1\,\,\,j\,\,\,\,j_2}_{m_1\,m\,m_2}
\,,
\ee 
where $\T^j_m$ is the $m$-th component of $\T^j$, $_{q}C^{\,j_1\,\,\,j\,\,\,\,j_2}_{m_1\,m\,m_2}$ is the $q$-WCG coefficient for coupling $j_1$ and $j$ to get $j_2$ and $N^j_{j_1j_2}$ is a constant independent of $m,m_1,m_2$.
\end{theorem}

The quantization map \eqref{eq:quantization_spinors} leads to the quantum spinors defined as
\be\ba{lllllll}
|t\ra=\mat{c}{e^{\f{\ka N_1}{4}}\zeta^\ka_0 \\ e^{-\f{\ka N_0}{4}}\zeta^\ka_1}
&\longrightarrow &
\bft^-=\mat{c}{\bft^-_-\\\bft^-_+}=\mat{c}{q^{\f{N_b}{4}}a \\q^{-\f{N_a}{4}}b}\,,
&\quad &
|t] =\mat{c}{-e^{-\f{\ka N_0}{4}}\bzeta^\ka_1 \\ e^{\f{\ka N_1}{4}}\bzeta^\ka_0}
&\longrightarrow &
\bft^+=\mat{c}{\bft^+_-\\ \bft^+_+} =\mat{c}{-b^\dagger q^{-\f{N_a+1}{4}}\\ a^\dagger q^{\f{N_b+1}{4}}}\,,
\\[0.4cm]
|\tau\ra =\mat{c}{e^{-\f{\ka N_1}{4}}\zeta_0^\ka \\ e^{\f{\ka N_0}{4}}\zeta^\ka_1}
&\longrightarrow &
\mt^-=\mat{c}{\mt^-_-\\ \mt^-_+}=\mat{c}{q^{-\f{N_b}{4}}a\\ q^{\f{N_a}{4}}b }\,,
 &\quad &
|\tau] = \mat{c}{- e^{\f{\ka N_0}{4}}\bzeta^\ka_1 \\ e^{-\f{\ka N_1}{4}}\bzeta_0^\ka}
&\longrightarrow &
\mt^+ =\mat{c}{\mt^+_-\\ \mt^+_+} =\mat{c}{-b^\dagger q^{\f{N_a+1}{4}} \\ a^\dagger q^{-\f{N_b+1}{4}}} \,,\\[0.4cm]
|\tt\ra =\mat{c}{e^{\f{\ka \Nt_1}{4}}\tzeta^\ka_0 \\ e^{-\f{\ka \Nt_0}{4}}\tzeta^\ka_1}
&\longrightarrow &
\bftt^-=\mat{c}{\bftt^-_-\\ \bftt^-_+}=\mat{c}{q^{\f{\Nt_b}{4}}\at \\ q^{-\f{\Nt_a}{4}}\bt}\,,
 &\quad &
|\tt]=\mat{c}{-e^{-\f{\ka \Nt_0}{4}}\btzeta^\ka_1 \\ e^{\f{\ka \Nt_1}{4}}\btzeta^\ka_0}
&\longrightarrow &
\bftt^+=\mat{c}{\bftt^+_-\\ \bftt^+_+}=\mat{c}{-\bt^\dagger q^{-\f{\Nt_a+1}{4}}\\ \at^\dagger q^{\f{\Nt_b+1}{4}}}\,,\\[0.4cm]
|\ttau\ra =\mat{c}{e^{-\f{\ka \Nt_1}{4}}\tzeta^\ka_0 \\ e^{\f{\ka \Nt_0}{4}}\tzeta^\ka_1}
&\longrightarrow &
\mtt^-=\mat{c}{\mtt^-_-\\\mtt^-_+}=\mat{c}{q^{-\f{\Nt_b}{4}}\at\\ q^{\f{\Nt_a}{4}}\bt }\,,
 &\quad &
|\ttau]=\mat{c}{ -e^{\f{\ka \Nt_0}{4}}\btzeta^\ka_1 \\ e^{-\f{\ka \Nt_1}{4}}\btzeta^\ka_0 }
&\longrightarrow &
\mtt^+ =\mat{c}{\mtt^+_-\\ \mtt^+_+}=\mat{c}{-\bt^\dagger q^{\f{\Nt_a+1}{4}} \\ \at^\dagger q^{-\f{\Nt_b+1}{4}}}\,.
\ea
\label{eq:def_quantum_spinor}
\ee
The spinors $t^\epsilon$ and $\tt^\epsilon$ are quantized as $\UQ$  spinor operators while the (braided) spinors $\tau^\epsilon$ and $\ttau^\epsilon$ are quantized as $\UQI$  spinor operators. Indeed, under the right adjoint action, these quantum spinors transform as desired:
\be \tabl{c}{
J_{\pm}\coact \bft^\epsilon_{\pm}= 0, \quad 
J_{\pm}\coact \bft^\epsilon_{\mp}=\bft^\epsilon_{\pm}, \quad 
K \coact \bft^\epsilon_{\pm}=q^{\mp \f14} \bft^\epsilon_\pm \,,\\[0.2cm]
J_{\pm}\coact \bftt^\epsilon_{\pm}= 0, \quad 
J_{\pm}\coact \bftt^\epsilon_{\mp}=\bftt^\epsilon_{\pm}, \quad 
K \coact \bftt^\epsilon_{\pm}=q^{\mp \f14} \bftt^\epsilon_\pm\,,\\[0.3cm]
J_{\pm}\bcoact \mt^\epsilon_{\pm}=0, \quad 
J_{\pm}\bcoact \mt^\epsilon_{\mp}= \mt^\epsilon_{\pm}, \quad 
K \bcoact \mt^\epsilon_{\pm}=q^{\mp \f14} \mt^\epsilon_\pm\,, \\[0.2cm]
J_{\pm}\bcoact \mtt^\epsilon_{\pm}= 0, \quad 
J_{\pm}\bcoact \mtt^\epsilon_{\mp}=\mtt_{\pm}, \quad 
K \bcoact \mtt^\epsilon_{\pm}=q^{\mp \f14} \mtt^\epsilon_\pm \,.
}\ee
As a consequence, the Wigner-Eckart theorem tells us that
\begin{subequations}
\begin{alignat}{4}
&\jml \bft^\epsilon_m \jmr =  
\delta_{j_{1}, j_2+\epsilon/2}\, \sqrt{[d_{j_1}]}\,
_{q}C^{j_1 \,\,\, \,\,\f12 \,\,\,\,\,j_2}_{m_1 \,-m \,m_2}
\,, 
\label{eq:matrix_t}
\\
&\jml  \mt^\epsilon_m \jmr =  
\delta_{j_{1}, j_2+\epsilon/2} \,\sqrt{[d_{j_1}]}\, 
_{q^{-1}}C^{j_1 \,\,\, \,\,\f12 \,\,\,\,\,j_2}_{m_1 \,-m \,m_2}
\,,   
\label{eq:matrix_tau}
\\
&\jml  \bftt^\epsilon_m \jmr =   
\delta_{j_{1}, j_2+\epsilon/2}\, \sqrt{[d_{j_1}]}\,
_{q}C^{j_1 \,\,\, \,\,\f12 \,\,\,\,\,j_2}_{m_1 \,-m \,m_2}
\,,   
\label{eq:matrix_tt}\\
&\jml  \mtt^\epsilon_m \jmr =   
\delta_{j_{1}, j_2+\epsilon/2} \,\sqrt{[d_{j_1}]}\, 
_{q^{-1}}C^{j_1 \,\,\, \,\,\f12 \,\,\,\,\,j_2}_{m_1 \,-m \,m_2}
\,.
\label{eq:matrix_ttau}
\end{alignat}
\end{subequations}
Therefore, as in the quantum fluxes, we again see both the $\UQ$ and $\UQI$ structures appearing upon quantization. We decorate the ribbon with spinor operators as in figure \ref{fig:reference}.  
 $\bft^\epsilon$ and $\bftt^\epsilon$ are the $\UQ$ quantum spinors, while $\mt^\epsilon$ and $\mtt^\epsilon$ are the $\UQI$ quantum spinors both in the sense of the right adjoint action.  The quantum spinor components satisfy the commutation relations
\be
\bft^\epsilon_-\bft^\epsilon_+=q^{-\f12}\bft^\epsilon_+\bft^\epsilon_-\,,\quad
\mt^\epsilon_- \mt^\epsilon_+=q^{\f12}\mt^\epsilon_+\mt^\epsilon_-\,,\quad
\bftt^\epsilon_- \bftt^\epsilon_+=q^{-\f12} \bftt^\epsilon_+ \bftt^\epsilon_-\,,\quad
\mtt^\epsilon_-\mtt^\epsilon_+=q^{\f12}\mtt^\epsilon_+\mtt^\epsilon_-\,,\quad
\epsilon=\pm\,.
\label{eq:commute_spinors}
\ee

\medskip 

We define the inner products of the spinors with a bilinear form $\Bq$ determined by the $q$-WCG coefficient $\pm\sqrt{[2]}\, _qC^{\demi \, \demi \, 0}_{m\; n \, 0}=\pm\delta_{m,-n}(-1)^{1/2-m}q^{m/2}$ with $q$ compatible with the spinor nature. $\Bq$ thus defines a (non-symmetric) metric on the spinors. We denote the inner products as spinor brackets in the following way
\begin{equation} \label{eq:spinor_inner_product}
\begin{aligned}
\la \bft |\bft\ra &:=
\Bq(\bft^+ ,\bft^- )=-\sqrt{[2]} \, _{q}C^{\demi \,\, \,\,\,\demi\,\,\, \, 0}_{m\; -m \, 0} \,\bft^+_{-m}\bft^-_{m}= [N]\,,\\
\la \mt|\mt\ra &:= 
\Bq(\mt^+ ,\mt^-)=- \sqrt{[2]} \, _{q^{-1}}C^{\demi \,\, \,\,\,\demi\,\,\, \, 0}_{m\; -m \, 0} \,\mt^+_{-m}\mt^-_{m}
= [N]\,,\\
\la \bftt|\bftt\ra &:=
\Bq(\bftt^+ ,\bftt^- )= -\sqrt{[2]} \, _{q}C^{\demi \,\, \,\,\,\demi\,\,\, \, 0}_{m\; -m \, 0} \,\bftt^+_{-m}\bftt^-_{m}
=  [\Nt]\,,\\
\la \mtt |\mtt\ra &:=
\Bqi(\mtt^+,\mtt^- )=-\sqrt{[2]} \, _{q^{-1}}C^{\demi \,\, \,\,\,\demi\,\,\, \, 0}_{m\; -m \, 0} \,\mtt^+_{-m}\mtt^-_{m}
=[\Nt]\,,
\end{aligned}
\end{equation}
as well as
\begin{equation}
\begin{aligned}
[ \bft | \bft ] &:=
\Bq(\bft^-,\bft^+ )=  \sqrt{[2]} \, _{q}C^{\demi \,\, \,\,\,\demi\,\,\, \, 0}_{m\; -m \, 0} \,\bft^-_{-m}\bft^+_{m}
= [N+2]\,,\\
[\mt|\mt] &:=
\Bqi(\mt^-,\mt^+)=  \sqrt{[2]} \, _{q^{-1}}C^{\demi \,\, \,\,\,\demi\,\,\, \, 0}_{m\; -m \, 0} \,\mt^-_{-m}\mt^+_{m}
=  [N+2]\,,\\
[\bftt|\bftt] &:=
\Bq(\bftt^-,\bftt^+)= \sqrt{[2]} \, _{q}C^{\demi \,\, \,\,\,\demi\,\,\, \, 0}_{m\; -m \, 0} \,\bftt^-_{-m}\bftt^+_{m}
=[\Nt+2]\,,
\\
[ \mtt | \mtt ] &:=
\Bqi(\mtt^-,\mtt^+ )=  \sqrt{[2]} \, _{q^{-1}}C^{\demi \,\, \,\,\,\demi\,\,\, \, 0}_{m\; -m \, 0} \,\mtt^-_{-m}\mtt^+_{m}
= [\Nt+2]\,,
\end{aligned}
\end{equation}
while it can be checked directly that the remaining vanish,
\begin{equation}
\begin{aligned}
[\bft|\bft\ra: &=\Bq(\bft^-,\bft^-)=0 = \Bq(\bft^+,\bft^+)=:\la \bft|\bft]\,,\\
[\mt|\mt\ra &:= \Bqi(\mt^-,\mt^-)=0=\Bqi(\mt^+,\mt^+)=:\la \mt|\mt]\,,\\
[\bftt|\bftt\ra &:=\Bq(\bftt^-,\bftt^-)=0=\Bq(\bftt^+,\bftt^+)=:\la \bftt|\bftt]\,,\\
[\mtt|\mtt\ra &:=\Bqi(\mtt^-,\mtt^-)=0=\Bqi(\mtt^+,\mbtt^+)=:\la \mtt|\mtt]\,.                       
\end{aligned}
\end{equation}
Unlike in the classical case, the norms of the spinors and their duals are not equal, $\la \cdot |\cdot\ra \neq [\cdot|\cdot]$, due to the non-commutativity \eqref{eq:commute_spinors} of the spinor components. Furthermore, one can get $[N+1]$ or $[\Nt+1]$ by the following inner products,
\begin{align}
[N+1]&=q^{-\f14} (\bft_-^-\bft_+^+-\bft^+_-\bft_+^-  ) 
= q^{\f14} ( \bft^+_+\bft^-_--\bft_+^- \bft_-^+ ) 
=q^{-\f14} (\mt_+^+\mt^-_- -\mt_+^-\mt^+_-)
=q^{\f14} (\mt^-_-\mt_+^+ -\mt^+_-\mt^-_+)\,, 
\label{eq:spinor_inner_product_2}
\\
[\Nt+1]&=q^{-\f14} (\bftt_-^-\bftt_+^+ - \bftt_-^+\bftt_+^- )
=q^{\f14} (\bftt_+^+\bftt_-^- -\bftt_+^-\bftt_-^+ )
=q^{-\f14} (\mtt_+^+\mtt_-^--\mtt_+^-\mtt_-^+ )
=q^{\f14} ( \mtt_-^-\mtt_+^+ -\mtt^+_-\mtt^-_+)\,.
\label{eq:spinor_inner_product_3}
\end{align}
They are actually those we will use to reconstruct the quantum holonomies. 

\subsection{Recovering the quantum holonomy-flux algebra}
\label{sec:quantum_holonomy_from_spinors}
Both the quantum fluxes and quantum holonomies can be built from the quantum spinors in a neat way as their classical counterparts \eqref{eq:u_from_spinor}.

\paragraph*{\textbf{Holonomies.}} We start with the following proposition.
\begin{prop}
Impose the norm matching constraint $N=\Nt$. 
Then the operator matrix $\uh=\mat{cc}{\uh_{--} & \uh_{-+} \\ \uh_{+-} & \uh_{++}}$ whose matrix elements are given by 
\be
\uh_{AB}=(-1)^{\f12-B}q^{\f{B}{2}}\sum_{\epsilon=\pm} \mt_A^\epsilon \bftt_{-B}^\epsilon \, \f{1}{[N+1]}
\label{eq:def_U_with_spinors}
\ee
is an $\SUQI(2)$ quantum matrix.
The operator matrix $\uth=\mat{cc}{\uth_{--} & \uth_{-+} \\ \uth_{+-} & \uth_{++}}$ whose matrix elements are given by
\be
\uth_{AB}=\f{1}{[\Nt+1]} (-1)^{\f12+B}q^{-\f{B}{2}}\sum_{\epsilon=\pm} \bft_A^\epsilon \mtt_{-B}^\epsilon 
\label{eq:def_Ut_with_spinors}
\ee
is an $\SU_{q}(2)$ quantum matrix.

In addition, together with the fluxes $L$ and $\Lt$ \eqref{eq:quantum_fluxes} defined in terms of the $\UQ$ generators given by the Jordan map \eqref{qJordanMap}, the holonomies defined this way satisfy the commutation relations \eqref{eq:RTT_all}.
\end{prop}
\begin{proof}
By repeatedly applying \eqref{eq:commute_spinors}--\eqref{eq:spinor_inner_product_3} and the commutation relation of the spinor components and the norm factor
\be
\f{1}{[N+1]}{\bf T}_m^\epsilon = {\bf T}_m^\epsilon \f{1}{[N+1+\epsilon]}\,,\quad
\f{1}{[\Nt+1]}\tilde{{\bf T}}_m^\epsilon =\tilde{{\bf T}}_m^\epsilon \f{1}{[\Nt+1+\epsilon]}\,,\quad
{\bf T}=\bft\,,\mt\,,\quad
\tilde{{\bf T}}=\bftt\,,\mtt\,,
\ee
one can compute that
\begin{align*}
&\uh_{--}\uh_{-+}=q^{\f12}\uh_{-+}\uh_{--}\,, \quad
\uh_{--}\uh_{+-}=q^{\f12}\uh_{+-}\uh_{--}\,, \quad
\uh_{-+}\uh_{++}=q^{\f12}\uh_{++}\uh_{-+}\,, \quad
\uh_{+-}\uh_{++}=q^{\f12}\uh_{++}\uh_{+-}\,, \\&
[\uh_{--} ,\uh_{++}] = -(q^{\f12}-q^{-\f12}) \uh_{-+}\uh_{+-}\,,\quad
[\uh_{-+} ,\uh_{+-}] = 0\,,\quad
\text{det}_{q^{-1}} \uh \equiv \uh_{--}\uh_{++} -q^{\f12} \uh_{-+}\uh_{+-} =\id\,,\\&
\uth_{--}\uth_{-+}=q^{-\f12}\uth_{-+}\uth_{--}\,, \quad
\uth_{--}\uth_{+-}=q^{-\f12}\uth_{+-}\uth_{--}\,, \quad
\uth_{-+}\uth_{++}=q^{-\f12}\uth_{++}\uth_{-+}\,, \quad
\uth_{+-}\uth_{++}=q^{-\f12}\uth_{++}\uth_{+-}\,, \\&
[\uth_{--} ,\uth_{++}] = (q^{\f12}-q^{-\f12}) \uth_{-+}\uth_{+-}\,,\quad
[\uth_{-+} ,\uth_{+-}] = 0\,,\quad
\text{det}_{q} \uth \equiv \uth_{--}\uth_{++} -q^{-\f12} \uth_{-+}\uth_{+-} =\id\,.
\end{align*}
Referring to Definition \ref{def:SUq2}, we conclude that $\uh$ is an $\SUQI(2)$ quantum matrix and $\uth$ is an $\SU_q(2)$ quantum matrix. 

\smallskip 

Using the Jordan map \eqref{qJordanMap}, the commutation relations between the $\UQ$ generators and the quantum spinors read
\begin{align}
\bft_\pm^\epsilon K = q^{\mp\f14}K \bft_\pm^\epsilon\,,\quad
\mt_\pm^\epsilon K = q^{\mp\f14}K \mt_\pm^\epsilon\,,\quad
\bftt_\pm^\epsilon \Kt = q^{\mp\f14}\Kt \bftt_\pm^\epsilon\,,\quad
\mtt_\pm^\epsilon \Kt = q^{\mp\f14}\Kt \mtt_\pm^\epsilon\,,\nn\\
\bft_\mp^\epsilon J_\pm -q^{\pm\f14}J_\pm \bft_\mp^\epsilon =K^{-1}\bft_\pm^\epsilon\,,\quad
\bft_\mp J_\mp =q^{\pm\f14}J_\mp\bft_\mp^\epsilon\,,\quad
\mt_\mp^\epsilon J_\pm - q^{\mp\f14}J_\pm \mt_\mp^\epsilon =K \mt_\pm^\epsilon\,,\quad
\mt_\mp^\epsilon J_\mp =q^{\mp\f14}J_\mp \mt_\mp^\epsilon\,, \\
\bftt_\mp^\epsilon\Jt_\pm -q^{\pm\f14}\Jt_\pm \bftt_\mp^\epsilon =\Kt^{-1}\bftt_\pm^\epsilon\,,\quad
\bftt_\mp \Jt_\mp =q^{\pm\f14}\Jt_\mp \bftt_\mp^\epsilon\,,\quad
\mtt_\mp^\epsilon \Jt_\pm - q^{\mp\f14}\Jt_\pm \mtt_\mp^\epsilon =\Kt \mtt_\pm^\epsilon\,,\quad
\mtt_\mp^\epsilon \Jt_\mp =q^{\mp\f14}\Jt_\mp \mtt_\mp^\epsilon\,,\nn
\end{align}
one can show that the commutation relations in \eqref{eq:RTT_all} are satisfied given the definition of the quantum holonomies \eqref{eq:def_U_with_spinors}, \eqref{eq:def_Ut_with_spinors} and the quantum fluxes \eqref{eq:quantum_fluxes}.
\end{proof}

\medskip
\paragraph*{\textbf{Flux vectors.}} 
We now reconstruct the quantization of the vectors $X$ and $X^\text{op}$ from \eqref{eq:def_X_Xop} in terms of the quantum spinors. They become $\UQ$ and $\UQI$ vector operators respectively, $\ie$spin 1 tensor operators.  The $\UQ$ quantum vectors can be built from the $\UQ$ spinors $\bft^{\epsilon}$ and $\bftt^\epsilon$ and the $\UQI$ vectors can be built from the $\UQI$ spinors $\mt^\epsilon$ and $\mtt^\epsilon$. 

Using the $q$-WCG coupling, one can define the $\UQ$ right adjoint vectors as \cite{Biedenharn:1996vv} 
\be
\bfX_{A} = \sum_{\substack{m,n=\pm\f12\\m+n=A}}\,_{q}C^{\,\,\f12\,\,\,\,\,\f12\,\,\,\,\,\,1}_{-m\,-n\,-A} \bft^+_m \bft^-_n \,,\quad A=0,\pm 1\,.
\ee
In components they read
\begin{align}
\bfX_0 &= _qC^{\,\f12\,\,\f12\,\,1}_{\f12\,-\f12\,0}\, \bft^+_-\bft^-_+ 
+\, _qC^{\,\,\,\f12\,\,\f12\,1}_{-\f12\,\f12\,0}\, \bft^+_+\bft^-_- 
=\f{1}{\sqrt{[2]}} \lb q^{\f12}J_+J_- - q^{-\f12} J_-J_+ \rb\,,  
\\ 
\bfX_{-1}&= \,_qC^{\f12\,\f12\,1}_{\,\f12\,\f12\,1}\, \bft^+_-\bft^-_-
=-J_-K^{-1}\,, \quad
\bfX_{1}=\, _qC^{\,\,\,\f12\,\,\,\,\f12\,\,\,\,\,1}_{-\f12\,-\f12\,-1}\, \bft^+_+\bft^-_+
=J_+K^{-1}\,.
\end{align}
It is easy to check that they  behave as a vector under the action of $\UQ$ 
\begin{equation}
J_\pm \coact \bfX_A =\sqrt{[1\mp A][1\pm A+1]} \,\bfX_{A\pm 1}\,,\quad
K\coact \bfX_A =q^{-\f{A}{2}} \bfX_A\,,
\end{equation}
so that the Wigner-Eckart theorem applies and gives the matrix elements of $\bfX_A$ in the irreducible representation $\cV^j$,
\begin{equation}
\la j,n| \bfX_A | j,m\ra =N_j \,_q C^{j\,\,\,\,1\,\,\,\,j}_{n\,-A\,m}\,,
\text{ with } N_j=\sqrt{\f{[2j][2j+2]}{[2]}}\,.
\end{equation}

Similarly, one defines the $\UQI$ vector as
\be
\Xop_{A} = \sum_{\substack{m,n=\pm\f12\\m+n=A}}\,_{q^{-1}}C^{\,\,\f12\,\,\,\,\,\f12\,\,\,\,\,\,\,1}_{-m\,-n\,-A} \mt^+_m \mt^-_n \,,\quad A=0,\pm 1\,,
\ee
whose components are
\begin{align}
\Xop_0 &= _{q^{-1}}C^{\f12\,\,\,\f12\,\,1}_{\,\f12\,-\f12\,0}\, \mt^+_-\mt^-_+ 
+ _{q^{-1}}C^{\,\,\,\f12\,\,\f12\,1}_{-\f12\,\f12\,0}\, \mt^+_+\mt^-_- 
=\f{1}{\sqrt{[2]}} \lb q^{-\f12}J_+J_- - q^{\f12} J_-J_+ \rb\,,  
\\
\Xop_{-1}&= _{q^{-1}}C^{\f12\,\f12\,1}_{\,\f12\,\f12\,1}\, \mt^+_-\mt^-_-
=-J_-K\,, \quad
\Xop_{1}= _{q^{-1}}C^{\,\,\,\f12\,\,\,\,\f12\,\,\,\,\,1}_{-\f12\,-\f12\,-1}\, \mt^+_+\mt^-_+
=J_+K\,.
\end{align}
They are indeed $\UQI$ vectors since
\begin{equation}
J_\pm \bcoact \Xop_A =\sqrt{[1\mp A][1\pm A+1]}\, \bfX_{A\pm 1}\,,\quad
K\bcoact \Xop_A =q^{-\f{A}{2}} \Xop_A\,,
\end{equation}
and from the Wigner-Eckart theorem,
\begin{equation}
\la j,n| \Xop_A | j,m\ra =N_j \,_{q^{-1}}C^{\,j\,\,\,1\,\,\,\,j}_{n\,-A\,m}\,,
\text{ with } N_j=\sqrt{\f{[2j][2j+2]}{[2]}}\,.
\end{equation}
One can see that $\bfX$ and $\Xop$ are the natural quantization the classical deformed vectors $\vec{X}$ and $\vec{X}^{\op}$ as defined in \eqref{eq:def_X_Xop}. The tilde sector of vectors $\bfXt$ and $\Xtop$ can also be built in the same way from $\bftt^\epsilon$ and $\mtt^\epsilon$ respectively. In addition, higher spin quantum vectors of $\UQ$ and $\UQI$ types can be built with the $q$-WCG coefficient in a similar method.

\subsection{Flipping the ribbon}\label{sec:flip}
In the following, we will omit the index $\epsilon$ on the spinor operators as it is not relevant for the present discussion. We introduce the operator $I$ associated to changing the orientation of an edge of $\Gamma$, which is a quantum version of $\iota$.

When changing the orientation of an edge, we have the following involutive transformation on the spinor operators
\begin{equation}
\mtt \rightarrow \mt, \quad \bftt \rightarrow \bft.  
\end{equation}
Since the tilde and non-tilde spinors are classically the same, and since the quantization map \eqref{eq:def_quantum_spinor} is the same for both, we can define
\begin{equation} \label{InvolutionQuantumSpinors}
I(\bft)=\bftt, \qquad I(\bftt) = \bft, \qquad I(\mt) = \mtt, \qquad I(\mtt) = \mt,
\end{equation}
and just like we did classically with $\iota$, we can lift $I$ to the $q$-bosons by setting
\begin{equation} \label{InvolutionqBosons}
I(a)=\tilde{a}, \qquad I(a^\dagger) = \tilde{a}^\dagger, \qquad I(b) = \tilde{b}, \qquad I(b^\dagger) = \tilde{b}^\dagger,
\end{equation}
and requiring that $I$ is an involution. By applying $I$ to \eqref{qJordanMap}, one finds
\begin{equation}
I(J_{\pm}) = \tilde{J}_{\pm},\qquad I(\tilde{J}_{\pm}) = J_{\pm}, \qquad I(K) = \tilde{K}, \qquad I(\tilde{K}) = K.
\end{equation}

It is then possible to find $I(L)$ in terms of $\tilde{L}$,
\begin{equation}
I(L) = \begin{pmatrix} \tilde{K}^{-1} & 0\\ -q^{\frac{1}{4}}(q^{\frac{1}{2}} - q^{-\frac{1}{2}}) \tilde{J}_+ & \tilde{K}\end{pmatrix} = S(\tilde{L})
\end{equation}
where $S$ is the antipode of $\UQ$. Similarly, one finds $I(\tilde{L}) = \overline{S}(L)$ with $\bS$ being the antipode of $\UQI$. Indeed, $S(\bS(L))\equiv L$ and $\bS(S(\Lt))\equiv \Lt$, consistently with the fact that $I$ is an involution.

The same can be applied to $U$. Parametrize the matrix elements of $\uh$ and $\uth$ as well as their antipode to be (See Definition \ref{def:SUq2} for definition of the Hopf algebra $\SU_q(2)$.)
\begin{align}
\uh&=\mat{cc}{\ah & \bh \\ \ch & \dhh}\in \SU_{q^{-1}}(2)\,,\quad\text{with}\quad
\ah\dhh-q^{\f12}\bh\ch=\id\,\quad
\bS(\uh)=\mat{cc}{\dhh & -q^{-\f12}\bh \\ -q^{\f12}\ch & \ah}\,,
\label{eq:antipide_of_SUqi2}\\
\uth&=\mat{cc}{\ath & \bth \\ \cth & \dth}\in \SU_q(2)\,,\quad\text{with}\quad
\ath\dth-q^{-\f12}\bth\cth=\id,\quad
S(\uth)=\mat{cc}{\dth & -q^{\f12}\bth \\ -q^{-\f12}\cth & \ath}\,,
\label{eq:antipide_of_SUq2}
\end{align}
where we have used $\bS$ to denote the antipode for $\SU_{q^{-1}}(2)$. It is natural to define the operator $I$ acting on the generators $\ah,\bh,\ch,\dhh$ of $\uh$ and generators $\ath,\bth,\cth,\dth$ of $\uth$ as
\be\ba{llll}
I(\ah)=\dth\,,\quad
&I(\bh)=-q^{\f12}\bth\,,\quad
&I(\ch)=-q^{-\f12}\cth\,,\quad
&I(\dhh)=\ath\,,\\[0.15cm]
I(\ath)=\dhh\,,\quad
&I(\bth)=-q^{-\f12}\bh\,,\quad
&I(\cth)=-q^{\f12}\ch\,,\quad
&I(\dth)=\ah\,,
\ea
\ee
where $I$ is indeed an involution. We then have
\be
I(\uh)=S(\uth)\,,\quad 
I(\uth)=\bS(\uh)\,.
\label{eq:I_on_U_Ut}
\ee
Recall that one can reconstruct these quantum holonomies in terms of the quantum spinors as in \eqref{eq:def_U_with_spinors} and \eqref{eq:def_Ut_with_spinors}, which we copy here 
\begin{equation*}
\uh_{AB}=(-1)^{\f12-B}q^{\f{B}{2}}\sum_{\epsilon=\pm} \mt_A^\epsilon \bftt_{-B}^\epsilon \, \f{1}{[N+1]}\in\SU_{q^{-1}}(2)\,,
\quad
\uth_{AB}=\f{1}{[\Nt+1]} (-1)^{\f12+B}q^{-\f{B}{2}}\sum_{\epsilon=\pm} \bft_A^\epsilon \mtt_{-B}^\epsilon 
\in\SU_{q}(2)\,.
\label{eq:def_U_Ut_with_spinors}
\end{equation*}
The matrix element of the antipodes of $\uh$ and $\uth$ defined in \eqref{eq:antipide_of_SUqi2} and \eqref{eq:antipide_of_SUq2} can be equivalently written as
\begin{align}
(\bS(\uh))_{AB}
&=(-1)^{B-A}q^{\f{A-B}{2}}\uh_{-B\,-A}
=\f{1}{[N+1]}(-1)^{\f12+B}q^{-\f{B}{2}}\sum_{\epsilon}\bftt_A^\epsilon\mt_{-B}^\epsilon \,,\\
(S(\uth))_{AB} 
&= (-1)^{A-B}q^{\f{B-A}{2}}\uth_{-B\,-A}
=(-1)^{\f12-B}q^{\f{B}{2}}\sum_{\epsilon}
\mtt_A^\epsilon \bft_{-B}^\epsilon \f{1}{[\Nt+1]}\,.
\end{align}
Then \eqref{eq:I_on_U_Ut} can be deduced from \eqref{InvolutionQuantumSpinors}.

Therefore, we have a complete map for quantum objects in terms of flipping the ribbons. We can then focus only on one orientation for a ribbon and use the involution map $I$ to deduce the results after change of orientation.


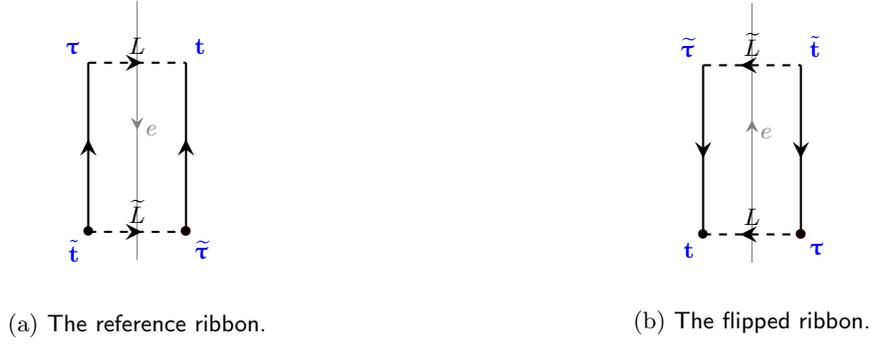
\begin{figure}[h!]
	\centering
	\begin{minipage}{0.45\textwidth}
\centering
	\begin{tikzpicture}[scale=1.5]
\coordinate (o) at (0,0);
\coordinate (a1) at ([shift=(150:0.5cm)]o);
\coordinate (a2) at ([shift=(30:0.5cm)]o);
\coordinate (b1) at ([shift=(90:1.5cm)]a1);
\coordinate (b2) at ([shift=(90:1.5cm)]a2);
\coordinate (a3) at ([shift=(-90:0.5cm)]o);
\coordinate (c1) at ([shift=(-150:1.5cm)]a3);
\coordinate (c2) at ([shift=(-150:1.5cm)]a1);
\coordinate (d1) at ([shift=(-30:1.5cm)]a3);
\coordinate (d2) at ([shift=(-30:1.5cm)]a2);

\coordinate (A) at ([shift=(90:2.3cm)]o);
\coordinate (C) at ([shift=(-30:2.3cm)]o);
\coordinate (B) at ([shift=(210:2.3cm)]o);

\draw[gray,decoration={markings,mark=at position 0.55 with {\arrow[scale=1.5,>=stealth]{<}}},postaction={decorate}] (o) -- node[right,pos=.5]{$e$}(A);

\draw[thick,dashed,decoration={markings,mark=at position 0.55 with {\arrow[scale=1.5,>=stealth]{>}}},postaction={decorate}] (a1) -- node[above,pos=.5]{$\lth$}(a2);

\draw[thick,dashed,decoration={markings,mark=at position 0.55 with {\arrow[scale=1.5,>=stealth]{>}}},postaction={decorate}] (b1) -- node[above,pos=.5]{$\lh$}(b2);

\draw[thick,decoration={markings,mark=at position 0.55 with {\arrow[scale=1.5,>=stealth]{<}}},postaction={decorate}] (b1) -- node[left,pos=.5]{$$}(a1);
\draw[thick,decoration={markings,mark=at position 0.55 with {\arrow[scale=1.5,>=stealth]{<}}},postaction={decorate}] (b2) -- node[right,pos=.5]{$$}(a2);

\draw[red] (a2) node{$\bullet$};
\draw (a1) node{$\bullet$};
\draw (a2) node{$\bullet$};
\draw[blue] (a1) node[below left]{$\bftt$};
\draw[blue] (a2) node[below right]{$\mtt$};	
\draw[blue] (b1) node[above left]{$\mt$};
\draw[blue] (b2) node[above right]{$\bft$};	
	\end{tikzpicture}
\subcaption{The reference ribbon.}
\label{fig:referencebis}
\end{minipage}  
\begin{minipage}{0.45\textwidth}
\centering
	\begin{tikzpicture}[scale=1.5]
\coordinate (o) at (0,0);
\coordinate (a1) at ([shift=(150:0.5cm)]o);
\coordinate (a2) at ([shift=(30:0.5cm)]o);
\coordinate (b1) at ([shift=(90:1.5cm)]a1);
\coordinate (b2) at ([shift=(90:1.5cm)]a2);
\coordinate (a3) at ([shift=(-90:0.5cm)]o);
\coordinate (c1) at ([shift=(-150:1.5cm)]a3);
\coordinate (c2) at ([shift=(-150:1.5cm)]a1);
\coordinate (d1) at ([shift=(-30:1.5cm)]a3);
\coordinate (d2) at ([shift=(-30:1.5cm)]a2);

\coordinate (A) at ([shift=(90:2.3cm)]o);
\coordinate (C) at ([shift=(-30:2.3cm)]o);
\coordinate (B) at ([shift=(210:2.3cm)]o);

\draw[gray,decoration={markings,mark=at position 0.55 with {\arrow[scale=1.5,>=stealth]{>}}},postaction={decorate}] (o) -- node[right,pos=.5]{$e$}(A);

\draw[thick,dashed,decoration={markings,mark=at position 0.55 with {\arrow[scale=1.5,>=stealth]{<}}},postaction={decorate}] (a1) -- node[above,pos=.5]{$\lh$}(a2);

\draw[thick,dashed,decoration={markings,mark=at position 0.55 with {\arrow[scale=1.5,>=stealth]{<}}},postaction={decorate}] (b1) -- node[above,pos=.5]{$\lth$}(b2);

\draw[thick,decoration={markings,mark=at position 0.55 with {\arrow[scale=1.5,>=stealth]{>}}},postaction={decorate}] (b1) -- node[left,pos=.5]{$$}(a1);
\draw[thick,decoration={markings,mark=at position 0.55 with {\arrow[scale=1.5,>=stealth]{>}}},postaction={decorate}] (b2) -- node[right,pos=.5]{$$}(a2);

\draw[red] (a2) node{$\bullet$};
\draw (a1) node{$\bullet$};
\draw (a2) node{$\bullet$};
\draw[blue] (a1) node[below left]{$\bft$};
\draw[blue] (a2) node[below right]{$\mt$};
\draw[blue] (b1) node[above left]{$\mtt$};
\draw[blue] (b2) node[above right]{$\bftt$};
	\end{tikzpicture}
\subcaption{The flipped ribbon.}
\label{fig:flipped}
\end{minipage} 
 \caption{Flipping the reference ribbon due to the change of orientation of the edge $e$ is equivalent to the spinor flip $\mtt \rightarrow \mt, \quad \bftt \rightarrow \bft.  
$   }
\end{figure}

\subsection{$\cR$-matrix as parallel transport}\label{sec:rmat//}
In the classical construction, the different spinors are related through parallel transport by the $\AN(2)$ holonomies. We will see that their quantum counterparts, the spinor operators, are related by $\AN_q(2)$ holonomies. We expect to have two possible cases, either lower triangular or upper triangular.

\medskip 

\paragraph*{\textbf{Parallel transport within a ribbon.} }
Let's start with the classical covariant and braided-covariant spinors of a \emph{single} ribbon related to one another by $\AN(2)$ parallel transport in \eqref{eq:braided-cov} and the first equation of \eqref{eq:tilde_braided_covariant}. At the quantum level, we have analogue relations
\begin{align} \label{eq:transport0}
\lh \mt^\epsilon &=\mat{c}{K^{-1}\mt_-^\epsilon \\
-(q^{\f34} - q^{-\f14})J_+\mt_-^\epsilon+K\mt_+^\epsilon }
=q^{\f{1+\epsilon}{4}}q^{\epsilon\f{N}{4}} \bft^\epsilon\,,
\quad 
\lth \bftt^\epsilon = 
\mat{c}{\Kt \bftt^\epsilon_- \\ (q^{\f14}-q^{-\f34})\Jt_+ \bftt^\epsilon_- + \Kt^{-1}\bftt_+^\epsilon}
=q^{-\epsilon \f{\Nt}{4}} q^{-\f{1+\epsilon}{4}}\mtt^\epsilon\,, \\ 
\bS(\lh)\bft^\epsilon &=\mat{c}{K\bft^\epsilon_- \\
(q^{\f14}-q^{-\f34})J_+\bft_-^\epsilon +K^{-1}\bft_+^\epsilon}
= q^{-\f{1+\epsilon}{4}}q^{-\epsilon \f{N}{4}}\mt^\epsilon\,,
\quad 
S(\lth)\mtt^\epsilon =
\mat{c}{\Kt^{-1}\mtt^\epsilon_- \\ -(q^{\f34}-q^{-\f14})\Jt_+ \mtt^\epsilon_- +\Kt \mtt_+^\epsilon }
=q^{\epsilon \f{\Nt}{4}} q^{\f{1+\epsilon}{4}}\bftt^\epsilon\,.
\label{eq:transport1}
\end{align}
One can take the complex conjugate of these relations and get equivalently,
\begin{subequations}
\begin{align}
(-1)^{\f12-A}q^{\f{A}{2}} \bft_{-A}^\epsilon &= q^{\epsilon\f{N}{4}} q^{\f{\epsilon-2}{4}} (-1)^{\f12-B}q^{-\f{B}{2}}\mt^\epsilon_{-B} \lb L^\dagger\rb_B{}^A   \,, \\
(-1)^{\f12-A}q^{-\f{A}{2}} \mt_{-A}^\epsilon &= q^{-\epsilon\f{N}{4}} q^{\f{2-\epsilon}{4}} (-1)^{\f12-B}q^{\f{B}{2}}\bft^\epsilon_{-B} 
\lb \bS(L)^\dagger\rb_B{}^A \,,\\
(-1)^{\f12-A}q^{-\f{A}{2}} \mtt_{-A}^\epsilon &= q^{-\epsilon\f{\Nt}{4}} q^{\f{2-\epsilon}{4}} (-1)^{\f12-B}q^{\f{B}{2}}\bftt^\epsilon_{-B} 
\lb \Lt^\dagger\rb_B{}^A \,,\\
(-1)^{\f12-A}q^{\f{A}{2}} \bftt_{-A}^\epsilon &= q^{\epsilon\f{\Nt}{4}} q^{\f{\epsilon-2}{4}} (-1)^{\f12-B}q^{-\f{B}{2}}\mtt^\epsilon_{-B} \lb S(\Lt)^\dagger\rb_B{}^A\,.
\end{align}
\label{eq:relate_spinors_conjugate}
\end{subequations}
To get \eqref{eq:relate_spinors_conjugate} from \eqref{eq:transport0} and \eqref{eq:transport1}, we have used the formulas for taking the complex conjugating of spinor components
\be\ba{ll}
\lb\bft^\epsilon_A \rb^\dagger =\epsilon (-1)^{\f12-A}q^{\f{A}{2}} \bft_{-A}^{-\epsilon}\,,&
\lb\mt^\epsilon_A \rb^\dagger=\epsilon (-1)^{\f12-A}q^{-\f{A}{2}} \mt_{-A}^{-\epsilon}\,,\\[0.15cm]
\lb\bftt^\epsilon_A \rb^\dagger =\epsilon (-1)^{\f12-A}q^{\f{A}{2}} \bftt_{-A}^{-\epsilon}\,,&
\lb\mtt^\epsilon_A \rb^\dagger=\epsilon (-1)^{\f12-A}q^{-\f{A}{2}} \mtt_{-A}^{-\epsilon}\,,
\ea
\ee
and the commutation relation of the factor $q^{\epsilon\f{N}{4}}$ or $q^{\epsilon\f{\Nt}{4}}$ with the spinor components
\be
q^{\epsilon\f{N}{4}}\bft^{\epsilon'}_A 
= q^{\f{\epsilon\epsilon'}{4}}\bft^{\epsilon'}_Aq^{\epsilon\f{N}{4}} \,,\quad
q^{\epsilon\f{N}{4}}\mt^{\epsilon'}_A 
= q^{\f{\epsilon\epsilon'}{4}}\mt^{\epsilon'}_Aq^{\epsilon\f{N}{4}} \,,\quad
q^{\epsilon\f{\Nt}{4}}\bftt^{\epsilon'}_A 
= q^{\f{\epsilon\epsilon'}{4}}\bftt^{\epsilon'}_Aq^{\epsilon\f{\Nt}{4} }\,,\quad
q^{\epsilon\f{\Nt}{4}}\mtt^{\epsilon'}_A 
= q^{\f{\epsilon\epsilon'}{4}}\mtt^{\epsilon'}_A q^{\epsilon\f{\Nt}{4}}\,.
\ee
This quantum version of the parallel transport works within a single ribbon, see Figure \ref{fig:referencebis}. Let us now consider what happens when dealing with more ribbons. 

\medskip

\paragraph*{\textbf{Spinors for many ribbons.}}

We are interested in defining spinor operators when dealing with many ribbon edges. We focus on a ribbon graph $\Gamma_{\text{rib}}$ where the graph $\Gamma$ is an $N_v$-valent vertex $v$ with $N_v$ edges ordered and labeled as $e_1$ and $e_N$ going counterclockwise.  The ribbon graph $\Gamma_{\text{rib}}$ is  an $N_v$-gon $R(v)$ surrounded by $N_v$ ribbon edges $R(e_n)$, $n \in \{1, \cdots, N_v\}$.
Once more, we do not consider the index $\epsilon$ which does not bring anything to the present discussion. For the ribbon edge $R(e_n)$, we introduce
\be
\mtt_n=\id\otimes\cdots \otimes\mtt \otimes \cdots\otimes\id,  \quad \bftt_n=\id\otimes\cdots \otimes\bftt \otimes \cdots\otimes\id. 
\ee
  These objects, $\mtt_n$ or $\bftt_n$, are built using permutations, starting respectively from $\mtt_1$ or $\bftt_1$. However, the permutation is not consistent with the coproduct if it is non co-commutative. Consequently, due to the non-co-commutativity of the coproducts of $\UQ$ and $\UQI$, these objects are not spinor operators, except $\mtt_1$ and $\bftt_1$. 
  
   \medskip 
   
  We now want to define  spinor operators, that is objects transforming covariantly under the $\UQ$ and $\UQI$ adjoint actions. 
  To make the distinction between the objects living on the $n^{th}$ leg, $\mtt_n$ or $\bftt_n$,  and the spinor operators,  we will denote ${}^{(n)}\mtt$ and  ${}^{(n)}\bftt$, the objects transforming respectively  as a $\UQI$ and $\UQ$ spinor operators.
  The construction of the spinor operators on different Hilbert spaces is usually done using the braiding induced by the $\cR$-matrix \cite{Rittenberg:1991tv}.

As a consequence the usual construction of spinor operators (or any tensor operators) is in terms of the $\cR$-matrix. There are two ways to define such a spinor operator. Explicitly, we use $\cR_{ij}^{-1}$ or $\cR_{ji}$ to define the $\UQI$ tensor operator ${}^{(n)}\mtt$.
\begin{subequations}
\bes \label{spinorRmatrixUQI}
{}^{(n)}\mtt_A  
&=&\cR_{n-1,n}^{-1}\cR_{n-2,n}^{-1}\cdots\cR_{2n}^{-1} \cR_{1n}^{-1} (\mtt_n)_A \cR_{1n}\cR_{2n}\cdots \cR_{n-2,n}\cR_{n-1,n}\otimes \id \otimes\cdots\\
\text{or}\quad{}^{(n)}\mtt_A
&=&\cR_{n,n-1}\cR_{n,n-2}\cdots\cR_{n2} \cR_{n1} (\mtt_n)_A \cR_{n1}^{-1}\cR_{n2}^{-1}\cdots \cR_{n,n-2}^{-1}\cR_{n,n-1}^{-1}\otimes \id \otimes\cdots\,.
\ees
\label{eq:n_mtt}
\end{subequations}
The two formulas of \eqref{eq:n_mtt} are proportional to each other with the proportionality coefficient being a function of the norms $N_1,\cdots,N_n$ which commutes with the $\UQ$ (or $\UQI$) generators. 
Similarly, we use $\cR_{ij}$ or $\cR_{ji}^{-1}$ to define the $\UQ$ tensor operator ${}^{(n)}\bftt$
\begin{subequations}
\bes \label{spinorRmatrixUQ}
{}^{(n)}\bftt_A 
& =&\cR_{n-1,n}\cR_{n-2,n}\cdots\cR_{2n}\cR_{1n} (\bftt_n)_A  \cR_{1n}^{-1}\cR_{2n}^{-1}\cdots \cR_{n-2,n}^{-1}\cR_{n-1,n}^{-1}\otimes \id \otimes\cdots\\
\text{or}\quad{}^{(n)}\bftt_A
&=&\cR_{n,n-1}^{-1}\cR_{n,n-2}^{-1}\cdots\cR_{n2}^{-1} \cR_{n1}^{-1} (\bftt_n)_A \cR_{n1}\cR_{n2}\cdots \cR_{n,n-2}\cR_{n,n-1}\otimes \id \otimes\cdots\,. 
\ees
\label{eq:n_bftt}
\end{subequations}

\medskip

We now show that these $\UQI$ spinors (resp. $\UQ$ spinors) can be equivalently obtained by using the quantum parallel transport induced by $\lth$ (resp. $S(\lth)$) or $S(\Lt)^\dagger$ (resp.  $\Lt^\dagger$).

\paragraph*{\textbf{Braiding as parallel transport.}}

Let us focus first on the case with all ribbon edges $R(e_n)$ oriented in the same way corresponding to incoming edges in the associated graph.  We focus on the $N_v$-gon $R(v)$. 

\begin{prop}\label{prop:R=//} 
The braiding induced by the $\cR$-matrix can be seen as a parallel transport.  
\begin{align}
&\begin{split}
{}^{(n)}\mtt_A  
&=\cR_{n-1,n}^{-1}\cR_{n-2,n}^{-1}\cdots\cR_{2n}^{-1} \cR_{1n}^{-1} (\mtt_n)_A \cR_{1n}\cR_{2n}\cdots \cR_{n-2,n}\cR_{n-1,n}\otimes \id \otimes\cdots\\
&=(\Lt\otimes \cdots\otimes \Lt\otimes \mtt_n)_A\otimes \id \otimes\cdots \\
&= \Lt_{A}{}^{A_2}\otimes \Lt_{A_2}{}^{A_3} \otimes \cdots \otimes \mtt_{A_{n-1}}\otimes \id \otimes\cdots\,,
 \end{split}  
 \label{rmat1}\\
&\text{or}  
\begin{split}
{}^{(n)}\mtt_A 
&=\cR_{n,n-1}\cR_{n,n-2}\cdots\cR_{n2} \cR_{n1} (\mtt_n)_A \cR_{n1}^{-1}\cR_{n2}^{-1}\cdots \cR_{n,n-2}^{-1}\cR_{n,n-1}^{-1}\otimes \id \otimes\cdots\\
&(S(\Lt)^\dagger\otimes \cdots\otimes S(\Lt)^\dagger\otimes \mtt_n)_A\otimes \id \otimes\cdots \\
&= (S(\Lt)^\dagger)_{A}{}^{A_2}\otimes (S(\Lt)^\dagger)_{A_2}{}^{A_3} \otimes \cdots \otimes \mtt_{A_{n-1}}\otimes \id \otimes\cdots\,,
\end{split}
\label{rmat2}\\[0.3cm]
&\begin{split}
{}^{(n)}\bftt_A 
&=\cR_{n-1,n}\cR_{n-2,n}\cdots\cR_{2n}\cR_{1n} (\bftt_n)_A  \cR_{1n}^{-1}\cR_{2n}^{-1}\cdots \cR_{n-2,n}^{-1}\cR_{n-1,n}^{-1}\otimes \id \otimes\cdots\\
&=(S(\Lt)\otimes \cdots\otimes S(\Lt)\otimes \bftt_n)_A\otimes \id \otimes\cdots \\
&= S(\Lt)_{A}{}^{A_2}\otimes S(\Lt)_{A_2}{}^{A_3} \otimes \cdots \otimes \bftt_{A_{n-1}}\otimes \id \otimes\cdots\,.
\end{split}
\label{rmat3} \\
&\text{or}
\begin{split}
 {}^{(n)}\bftt_A 
 &=\cR_{n,n-1}^{-1}\cR_{n,n-2}^{-1}\cdots\cR_{n2}^{-1} \cR_{n1}^{-1} (\bftt_n)_A \cR_{n1}\cR_{n2}\cdots \cR_{n,n-2}\cR_{n,n-1}\otimes \id \otimes\cdots\\
&(\Lt^\dagger\otimes \cdots\otimes \Lt^\dagger\otimes \bftt_n)_A\otimes \id \otimes\cdots \\
&= (\Lt^\dagger)_{A}{}^{A_2}\otimes (\Lt^\dagger)_{A_2}{}^{A_3} \otimes \cdots \otimes \bftt_{A_{n-1}}\otimes \id \otimes\cdots\,.
\end{split}
\label{rmat4}
\end{align}
\end{prop}

\begin{proof}
For notational convenience, we remove the tildes of the generators of $\UQ$ in the tilde sector. We consider \eqref{rmat1} at $n=2$. Then from the last line,
\begin{equation}
    ^{(2)}\mtt = \begin{pmatrix} K\otimes \mtt_-\\ q^{-\frac{1}{4}}(q^{\frac{1}{2}} - q^{-\frac{1}{2}}) J_+\otimes \mtt_- + K^{-1}\otimes \mtt_+\end{pmatrix}\,.
\end{equation}
We will show that the first line, $\ie$$\cR_{12}^{-1}(\id\otimes \mtt)\cR_{12}$ gives the same object. By using \eqref{qJordanMap} and \eqref{eq:def_quantum_spinor} to express the generators of $\UQ$ and the spinors in terms of the $q$-harmonic oscillators, we find
\begin{equation}
    J_z\mtt_+ = \mtt_+\bigl(J_z+\frac{1}{2}\bigr),\qquad J_+ \mtt_+ = q^{-\frac{1}{4}}\mtt_+ J_+,\qquad J_-\mtt_+ = q^{-\frac{1}{4}}\bigl(\mtt_+J_- - K\mtt_-)
\end{equation}
and
\begin{equation}
    J_z\mtt_- = \mtt_-\bigl(J_z-\frac{1}{2}\bigr),\qquad J_- \mtt_- = q^{\frac{1}{4}}\mtt_- J_-,\qquad J_+\mtt_- = q^{\frac{1}{4}}\bigl(\mtt_-J_+ - K\mtt_+).
\end{equation}
It leads to the commutation relations
\begin{equation}
    [KJ_\pm,\mtt_\mp] = -q^{\pm\f14}K^2\mtt_\pm,\qquad
    [KJ_\pm,\mtt_\pm ]=0.
\end{equation}
Consider the first line of \eqref{rmat1} for $A=-$, then
\begin{equation}
\begin{aligned}
\cR_{12}^{-1}(\id \otimes \mtt_-)
&= q^{-J_z \otimes J_z} \sum_{n=0}^{\infty} 
\f{(1-q)^n}{[n]!} q^{-\f{n(n-1)}{4}} \lb K^{-1}J_+ \rb^n \otimes \lb KJ_- \rb^n \mtt_-\\
&=q^{-J_z \otimes J_z} \sum_{n=0}^{\infty} 
\f{(1-q)^n}{[n]!} q^{-\f{n(n-1)}{4}} \lb K^{-1}J_+ \rb^n \otimes  \mtt_- \lb KJ_- \rb^{n} \\
&=q^{-J_z \otimes J_z} (\id\otimes  \mtt_-)\sum_{n=0}^{\infty} 
\f{(1-q)^n}{[n]!} q^{-\f{n(n-1)}{4}} \lb K^{-1}J_+ \rb^n \otimes  \lb KJ_- \rb^{n} \\
&=(\id\otimes  \mtt_-) q^{-J_z \otimes (J_z-\f12)}\sum_{n=0}^{\infty} 
\f{(1-q)^n}{[n]!} q^{-\f{n(n-1)}{4}} \lb K^{-1}J_+ \rb^n \otimes  \lb KJ_- \rb^{n} \\
&=(K\otimes \mtt_-)\cR_{12}^{-1} 
\end{aligned}
\end{equation}
as desired.

Computing $\cR_{12}^{-1}(\id \otimes \mtt_+)$ takes more work as $KJ_-$ and $\mtt_+$ do not commute. Indeed, each time we put $KJ_-$ to the right of $\mtt_+$, we get an extra term  $-q^{\f12}K^2\mtt_-$. This gives
\begin{equation}
    (KJ_-)^n\mtt_+ = \mtt_+ (KJ_-)^n -q^{-\frac{1}{4}}\sum_{k=0}^{n-1}K^2\mtt_- q^k(KJ_-)^{n-1} = \mtt_+ (KJ_-)^n -q^{-\frac{1}{4}} \frac{1-q^n}{1-q} K^2\mtt_- (KJ_-)^{n-1}
\end{equation}
by using $J_-^{k}K^2=q^k K^2J_-^k$. We can thus write
\begin{equation}
\begin{aligned}
\cR_{12}^{-1}(\id \otimes \mtt_+)
&=q^{-J_z \otimes J_z} \lb \id\otimes \mtt_+ \rb\sum_{n=0}^{\infty} 
\f{(1-q)^n}{[n]!} q^{-\f{n(n-1)}{4}} \lb K^{-1}J_+ \rb^n \otimes \lb KJ_- \rb^n
-q^{-\f14}q^{-J_z \otimes J_z} 
\lb K^{-1}J_+\otimes K^2\mtt_- \rb\\
&\sum_{n=0}^{\infty}  
\f{(1-q)^{n-1}}{[n-1]!} q^{-\f{(n-1)(n-2)}{4}} \f{(1-q)}{[n]} q^{-\f{(n-1)}{2}} \f{1-q^n}{1-q}
\lb (K^{-1}J_+)^{n-1} \otimes (KJ_-)^{n-1}\rb\\
&= \lb \id\otimes \mtt_+ \rb q^{-J_z \otimes (J_z+\f12)}\sum_{n=0}^{\infty} 
\f{(1-q)^n}{[n]!} q^{-\f{n(n-1)}{4}} \lb K^{-1}J_+ \rb^n \otimes \lb KJ_- \rb^n\\
&+(q^{\f34}-q^{-\f14})\lb K^{-1}J_+\otimes K^2\mtt_- \rb q^{-(J_z+1)\otimes(J_z-\f12)}
\sum_{n=1}^{\infty}  
\f{(1-q)^{n-1}}{[n-1]!} q^{-\f{(n-1)(n-2)}{4}} \lb (K^{-1}J_+)^{n-1} \otimes (KJ_-)^{n-1}\rb\\
&=(K^{-1}\otimes \mtt_+)\cR_{12}^{-1}
+q^{\f12}(q^{\f34}-q^{-\f14})\lb K^{-1}J_+K \otimes K^2\mtt_-K^{-2} \rb\cR_{12}^{-1}\\
&=\lb K^{-1}\otimes \mtt_+ + (q^{\f14}-q^{-\f34}) J_+\otimes \mtt_- \rb \cR_{12}^{-1}
\equiv (\Lt\otimes \mtt)_+ \cR_{12}^{-1}\,.
\end{aligned}
\end{equation}

The generalization to any $n$ is straightforward as
\be\begin{split}
^{(n)}\mtt_A 
&=\cR_{n-1,n}^{-1}\cR_{n-2,n}^{-1}\cdots \cR_{2n}^{-1} R_{1n}^{-1} (\mtt_n)_A R_{1n}\cR_{2n}\cdots \cR_{n-2,n}\cR_{n-1,n}\\
&=(\Lt_{A}{}^{A_2}\otimes \id\otimes\cdots) \cR_{n-1,n}^{-1}\cR_{n-2,n}^{-1}\cdots \cR_{2n}^{-1}(\mtt_n)_{A_2}\cR_{2n}\cdots \cR_{n-2,n}\cR_{n-1,n} \\
&=(\Lt_{A}{}^{A_2} \otimes \Lt_{A_2}{}^{A_3}\otimes\id\otimes\cdots )\cR_{n-1,n}^{-1}\cR_{n-2,n}^{-1}\cdots \cR_{3n}^{-1} (\mtt_n)_{A_3}\cR_{3n}\cdots \cR_{n-2,n}\cR_{n-1,n}\\
&=\cdots
=\Lt_{A}{}^{A_2}\otimes \Lt_{A_2}{}^{A_3} \otimes \cdots \otimes \mtt_{A_{n-1}}\otimes \id \otimes\cdots\,.
\end{split}\ee
Therefore, we have proved \eqref{rmat1}. Equations \eqref{rmat2}-\eqref{rmat4} can be proven using the same method.
\end{proof}

\medskip

\paragraph*{\textbf{Geometric interpretation.}} 
\begin{figure}[h!]
	\centering
\begin{minipage}{0.45\textwidth}
\centering
	\begin{tikzpicture}[scale=1.5]
\coordinate (o) at (0,0);
\coordinate (a1) at ([shift=(150:0.5cm)]o);
\coordinate (a2) at ([shift=(30:0.5cm)]o);
\coordinate (b1) at ([shift=(90:1.5cm)]a1);
\coordinate (b2) at ([shift=(90:1.5cm)]a2);
\coordinate (a3) at ([shift=(-90:0.5cm)]o);
\coordinate (c1) at ([shift=(-150:1.5cm)]a3);
\coordinate (c2) at ([shift=(-150:1.5cm)]a1);
\coordinate (d1) at ([shift=(-30:1.5cm)]a3);
\coordinate (d2) at ([shift=(-30:1.5cm)]a2);

\coordinate (A) at ([shift=(90:2.3cm)]o);
\coordinate (C) at ([shift=(-30:2.3cm)]o);
\coordinate (B) at ([shift=(210:2.3cm)]o);

\draw[gray,decoration={markings,mark=at position 0.55 with {\arrow[scale=1.5,>=stealth]{<}}},postaction={decorate}] (o) -- node[right,pos=.5]{$e_1$}(A);
\draw[gray,decoration={markings,mark=at position 0.55 with {\arrow[scale=1.5,>=stealth]{<}}},postaction={decorate}] (o) -- node[below,pos=.4]{$e_2$}(B);

\draw[thick,dashed,decoration={markings,mark=at position 0.55 with {\arrow[scale=1.5,>=stealth]{>}}},postaction={decorate}] (a1) -- node[above,pos=.5]{$\lth$}(a2);

\draw[thick,dashed,decoration={markings,mark=at position 0.55 with {\arrow[scale=1.5,>=stealth]{>}}},postaction={decorate}] (a3) -- node[left,pos=.4]{$\lth$}(a1);

\draw[thick,dashed,decoration={markings,mark=at position 0.55 with {\arrow[scale=1.5,>=stealth]{<}}},postaction={decorate}] (b1) -- node[above,pos=.5]{$\lh$}(b2);
\draw[thick,dashed,decoration={markings,mark=at position 0.55 with {\arrow[scale=1.5,>=stealth]{<}}},postaction={decorate}] (c1) -- node[left,pos=.5]{$\lh$}(c2);

\draw[thick,decoration={markings,mark=at position 0.55 with {\arrow[scale=1.5,>=stealth]{<}}},postaction={decorate}] (b1) -- node[left,pos=.5]{$$}(a1);
\draw[thick,decoration={markings,mark=at position 0.55 with {\arrow[scale=1.5,>=stealth]{<}}},postaction={decorate}] (b2) -- node[right,pos=.5]{$$}(a2);

\draw[thick,decoration={markings,mark=at position 0.55 with {\arrow[scale=1.5,>=stealth]{<}}},postaction={decorate}] (c1) -- node[below,pos=.5]{$$}(a3);
\draw[thick,decoration={markings,mark=at position 0.55 with {\arrow[scale=1.5,>=stealth]{<}}},postaction={decorate}] (c2) -- node[above,pos=.5]{$$}(a1);

\draw[red] (a2) node{$\bullet$};
\draw[blue] (a2) node[above right]{$\mtt_1$};
\draw (a1) node{$\bullet$};

\draw[blue] (a1) node[below]{$\mtt_2$};
	\end{tikzpicture}
 \subcaption{}
 \label{fig:orientation-choice_a}
\end{minipage}	
\quad
	\begin{minipage}{0.45\textwidth}
\centering
	\begin{tikzpicture}[scale=1.5]
\coordinate (o) at (0,0);
\coordinate (a1) at ([shift=(150:0.5cm)]o);
\coordinate (a2) at ([shift=(30:0.5cm)]o);
\coordinate (b1) at ([shift=(90:1.5cm)]a1);
\coordinate (b2) at ([shift=(90:1.5cm)]a2);
\coordinate (a3) at ([shift=(-90:0.5cm)]o);
\coordinate (c1) at ([shift=(-150:1.5cm)]a3);
\coordinate (c2) at ([shift=(-150:1.5cm)]a1);
\coordinate (d1) at ([shift=(-30:1.5cm)]a3);
\coordinate (d2) at ([shift=(-30:1.5cm)]a2);

\coordinate (A) at ([shift=(90:2.3cm)]o);
\coordinate (C) at ([shift=(-30:2.3cm)]o);
\coordinate (B) at ([shift=(210:2.3cm)]o);

\draw[gray,decoration={markings,mark=at position 0.55 with {\arrow[scale=1.5,>=stealth]{<}}},postaction={decorate}] (o) -- node[right,pos=.5]{$e_1$}(A);

\draw[gray,decoration={markings,mark=at position 0.55 with {\arrow[scale=1.5,>=stealth]{<}}},postaction={decorate}] (o) -- node[below,pos=.5]{$e_2$}(C);

\draw[thick,dashed,decoration={markings,mark=at position 0.55 with {\arrow[scale=1.5,>=stealth]{>}}},postaction={decorate}] (a1) -- node[above,pos=.5]{$\lth$}(a2);
\draw[thick,dashed,decoration={markings,mark=at position 0.55 with {\arrow[scale=1.5,>=stealth]{>}}},postaction={decorate}] (a2) -- node[right,pos=.5]{$\lth$}(a3);

\draw[thick,dashed,decoration={markings,mark=at position 0.55 with {\arrow[scale=1.5,>=stealth]{>}}},postaction={decorate}] (b1) -- node[above,pos=.5]{$\lh$}(b2);

\draw[thick,dashed,decoration={markings,mark=at position 0.55 with {\arrow[scale=1.5,>=stealth]{<}}},postaction={decorate}] (d1) -- node[right,pos=.5]{$\lh$}(d2);

\draw[thick,decoration={markings,mark=at position 0.55 with {\arrow[scale=1.5,>=stealth]{<}}},postaction={decorate}] (b1) -- node[left,pos=.5]{$$}(a1);
\draw[thick,decoration={markings,mark=at position 0.55 with {\arrow[scale=1.5,>=stealth]{<}}},postaction={decorate}] (b2) -- node[right,pos=.5]{$$}(a2);

\draw[thick,decoration={markings,mark=at position 0.55 with {\arrow[scale=1.5,>=stealth]{<}}},postaction={decorate}] (d1) -- node[below,pos=.5]{$$}(a3);
\draw[thick,decoration={markings,mark=at position 0.55 with {\arrow[scale=1.5,>=stealth]{<}}},postaction={decorate}] (d2) -- node[above,pos=.5]{$$}(a2);

\draw[red] (a1) node{$\bullet$};
\draw (a2) node{$\bullet$};
\draw[blue] (a1) node[above right]{$\bftt_1$};
\draw[blue] (a2) node[below right]{$\bftt_2$};		
	\end{tikzpicture}
 \subcaption{}
 \label{fig:orientation-choice_b}
\end{minipage} 
\caption{The choice of cilium is given by the red bullet. In the \ref{fig:orientation-choice_a}, the orientation is anti-clockwise, while in \ref{fig:orientation-choice_b}, the orientation is clock-wise. This choice matters since we usually order the tensor product from left to right. }
\label{fig:orientation-choice}
\end{figure}
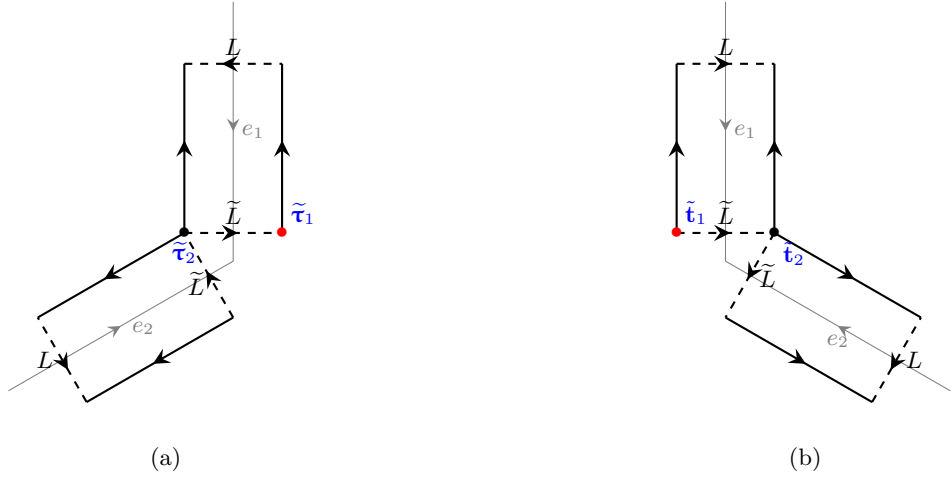

We have just shown that the braiding induced by the $\cR$-matrix can be explicitly written as a parallel transport along the ribbons using $AN_q(2)$ or $AN_{q^{-1}}(2)$ holonomies\footnote{Recall the matrix elements of $AN_q(2)$ and $AN_{q^{-1}}(2)$ are given by the generators of $\UQ$.}.  Indeed, equations \eqref{rmat1} -- \eqref{rmat4} tell us that the algebraic definition of a tensor operator written in terms of the $\cR$-matrix can be replaced by a definition which has a very natural geometrical interpretation when working with ribbons. 

Let's illustrate the geometrical definition of the tensor operator $^{(n)}\mtt$ given in \eqref{rmat1} in terms of parallel transports by $\lth$'s.  We put consecutively the ribbon edges, so that they share a vertex.  Let us deal again with the case where all the links are incoming. The construction is illustrated in Figure \ref{fig:orientation-choice}. 

The first step consists in identifying a reference point. This corresponds to choosing a \textit{cilium}. We naturally choose the reference point to sit on the ribbon edge $R(e_1)$. The construction of the spinor operators will depend on the orientation chosen for the ordering of the ribbon edges: counterclockwise or clockwise starting from $R(e_1)$. 
Indeed, the source point can be the left- or right-end point. 
(Left- or right-end point is specified by sitting at the vertex in $\Gamma$ and looking towards the outgoing direction of the relevant edge.)
Let us choose first the right end point to be our cilium as in  Figure~\ref{fig:orientation-choice_a} (the vertex in red). This means that $^{(1)}\mtt$ is the reference spinor. 
We choose to order the ribbons counter-clockwise which is the orientation consistent with the definition of the spinors given in Proposition \ref{prop:R=//}.
 
Indeed, the parallel transport by $\lth$ indicates that we take $\mtt_2$ -- which sits at the left-end point of $R(e_1)$ since the right-end point of $R(e_2)$ is identified with the left-end point of $R(e_1)$ -- and transport it to the reference point.

We proceed recursively with other ribbons. The object $\mtt_3$ sitting at the right-end point of $R(e_3)$ which is identified with the left end point of $R(e_2)$. We can transport $\mtt_3$ using $\lth $ to $^{(2)}\mtt$, and so on and so forth.

Therefore, the geometrical construction of the spinor operator $^{(n)}\mtt$ is obtained by parallel transporting $\mtt_n$, which sits at the right-end point of ribbon $R(e_n)$, along the ribbon short sides using the $\lth$'s to go from the right-end point to the left-end point of each ribbon until reaching the reference point (the right-end point of $R(e_1)$).

\medskip 

If instead we choose the cilium to be at the left-end point of ribbon 1, this means we use as a reference $\bftt$. This means that we order/add ribbons now in a clockwise manner.  This is illustrated in the Figure~\ref{fig:orientation-choice_b}.

\medskip

Now let us discuss the case when the edges do not have the same orientations. 

\medskip 

\paragraph*{\textbf{Flipping ribbons, again. }} 
\begin{figure}[h!]
	\centering
\begin{minipage}{0.45\textwidth}
\centering
	\begin{tikzpicture}[scale=1.5]
\coordinate (o) at (0,0);
\coordinate (a1) at ([shift=(150:0.5cm)]o);
\coordinate (a2) at ([shift=(30:0.5cm)]o);
\coordinate (b1) at ([shift=(90:1.5cm)]a1);
\coordinate (b2) at ([shift=(90:1.5cm)]a2);
\coordinate (a3) at ([shift=(-90:0.5cm)]o);
\coordinate (c1) at ([shift=(-150:1.5cm)]a3);
\coordinate (c2) at ([shift=(-150:1.5cm)]a1);
\coordinate (d1) at ([shift=(-30:1.5cm)]a3);
\coordinate (d2) at ([shift=(-30:1.5cm)]a2);

\coordinate (A) at ([shift=(90:2.3cm)]o);
\coordinate (C) at ([shift=(-30:2.3cm)]o);
\coordinate (B) at ([shift=(210:2.3cm)]o);

\draw[gray,decoration={markings,mark=at position 0.55 with {\arrow[scale=1.5,>=stealth]{<}}},postaction={decorate}] (o) -- node[right,pos=.5]{$e_1$}(A);
\draw[gray,decoration={markings,mark=at position 0.55 with {\arrow[scale=1.5,>=stealth]{<}}},postaction={decorate}] (o) -- node[below,pos=.4]{$e_2$}(B);

\draw[thick,dashed,decoration={markings,mark=at position 0.55 with {\arrow[scale=1.5,>=stealth]{>}}},postaction={decorate}] (a1) -- node[above,pos=.5]{$\lth$}(a2);

\draw[thick,dashed,decoration={markings,mark=at position 0.55 with {\arrow[scale=1.5,>=stealth]{>}}},postaction={decorate}] (a3) -- node[left,pos=.4]{$\lth$}(a1);

\draw[thick,dashed,decoration={markings,mark=at position 0.55 with {\arrow[scale=1.5,>=stealth]{>}}},postaction={decorate}] (b1) -- node[above,pos=.5]{$\lh$}(b2);
\draw[thick,dashed,decoration={markings,mark=at position 0.55 with {\arrow[scale=1.5,>=stealth]{>}}},postaction={decorate}] (c1) -- node[left,pos=.5]{$\lh$}(c2);

\draw[thick,decoration={markings,mark=at position 0.55 with {\arrow[scale=1.5,>=stealth]{<}}},postaction={decorate}] (b1) -- node[left,pos=.5]{$$}(a1);
\draw[thick,decoration={markings,mark=at position 0.55 with {\arrow[scale=1.5,>=stealth]{<}}},postaction={decorate}] (b2) -- node[right,pos=.5]{$$}(a2);

\draw[thick,decoration={markings,mark=at position 0.55 with {\arrow[scale=1.5,>=stealth]{<}}},postaction={decorate}] (c1) -- node[below,pos=.5]{$$}(a3);
\draw[thick,decoration={markings,mark=at position 0.55 with {\arrow[scale=1.5,>=stealth]{<}}},postaction={decorate}] (c2) -- node[above,pos=.5]{$$}(a1);

\draw[red] (a2) node{$\bullet$};
\draw[blue] (a2) node[above right]{$\mtt_1$};
\draw (a1) node{$\bullet$};

\draw[blue] (a1) node[below]{$\mtt_2$};
	\end{tikzpicture}
 \subcaption{We transport $\mtt_2$ using $\lth$ to the cilium (in red) to define a spinor ${}^{{(2)}}\mtt$. }
 \label{fig:1in2in}
\end{minipage}	
\begin{minipage}{0.45\textwidth}
\centering
	\begin{tikzpicture}[scale=1.5]
\coordinate (o) at (0,0);
\coordinate (a1) at ([shift=(150:0.5cm)]o);
\coordinate (a2) at ([shift=(30:0.5cm)]o);
\coordinate (b1) at ([shift=(90:1.5cm)]a1);
\coordinate (b2) at ([shift=(90:1.5cm)]a2);
\coordinate (a3) at ([shift=(-90:0.5cm)]o);
\coordinate (c1) at ([shift=(-150:1.5cm)]a3);
\coordinate (c2) at ([shift=(-150:1.5cm)]a1);
\coordinate (d1) at ([shift=(-30:1.5cm)]a3);
\coordinate (d2) at ([shift=(-30:1.5cm)]a2);

\coordinate (A) at ([shift=(90:2.3cm)]o);
\coordinate (C) at ([shift=(-30:2.3cm)]o);
\coordinate (B) at ([shift=(210:2.3cm)]o);

\draw[gray,decoration={markings,mark=at position 0.55 with {\arrow[scale=1.5,>=stealth]{>}}},postaction={decorate}] (o) -- node[right,pos=.5]{$e_1$}(A);
\draw[gray,decoration={markings,mark=at position 0.55 with {\arrow[scale=1.5,>=stealth]{<}}},postaction={decorate}] (o) -- node[below,pos=.4]{$e_2$}(B);

\draw[thick,dashed,decoration={markings,mark=at position 0.55 with {\arrow[scale=1.5,>=stealth]{<}}},postaction={decorate}] (a1) -- node[above,pos=.5]{$\lh$}(a2);

\draw[thick,dashed,decoration={markings,mark=at position 0.55 with {\arrow[scale=1.5,>=stealth]{>}}},postaction={decorate}] (a3) -- node[left,pos=.4]{$\lth$}(a1);

\draw[thick,dashed,decoration={markings,mark=at position 0.55 with {\arrow[scale=1.5,>=stealth]{<}}},postaction={decorate}] (b1) -- node[above,pos=.5]{$\lth$}(b2);
\draw[thick,dashed,decoration={markings,mark=at position 0.55 with {\arrow[scale=1.5,>=stealth]{>}}},postaction={decorate}] (c1) -- node[left,pos=.5]{$\lh$}(c2);

\draw[thick,decoration={markings,mark=at position 0.55 with {\arrow[scale=1.5,>=stealth]{>}}},postaction={decorate}] (b1) -- node[left,pos=.5]{$$}(a1);
\draw[thick,decoration={markings,mark=at position 0.55 with {\arrow[scale=1.5,>=stealth]{>}}},postaction={decorate}] (b2) -- node[right,pos=.5]{$$}(a2);

\draw[thick,decoration={markings,mark=at position 0.55 with {\arrow[scale=1.5,>=stealth]{<}}},postaction={decorate}] (c1) -- node[below,pos=.5]{$$}(a3);
\draw[thick,decoration={markings,mark=at position 0.55 with {\arrow[scale=1.5,>=stealth]{<}}},postaction={decorate}] (c2) -- node[above,pos=.5]{$$}(a1);

\draw[red] (a2) node{$\bullet$};
\draw[blue] (a2) node[above right]{$\mt_1$};
\draw (a1) node{$\bullet$};

\draw[blue] (a1) node[below]{$\mtt_2$};
	\end{tikzpicture}
 \subcaption{We transport $\mtt_2$ using $\bS(\lh)=\lth$ to the cilium (in red) to recover a spinor ${}^{{(2)}}\mtt$. }
 \label{fig:1out2in}
\end{minipage}	
\quad  
\begin{minipage}{0.45\textwidth}
\centering
	\begin{tikzpicture}[scale=1.5]
\coordinate (o) at (0,0);
\coordinate (a1) at ([shift=(150:0.5cm)]o);
\coordinate (a2) at ([shift=(30:0.5cm)]o);
\coordinate (b1) at ([shift=(90:1.5cm)]a1);
\coordinate (b2) at ([shift=(90:1.5cm)]a2);
\coordinate (a3) at ([shift=(-90:0.5cm)]o);
\coordinate (c1) at ([shift=(-150:1.5cm)]a3);
\coordinate (c2) at ([shift=(-150:1.5cm)]a1);
\coordinate (d1) at ([shift=(-30:1.5cm)]a3);
\coordinate (d2) at ([shift=(-30:1.5cm)]a2);

\coordinate (A) at ([shift=(90:2.3cm)]o);
\coordinate (C) at ([shift=(-30:2.3cm)]o);
\coordinate (B) at ([shift=(210:2.3cm)]o);

\draw[gray,decoration={markings,mark=at position 0.55 with {\arrow[scale=1.5,>=stealth]{<}}},postaction={decorate}] (o) -- node[right,pos=.5]{$e_1$}(A);
\draw[gray,decoration={markings,mark=at position 0.55 with {\arrow[scale=1.5,>=stealth]{>}}},postaction={decorate}] (o) -- node[below,pos=.4]{$e_2$}(B);

\draw[thick,dashed,decoration={markings,mark=at position 0.55 with {\arrow[scale=1.5,>=stealth]{>}}},postaction={decorate}] (a1) -- node[above,pos=.5]{$\lth$}(a2);

\draw[thick,dashed,decoration={markings,mark=at position 0.55 with {\arrow[scale=1.5,>=stealth]{<}}},postaction={decorate}] (a3) -- node[left,pos=.4]{$\lh$}(a1);

\draw[thick,dashed,decoration={markings,mark=at position 0.55 with {\arrow[scale=1.5,>=stealth]{>}}},postaction={decorate}] (b1) -- node[above,pos=.5]{$\lh$}(b2);
\draw[thick,dashed,decoration={markings,mark=at position 0.55 with {\arrow[scale=1.5,>=stealth]{<}}},postaction={decorate}] (c1) -- node[left,pos=.5]{$\lth$}(c2);

\draw[thick,decoration={markings,mark=at position 0.55 with {\arrow[scale=1.5,>=stealth]{<}}},postaction={decorate}] (b1) -- node[left,pos=.5]{$$}(a1);
\draw[thick,decoration={markings,mark=at position 0.55 with {\arrow[scale=1.5,>=stealth]{<}}},postaction={decorate}] (b2) -- node[right,pos=.5]{$$}(a2);

\draw[thick,decoration={markings,mark=at position 0.55 with {\arrow[scale=1.5,>=stealth]{>}}},postaction={decorate}] (c1) -- node[below,pos=.5]{$$}(a3);
\draw[thick,decoration={markings,mark=at position 0.55 with {\arrow[scale=1.5,>=stealth]{>}}},postaction={decorate}] (c2) -- node[above,pos=.5]{$$}(a1);

\draw[red] (a2) node{$\bullet$};
\draw[blue] (a2) node[above right]{$\mtt_1$};
\draw (a1) node{$\bullet$};
\draw[blue] (a1) node[below]{$\mt_2$};
	\end{tikzpicture}
 \subcaption{We transport $\mt_2$ using $\lth$ to the cilium (in red) to recover spinor ${}^{{(2)}}\mt$. }
 \label{fig:1in2out}
\end{minipage}	
\quad 
\begin{minipage}{0.45\textwidth}
\centering
	\begin{tikzpicture}[scale=1.5]
\coordinate (o) at (0,0);
\coordinate (a1) at ([shift=(150:0.5cm)]o);
\coordinate (a2) at ([shift=(30:0.5cm)]o);
\coordinate (b1) at ([shift=(90:1.5cm)]a1);
\coordinate (b2) at ([shift=(90:1.5cm)]a2);
\coordinate (a3) at ([shift=(-90:0.5cm)]o);
\coordinate (c1) at ([shift=(-150:1.5cm)]a3);
\coordinate (c2) at ([shift=(-150:1.5cm)]a1);
\coordinate (d1) at ([shift=(-30:1.5cm)]a3);
\coordinate (d2) at ([shift=(-30:1.5cm)]a2);

\coordinate (A) at ([shift=(90:2.3cm)]o);
\coordinate (C) at ([shift=(-30:2.3cm)]o);
\coordinate (B) at ([shift=(210:2.3cm)]o);

\draw[gray,decoration={markings,mark=at position 0.55 with {\arrow[scale=1.5,>=stealth]{>}}},postaction={decorate}] (o) -- node[right,pos=.5]{$e_1$}(A);
\draw[gray,decoration={markings,mark=at position 0.55 with {\arrow[scale=1.5,>=stealth]{>}}},postaction={decorate}] (o) -- node[below,pos=.4]{$e_2$}(B);

\draw[thick,dashed,decoration={markings,mark=at position 0.55 with {\arrow[scale=1.5,>=stealth]{<}}},postaction={decorate}] (a1) -- node[above,pos=.5]{$\lh$}(a2);
\draw[thick,dashed,decoration={markings,mark=at position 0.55 with {\arrow[scale=1.5,>=stealth]{<}}},postaction={decorate}] (a3) -- node[left,pos=.4]{$\lh$}(a1);

\draw[thick,dashed,decoration={markings,mark=at position 0.55 with {\arrow[scale=1.5,>=stealth]{<}}},postaction={decorate}] (b1) -- node[above,pos=.5]{$\lth$}(b2);
\draw[thick,dashed,decoration={markings,mark=at position 0.55 with {\arrow[scale=1.5,>=stealth]{<}}},postaction={decorate}] (c1) -- node[left,pos=.5]{$\lth$}(c2);

\draw[thick,decoration={markings,mark=at position 0.55 with {\arrow[scale=1.5,>=stealth]{>}}},postaction={decorate}] (b1) -- node[left,pos=.5]{$$}(a1);
\draw[thick,decoration={markings,mark=at position 0.55 with {\arrow[scale=1.5,>=stealth]{>}}},postaction={decorate}] (b2) -- node[right,pos=.5]{$$}(a2);

\draw[thick,decoration={markings,mark=at position 0.55 with {\arrow[scale=1.5,>=stealth]{>}}},postaction={decorate}] (c1) -- node[below,pos=.5]{$$}(a3);
\draw[thick,decoration={markings,mark=at position 0.55 with {\arrow[scale=1.5,>=stealth]{>}}},postaction={decorate}] (c2) -- node[above,pos=.5]{$$}(a1);

\draw[red] (a2) node{$\bullet$};
\draw[blue] (a2) node[above right]{$\mt_1$};
\draw (a1) node{$\bullet$};

\draw[blue] (a1) node[below]{$\mt_2$};
	\end{tikzpicture}
 \subcaption{We transport $\mt_2$ using $\bS(\lh)=\lth$ to the cilium (in red) to define a spinor ${}^{(2)}\mt$. }
 \label{fig:1out2out}
\end{minipage}	
\caption{The choice of cilium, the right end-point of ribbon 1, is given by the red bullet. In each case, we transport the relevant spinor living on the right end-point of ribbon 2 using the holonomy in ribbon 1. We recover the same spinor in each case as in the un-flipped case.  }
\label{fig:orientation-choice_2}
\end{figure}
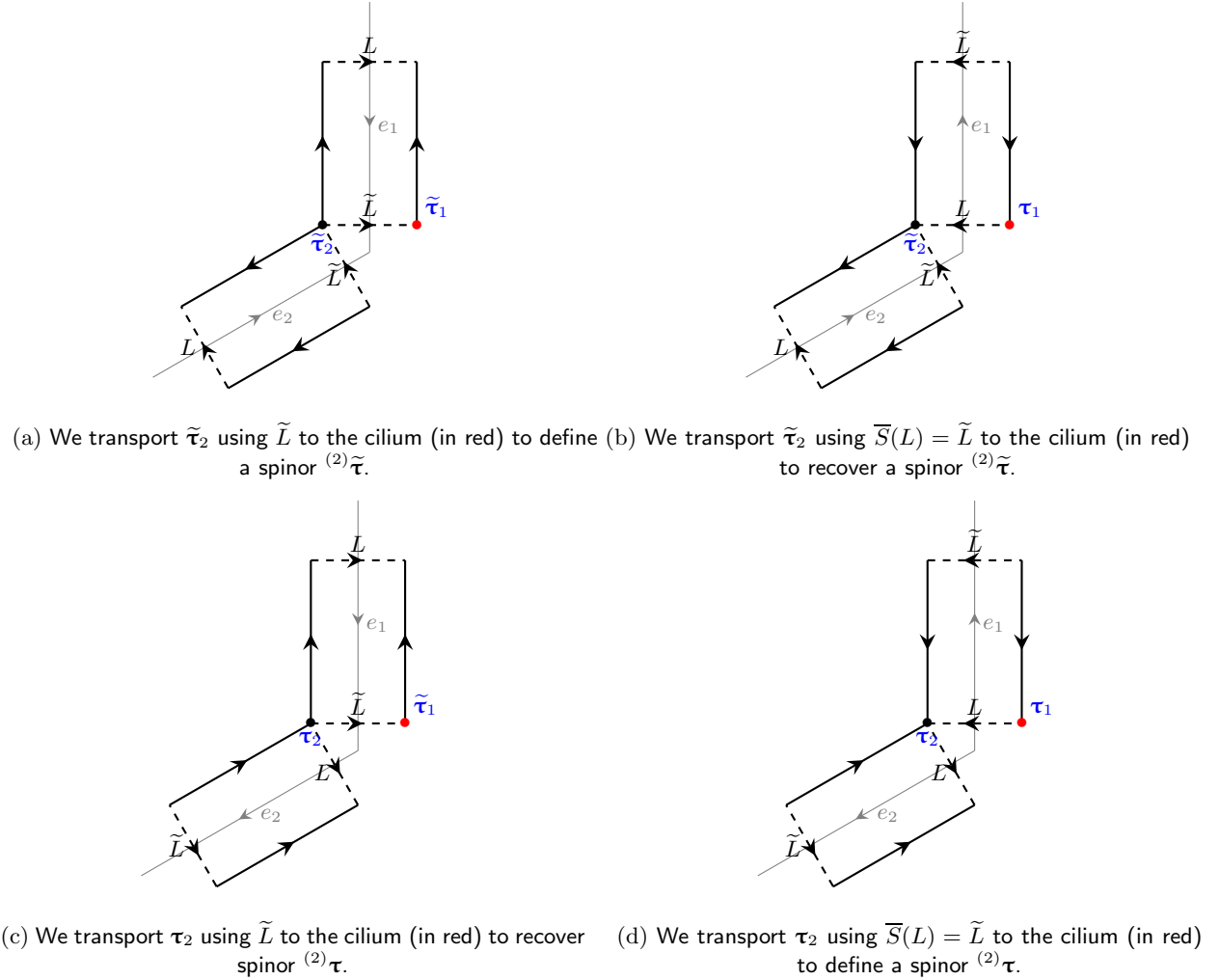
We again drop the $\epsilon$ spinor decoration since it does not bring anything to the present discussion.  As discussed in section \ref{sec:flip}, when we flip the orientation of the ribbon, the exchange of variables is performed by $I$ such that
\be
I(\bft) = \bftt, \quad I(\mt) = \mtt, \quad I(\lh) = S(\lth), \quad I(\lth) = \bS(\lh). 
\ee
When flipping the orientation of an edge in Proposition \ref{prop:R=//}, it is thus enough to apply the operator $I$, but only to the factor of the tensor product which corresponds to this edge.

For instance, consider $n=2$ and reverse the orientation of the edge 2 only (not 1). Then applying $I$ on ribbon 2 (which we henceforth denote $I_2$) to the last line of \eqref{rmat1} gives
\begin{equation}
    ^{(2)} \mt_A = I_2(\tilde{L}_A{}^B\otimes \mtt_B) = \tilde{L}_A{}^B\otimes I(\mtt_B) = \tilde{L}_A{}^B\otimes \mt_B
    \label{eq:c_case_tau}
\end{equation}
and to the last line of \eqref{rmat3},
\begin{equation}
    ^{(2)} \bft_A= S(\tilde{L})_A{}^B \otimes \bft_B.
     \label{eq:c_case_t}
\end{equation}

The geometric picture is as follows. The first relation \eqref{eq:c_case_tau} consists in the case where the cilium is at the right-end point. Because ribbon 2 is flipped, we have $\mt_2$ that stands at the right-end point of ribbon 2 which is identified with the left-end point of ribbon 1. We then parallel transport $\mt_2$ using $\tilde{L}$ on the sector 1 (See Figure~\ref{fig:1in2out}). The same applies for $^{(2)} \bft_A$, when the cilium is taken as the left-end point. 

\smallskip

Consider now the case where it is ribbon 1 which is flipped (outgoing) but ribbon 2 is \textit{not} (it is incoming), see Figure \ref{fig:1out2in}. We thus apply $I$ to the first factor of the tensor product in the last lines of \eqref{rmat1} and \eqref{rmat3},
\begin{equation}
    ^{(2)} \mtt_A = I_1(\tilde{L}_A{}^B\otimes \mtt_B) = \overline{S}(L)_A{}^B\otimes \mtt_B,\qquad ^{(2)} \bftt_A= S(I(\tilde{L}))_A{}^B \otimes \bftt_B = L_A{}^B \otimes \bftt_B.
\end{equation}
In the first case, we take the cilium to be the right-end point of ribbon 1, which is decorated by the spinor $\mt_1$. On the right-end point of ribbon 2, identified with the left-end point of ribbon 1, we have $\mtt_2$. We can define a spinor operator  
by transporting  $\mtt_2$ to the cilium through $\bS(L)$,
that is $(\bS(L))_A{}^B\otimes \mtt_B$.

\smallskip

When both ribbon 1 and ribbon 2 are flipped, see figure \ref{fig:1out2out}, we use the map $I_{12}$ which flips the sectors 1 and 2.  
As we just discussed, we can define the spinor
\begin{equation}
    ^{(2)} \mt_A = I_{12}(\tilde{L}_A{}^B\otimes \mtt_B) = \overline{S}(L)_A{}^B\otimes \mt_B,
    \qquad ^{(2)} \bft_A= I_{12}(S((\tilde{L}))_A{}^B \otimes \bftt_B )= L_A{}^B \otimes \bft_B.
    \label{eq:d_case}
\end{equation}
We still take the right-end point of ribbon 1 as the reference point, we have now $\mt_1$ sitting at the cilium. At the right-end point of the ribbon 2, coinciding with the left-end point of ribbon 1, we have $\mt_2$. We can define a spinor operator  
by transporting $\mt_2$ to the cilium through $\bS(L)$,
that is $(\bS(L))_A{}^B\otimes \mt_B$.

\smallskip

To summarize, 
\textit{the definition of the spinor operator on different ribbons does not depend on the orientation of the edges}, since for example $\tilde{L}$ and $\bS(L)$ are the same operators and so are $\mtt$ and $\mt$. So \eqref{eq:d_case} is the same as \eqref{eq:c_case_tau} and \eqref{eq:c_case_t}.

\subsection{Observables}

We will now proceed to the quantization of the observables defined in Section \ref{sec:scalar_product}. The first part of this subsection has already appeared in  \cite{Dupuis:2013haa,Dupuis:2013lka,Bonzom:2014bua}. The spinors are promoted to spinor operators as we have discussed previously. The scalar product is obtained by contracting with a Clebsch-Gordan coefficients projecting the tensor product of two spin 1/2 representations to the trivial representation.

\begin{prop}\label{prop:obs1}
The quantization of the general observable \eqref{eq:Eij_classical-bis} living on the edges $e_i$ and $e_j$ with $i\leq j$ is given by, up to some overall normalization constant,
\be
\E^{\epsilon_i,\epsilon_j}_{e_ie_j}
=\begin{cases}
\sum_{A} (-1)^{\f12+A}q^{-\f{A}{2}}\,^{(i)}\mtt_{-A}^{\epsilon_i}\,^{(j)}\mtt_A^{\epsilon_j} &\text{for $o_{i}=o_{j}=-1$}\\
\sum_{A} (-1)^{\f12+A}q^{-\f{A}{2}}\,^{{(i)}}{\mt_{-A}^{\epsilon_i}}\,^{(j)}\mtt_A^{\epsilon_j}   & \text{for $o_{i}=-o_{j}=-1$}\\
\sum_A (-1)^{\f12+A} q^{-\f{A}{2}}\,^{{(i)}}\mtt_{-A}^{\epsilon_i}\,^{(j)}{\mt_A^{\epsilon_j}} &  \text{for $o_{i}=-o_{j}=1$}\\
\sum_A (-1)^{\f12+A} q^{-\f{A}{2}}\,^{(i)}{\mt_{-A}^{\epsilon_i}}\,^{(j)}{\mt_A^{\epsilon_j}} & \text{for $o_{i}=o_{j}=1$}
\end{cases}\,.
\label{eq:def_qEij}
\ee
\end{prop}

Since the quantum operators $\mt^\epsilon$ and $\mtt^{\epsilon}$ have the same matrix element, or as we discussed in section \ref{sec:flip} the spinors are invariant under the flip of the ribbon, \textit{the observables for the different orientations in Proposition \ref{prop:obs1} are actually the same}\footnote{We remind the readers that the observable defined in \eqref{eq:def_qEij} is not the same as in \cite{Bonzom:2021ham} for different orientations. Here the $\E_{ij}$'s are defined in the same way for different orientations of $e_i$ and $e_j$, while they are defined differently in \cite{Bonzom:2021ham} for a uniform action on the intertwiners for different orientation cases.}.
A natural question to enquire is the algebra that they satisfy, if they satisfy one. One can indeed check that if we were to build observables from the fluxes, the algebra of observables would not close (even with no quantum deformation \cite{Girelli:2005ii}). The great advantage of using spinor variables is that they provide a closed algebra of observables \cite{Girelli:2005ii, Freidel:2009ck, Girelli:2017dbk}. In the non-deformed case, the algebra of observables is given in terms of the $\so^*(2n)$ Lie algebra \cite{Girelli:2017dbk} where $n$ here stands for the number of edges meeting at the vertex of $\Gamma$.

If we denote the generators of $\so^*(2n)$ by  $e_{ij},f_{ij},\tilde{f}_{ij}\,,i,j = 1,\cdots,n$, their commutation relations are
\be\begin{split}
&[e_{ij},e_{kl}] = \delta_{jk}e_{il} - \delta_{il}e_{kj}\,,\quad
[e_{ij},f_{kl}] = \delta_{il}f_{jk} - \delta_{ik}f_{jl}\,,\quad
[e_{ij},\ft_{kl}] = \delta_{jk}\ft_{il} - \delta_{jl}\ft_{ik}\,,\\
&[f_{ij},\ft_{kl}] = \delta_{jl}e_{ki} + \delta_{ik}e_{lj} - \delta_{jk}e_{li} - \delta_{il} e_{kj}\,,\quad
[f_{ij},f_{kl}] = [\ft_{ij},\ft_{kl}] = 0\,.
\end{split}
\label{eq:so*2n}
\ee 
We can identify $\u(n)$ as a Lie sub-algebra generated by $\{e_{ij}\}$. 

\medskip 

We want to show now that a similar statement holds in the deformed case, $\ie$we have a deformation of the  $\so^*(2n)$ algebra which contains a deformation of the $\u(n)$ algebra. The deformation of the $\u(n)$ algebra was already identified in \cite{Dupuis:2013lka} using the $\cR$-matrix formalism. We extend here the construction to have the full deformation of $\so^*(2n)$. We are first going to recover the deformed substructure $\UUQn$ then the full deformed algebra $\UQso$. 

\medskip

Given a semi-simple Lie algebra, its deformation is given in terms of the Serre-Chevalley relations \cite{Chari1:995guide}. The (Cartan-Weyl) generators are constructed by induction.

We have constructed a set of observables using the spinor parametrization. As we discussed, we can obtain different parametrizations because we can use different types of parallel transport, either $L$ or $S(L)^{\dagger}$. Hence in terms of the spinor parameterization, we also have some arbitrariness in terms of the explicit expression of the observables.  We know that at the classical level these observables form the algebra $\so^*(2n)$. Hence we could apply the Serre-Chevalley induction for the deformed case. The goal is then to relate this construction to the parameterization in terms of the spinors. \textit{We are going to show that the Serre-Chevalley construction picks exclusively the parallel transport induced by $S(L)^{\dagger}$. }
Let us recall more details on the Serre-Chevalley induction process to fix the notations. 

\medskip

The definition of the $\UUQn$ from the Cartan-Weyl generators $\cE_{ij}$ is as follows \cite{Biedenharn:1996vv}. We first specify the Chevalley set of generators containing $n-1$ raising, $n-1$ lowering and $n-1$ diagonal generators, denoted respectively as $\cE_{i,i+1}$, $\cE_{i,i-1}$, and $\cE_i$, which satisfy the following commutation relations
\be\label{eq:Uq_uN_algebra}
[\cE_i,\cE_j]=0\,,\quad
[\cE_i,\cE_{j,j+1}] = (\delta_{ij}-\delta_{i,j+1})\cE_{j,j+1}\,,\quad
[\cE_i,\cE_{j+1,j}] = (\delta_{i,j+1}-\delta_{ij})\cE_{j+1,j}\,,\quad
[\cE_{i,i+1},\cE_{j+1,j}] = \delta_{ij} [\cE_i-\cE_{i+1}]\,.
\ee
The remaining Cartan-Weyl generators $\cE_{ij}$ and $\cE_{ji}$ with $j>i+1$ are defined recursively as follows.
\begin{subequations}
\begin{align}
\cE_{ij}&:= q^{\f{N_{j-1}}{2}}\lb \cE_{i,j-1}\cE_{j-1,j} - q^{\f12}\cE_{j-1,j}\cE_{i,j-1} \rb\,, \\
\cE_{ji}&:= q^{-\f{N_{j-1}}{2}}\lb \cE_{j,j-1}\cE_{j-1,i} - q^{-\f12}\cE_{j-1,i}\cE_{j,j-1} \rb\,. 
\end{align}
\label{eq:Chevalley_to_CartanWeyl}
\end{subequations}
By the Jordan map, the Chevalley set can be defined in terms of the $q$-bosons $(a_i,a^\dagger_i,b_i,b^\dagger_i)$: 
\be
\cE_{i,i+1}=a^\dagger_i a_{i+1} q^{\f{N_{b_i}-N_{b_{i+1}}}{4}} + b_i^\dagger b_{i+1} q^{\f{-N_{a_i}+N_{a_{i+1}}}{4}}\,,\quad
\cE_{i+1,i}=a_i a^\dagger_{i+1} q^{\f{N_{b_i}-N_{b_{i+1}}}{4}} + b_i b^\dagger_{i+1} q^{\f{-N_{a_i}+N_{a_{i+1}}}{4}}\,,\quad
\cE_i=N_i+1\,,
\label{eq:Eij_in_bosons}
\ee
and the other generators in terms of the $q$-bosons can be deduced from \eqref{eq:Chevalley_to_CartanWeyl}. 
 It is apparent that the definitions \eqref{eq:Eij_in_bosons} and \eqref{eq:Chevalley_to_CartanWeyl} of the quantum operators $\cE_{ij}$ and their quantum algebra given in \eqref{eq:Uq_uN_algebra} are the quantized version of the definitions \eqref{eq:qSO*2n_generators_classical_1}, \eqref{eq:qSO*2n_generators_classical_2} and \eqref{eq:qSO*_generators_ij_4}, \eqref{eq:qSO*_generators_ij_2} of the quadratic invariant observables $\fe_{ij}$ and their Poisson algebra \eqref{eq:qUn_Poisson} respectively. In particular, the quantum and Poisson algebras are related by $[\cE_{ij},\cE_{kl}]=i\hbar\{\fe_{ij},\fe_{kl}\}+O(\hbar^2)$. 
We can then identify directly the relations between the $\UUQn$ Chevalley set of generators and the quadratic operators constructed from the deformed quantum spinors. They simply are:
\be
\E_{i,i+1}^{+,-}=\cE_{i,i+1}\,,\quad 
\E_{i,i+1}^{-,+}=\cE_{i+1,i}\,,\quad
\E_{i,i}^{+,-}=[\cE_i-1]\,,\quad
\E_{i,i}^{-,+}=[\cE_i+1]
\,.
\ee
For the remaining Cartan-Weyl generators in terms of the quantum spinors, one can make use of the quantum fluxes to connect the spinors from distanced sites. The result is given in the following proposition.
\begin{prop}
\label{prop:UQn}
The Cartan-Weyl generators $\cE_{i,i+p}$ and $\cE_{i+p,i}$ of $\UUQn$ for any $p\in \N^+$ can be expressed with the quantum spinors at sites $i$ and $i+p$ and the quantum fluxes for ribbon edges connecting them. Explicitly, they can be written as
\begin{align}
\cE_{i,i+p}&=q^{\f{\sum_{k=1}^{p-1}N_{i+k}}{4}}
\sum_{\substack{A_i,A_{i+1},\\\cdots,A_{i+p}}} (-1)^{\f12+A_i}q^{\f{A_i}{2}} \bftt^+_{i,-A_i}
\prod_{k=1}^{p-1} (S(\Lt_{i+k})^\dagger)_{A_{i+k-1}}{}^{A_{i+k}}\mtt^-_{i+p,A_{i+p-1}}\,,
\label{eq:CW_raising_with_spinors}
\\
\cE_{i+p,i}&=q^{-\f{\sum_{k=1}^{p-1}N_{i+k}}{4}}
\sum_{\substack{A_i,A_{i+1},\\\cdots,A_{i+p}}} (-1)^{\f12-A_i}q^{\f{A_i}{2}} \bftt^-_{i,-A_i}
\prod_{k=1}^{p-1} (S(\Lt_{i+k})^\dagger)_{A_{i+k-1}}{}^{A_{i+k}}\mtt^+_{i+p,A_{i+p-1}}\,.
\label{eq:CW_lower_with_spinors}
\end{align}
\end{prop}
\begin{proof}
Notice that the following relations are satisfied.
\begin{subequations}
\begin{align}
\mtt_{i,A}^-\bftt_{i,B}^+ -q^{\f12}\bftt^+_{i,B}\mtt^-_{i,A}
&=q^{-\f{N_i}{4}}(-1)^{\f12-B}q^{\f{B}{2}}\lb S(\Lt_i)^\dagger\rb_{A}{}^{-B}\equiv q^{-\f{N_i}{4}}(-1)^{\f12+A}q^{-\f{A}{2}}\lb\Lt^\dagger_i\rb_{B}{}^{-A}\,, \\
\bftt^-_{i,A}\mtt^+_{i,B}-q^{-\f12}\mtt^+_{i,B}\bftt^-_{i,A} 
&=q^{\f{N_i}{4}} (-1)^{\f12+A}q^{\f{A}{2}} \lb S(\Lt_i)^\dagger\rb_{B}{}^{-A}
= q^{\f{N_i}{4}}(-1)^{\f12-B}q^{-\f{B}{2}}\lb \Lt^\dagger_i\rb_{A}{}^{-B}\,.
\end{align}
\label{eq:t_tau_to_L}
\end{subequations}
Using the scalar operator of two spinors at the same corner to define the $\UUQn$ generator
\be
\cE_{i,i+1}=\sum_{A_i=\pm\f12} (-1)^{\f12+A_i}q^{\f{A_i}{2}} \bftt^+_{i,-A_i}\mtt^-_{i+1,A_i}\,,\quad
\cE_{i+1,i}=\sum_{A_i=\pm\f12} (-1)^{\f12-A_i}q^{\f{A_i}{2}} \bftt^-_{i,-A_i}\mtt^+_{i+1,A_i}\,,
\ee
and the induction, one can show the validity of \eqref{eq:CW_raising_with_spinors} and \eqref{eq:CW_lower_with_spinors}.
\end{proof}

We extend the construction to include all the different types of observables and verify that the observables $\E^{\epsilon_i,\epsilon_j}_{ij}$ are the generators of $\UQso$, which is the $q$-deformation of the algebra $\so^*(2n)$ \cite{Girelli:2017dbk}. 
Denote for different sectors $\epsilon_i$ and $\epsilon_j$ for the quadratic operator $\E^{\epsilon_i,\epsilon_j}_{ij}$ as
\be\ba{llll}
\cE_{i,i}\equiv\cE_i := N_i +1\,,\quad &
\cE_{i,i+p}:= \E^{+,-}_{i,i+p}\,,\quad &
\cE_{i+p,i}:= \E^{-,+}_{i,i+p}\,,\quad \\[0.3cm]
\cF_{i,i+p}:= \E^{-,-}_{i,i+p}\,,\quad &
\cF_{i+p,i}:= -\cF_{i,i+p} \,,\quad &
\cFt_{i,i+p}:= -\E^{+,+}_{i,i+p}\,,\quad &
\cFt_{i+p,i}:= -\cFt_{i,i+p}\,.
\ea
\label{eq:Uq_so*2n_generators}
\ee  
\begin{prop}
\label{prop:F_recursion}
The operators $\cF_{i,i+p}$ and $\cFt_{i,i+p}$ with $p>1$ defined in \eqref{eq:Uq_so*2n_generators} satisfy the recursion relations in terms of $\cE_{ij}$ as follows.
\begin{subequations}
\begin{align}
\cF_{i,i+p} &=  \lb \cF_{i,i+p-1}\cE_{i+p-1,i+p} -q^{\f12} \cE_{i+p-1,i+p}\cF_{i,i+p-1} \rb 
=  q^{-\f{N_{i+1}}{2}} \lb \cF_{i+1,i+p}\cE_{i+1,i} -q^{-\f12}\cE_{i+1,i}\cF_{i+1,i+p}\rb \,, \\ 
\cFt_{i,i+p} &=  \lb \cE_{i+p,i+p-1}\cFt_{i,i+p-1}-q^{-\f12} \cFt_{i,i+p-1}\cE_{i+p,i+p-1} \rb 
= q^{\f{N_{i+1}}{2}} \lb \cE_{i,i+1}\cFt_{i+1,i+p} - q^{\f12}\cFt_{i+1,i+p}\cE_{i,i+1} \rb\,.
\end{align}
\label{eq:F_recursion}
\end{subequations}
The operators $\cE_{i,i+1}$, $\cF_{i,i+1}$ and $\cFt_{i,i+1}$ defined in \eqref{eq:Uq_so*2n_generators} form the generators of $\UQso$ which is a closed algebra. These generators satisfy \eqref{eq:Uq_uN_algebra} and the following commutation relations.
\be\begin{split}
&\cE_{i,i+1}\cF_{j,j+1}- q^{-\f12}\cF_{j,j+1}\cE_{i,i+1}  = \delta_{i,j+1}  \cF_{i+1,i-1}\,,\quad
q^{-\f12}\cE_{i+1,i}\cF_{j,j+1}-\cF_{j,j+1}\cE_{i+1,i} = - \delta_{i,j-1}q^{\f{\cE_{i+1}}{2}}\cF_{i,i+2}\,,\\
&\cE_{i,i+1}\cFt_{j,j+1} - q^{\f12}\cFt_{j,j+1}\cE_{i,i+1} = \delta_{i,j-1}q^{-\f{\cE_{i+1}}{2}} \cFt_{i,i+2}\,,\quad 
 q^{\f12}\cE_{i+1,i}\cFt_{j,j+1} -\cFt_{j,j+1}\cE_{i+1,i} = -\delta_{i,j+1} \cFt_{i+1,i-1}\,,\\
&[\cE_i,\cF_{j,j+1}] = -(\delta_{ij}+\delta_{i,j+1})\cF_{j,j+1}\,,\quad
[\cE_i,\cFt_{j,j+1}] = (\delta_{ij}+\delta_{i,j+1})\cFt_{j,j+1}\,,\\
&[\cF_{i,i+1},\cFt_{j,j+1}] = \delta_{ij}\lb[\cE_i+\cE_{i+1}] \rb -\delta_{i,j-1} \cE_{i+2,i} - \delta_{i,j+1} \cE_{i-1,i+1}\,,\quad
[\cF_{i,i+1},\cF_{j,j+1}]=[\cFt_{i,i+1},\cFt_{j,j+1}]=0\,.
\end{split}
\label{eq:Uq_so*2n_algebra}
\ee
\end{prop}
The first two lines can be seen directly from \eqref{eq:F_recursion}. The rest of the commutation relations can be calculated with the definition \eqref{eq:Uq_so*2n_generators} of the generators and the relation between the spinors and the flux as shown in \eqref{eq:t_tau_to_L}. 
The commutation relations \eqref{eq:Uq_so*2n_algebra} are quantum versions of the Poisson algebra \eqref{eq:qSO*2n_Poisson_algebra} and are consistent with \eqref{eq:so*2n} when $q\rightarrow 1$. In this sense, we view the operators $\cE_{i,i+1}$, $\cF_{i,i+1}$ and $\cFt_{i,i+1}$ as the generators of $\UQso$.

\section*{Conclusion}

In this article, we have considered the framework of deformed lattice gauge theory introduced in \cite{Bonzom:2014wva,Bonzom:2014bua}, both classically and quantumly. Our key focus was the definition of a complete set of local observables which are defined for any pairs of edges incident to a vertex. At the classical level, they are defined using the spinors first introduced in \cite{Dupuis:2014fya}, while the quantum aspect was touched in \cite{Dupuis:2013lka}. Any functions of the standard holonomies and fluxes can be written in terms of those spinors, hence any observables (invariant functions). 

In this paper, we have performed the full quantization of the spinors into spinor operators, and we have proved that it is possible to contruct the quantum holonomy and flux operators from them. The quantization relies on the structure of both $\UQ$ and $\UQI$ (and $\SU_q(2)$ and $\SUQI(2)$).

We were thus able to quantize the local observables. In particular, they are invariant objects at the quantum level. Interestingly, we noticed that the conjugation by the quantum $\cR$-matrix, which is used to build tensor operators on tensor products, can instead be implemented as parallel transport by the variables $L$ which are around the vertex. While it may not come as a surprise for experts in integrable systems, where the $L$ operators ($T$ operators in the standard notation of integrable systems) and the $\cR$-matrix come from \cite{saleur}, we find that this observation provides a neat geometric interpretation to the use of the $\cR$-matrix in the gauge theory setting. It also simplifies explicit calculations, as the $\cR$-matrix can thus be bypassed. This was already noticed in \cite{Bonzom:2014bua} and further used in \cite{Bonzom:2021ham}.

Around each vertex of the lattice, we have shown that the set of quantum local observables forms a deformation of the algebra $\mathfrak{so}^*(2n)$, with a $\mathcal{U}_q(\mathfrak{u}(n))$ sub-algebra. This is obtained by proving the Serre-Chevalley relations, which as we found, picks the parallel transport by $S(L)^\dagger$ to implement the conjugation by the $\cR$-matrix.

As a first application of this setup, we have equipped the gauge theory with the dynamics of 3D quantum gravity with a cosmological constant in \cite{Bonzom:2021ham}. Indeed, we considered the Gauss law, which enforces restriction to observables, and the Hamiltonian constraints as a dynamics. The latter  are matrix elements of the holonomies around faces in the spinor basis. They can be rewritten as sums of products of the present local observables over the vertices which are along faces. We have then performed the quantization of the Gauss law and of the Hamiltonian constraints. They give rise to difference equations in the spin network basis, from which we were able to derive the building blocks of the Turaev-Viro model as the changes of the coefficients in the spin network basis under Pachner moves. 
There are even more interesting follow-ups we could consider. 

\paragraph*{Generalization of the spinor formalism. } The spinor formalism is tied to the specific choice of group we considered, namely $\SU(2)$ and its deformation. It would be interesting to explore in which way the algebra of observables extends for a general Hopf algebra. More specifically, one could use  a specific class of representations (such as the fundamental representation for $\SU(2)$ and its deformation) to construct the notion of local observables $\ie$quantities defined on vertices which are invariant under the action of the dual Hopf algebra. It would be interesting to develop this in the finite dimensional case with finite groups for example. (The construction was already done in the undeformed, non-compact group case $\SU(1,1)$  \cite{Girelli:2017lfn}.)

\paragraph*{Application to Yang-Mills or Kitaev models.} The local observables we have introduced come as a deformation of local observables which were found in the context of loop quantum gravity. They have been extensively used to get a better understanding of the quantum nature of discrete geometries, the dynamical aspects \cite{Dupuis:2013haa,Dupuis:2013lka,Bonzom:2011nv}. 
It would be interesting to see how this approach could be relevant for other frameworks which also rely on the lattice gauge theory setup. As a first example, we would be interested in exploring how we can reformulate the Hamiltonian of the Kitaev model in terms of such observables. This was already proposed in \cite{Bonzom:2011nv} where the authors used coherent states in the flat case. With a proper choice of (quantum) group, the Kitaev model can be seen as a model of 3d gravity with particle excitations. Therefore such a reformulation would provide some interesting insights on how to include matter (spin or mass excitations) within the dynamics in 3D gravity.

\section*{Acknowledgements}

The authors would like to thank Etera Livine for early participation of this work. This research was supported in part by Perimeter Institute for Theoretical Physics. Research at Perimeter Institute is supported by the Government of Canada through the Department of Innovation, Science and Economic Development Canada and by the Province of Ontario through the Ministry of Research, Innovation and Science. QP is supported by a NSERC Discovery grant awarded to MD. VB is partially supported by the ANR-20-CE48-0018 "3DMaps" grant, and by the ANR-21-CE48-0017 LambdaComb grant. The University of Waterloo and the Perimeter Institute for Theoretical Physics are located in the traditional territory of the Neutral, Anishnaabe and Haudenosaunee peoples.

\appendix
\renewcommand\thesection{\Alph{section}}

\section{Explicit Poisson brackets for Heisenberg double $\SL(2,\bC)$}
\label{app:Poisson_bracket_SL2C}

In this appendix, we collect the Poisson brackets for the $\SU(2)$ holonomies $(u,\ut)$ and the $\AN(2)$ fluxes $(\ell,\lt)$ of the phase space described in Section \ref{sec:holonomy_flux_phase_space}. The Poisson brackets read
\be\ba{llll}
\{\ell_1,\ell_2\}=-[\rT,\ell_1\ell_2]\,,&\{\ell_1,u_2\}=-\ell_1\rT u_2\,,& \{u_1,\ell_2\}=\ell_2r u_1\,,& \{u_1,u_2\}=-[r,u_1u_2]\,,\\[0.2cm]
\{\lt_1,\lt_2\}=[\rT,\lt_1\lt_2]\,,& \{\lt_1,\ut_2\}=-\ut_2 \rT \lt_1\,,& \{\ut_1,\lt_2\}=\ut_1 r \lt_2\,,& \{\ut_1,\ut_2\}=[r,\ut_1\ut_2]\,,\\[0.2cm]
\{\ell_1,\ut_2\}=-\rT \ell_1\ut_2\,,
&\{\lt_1,u_2\}=-\lt_1u_2\rT \,,
&\{u_1,\lt_2\}=\lt_2u_1r\,,
&\{\ut_1,\ell_2\}=r\ut_1\ell_2\,,\\[0.2cm]
\{\lt_1,\ell_2\}=0\,,
&\{\ut_1,u_2\}=0\,.
\ea
\label{eq:poisson_all}
\ee
It is important to note that \eqref{eq:poisson_all} is not enough to describe the full Poisson structure. Notice that the $\an(2)$ Lie algebra is preserved under 
$\tau^i\rightarrow (\tau^i)^{\dagger}$, one can switch $r\rightarrow r^\dagger = -\rT$ in \eqref{eq:poisson_all} and write the Poisson brackets
\be\ba{llll}
\left|\ba{l}
\{\ell^\dagger_1,\ell_2 \}=-\ell^\dagger_1 r \ell_2+\ell_2 r \ell^\dagger_1\,,\\
\{\ell_1,\ell^\dagger_2 \}= -\ell_1 \rT \ell^\dagger_2 + \ell^\dagger_2 \rT \ell_1 \,,
\ea\right.
\qquad
&\left|\ba{l}
\{\ell^\dagger_1,\ell^\dagger_2\}={[}\rT,\ell^\dagger_1 \ell^\dagger_2{]}\,,\\
\{\ell^\dagger_1,\lt_2\}=0\,,
\ea\right.
\qquad
&\left|\ba{l}
\{\ell^\dagger_1,u_2\}=-r \ell^\dagger_1 u_2\,,\\
\{u_1, \ell^\dagger_2\} =  \rT u_1 \ell^\dagger_2\,,
\ea\right.
\qquad
&\left|\ba{l}
\{\ell^\dagger_1,\ut_2\}=-\ell^\dagger_1 r \ut_2\,,\\
\{\ut_1,\ell^\dagger_2\}=\ell^\dagger_2 \rT \ut_1\,,
\ea\right.
\ea\nn
\ee
\be
\ba{llll}
&\left|\ba{l}
\{\tilde{\ell}^\dagger,\tilde{\ell}_2\}=\tilde{\ell}^\dagger_1 r \tilde{\ell}_2 - \tilde{\ell}_2 r \tilde{\ell}^\dagger_1\,,\\
\{\tilde{\ell}_1,\tilde{\ell}^\dagger_2\}=\tilde{\ell}_1 \rT \tilde{\ell}^\dagger_2- \tilde{\ell}^\dagger_2 \rT \tilde{\ell}_1\,,
\ea\right.
\qquad
\left|\ba{l}
\{\tilde{\ell}^\dagger_1,\ell_2\}=0\,,\\
\{\tilde{\ell}^\dagger_1,\tilde{\ell}^\dagger_2\}=-{[}\rT,\tilde{\ell}^\dagger_1 \tilde{\ell}^\dagger_2{]} \,,
\ea\right.
\qquad
&\left|\ba{l}
\{\tilde{\ell}^\dagger_1,u_2\}= -u_2 r \tilde{\ell}^\dagger_1\,,\\
\{u_1,\tilde{\ell}^\dagger_2\}=u_1 \rT \tilde{\ell}^\dagger_2\,,
\ea\right.
\qquad
&\left|\ba{l}
\{\tilde{\ell}^\dagger_1,\ut_2\}=-\tilde{\ell}^\dagger_1 \ut_2 r\,,\\
\{\ut_1,\tilde{\ell}^\dagger_2\}=\ut_1 \tilde{\ell}^\dagger_2 \rT \,.
\ea\right.
\ea
\label{eq:poisson_dagger}
\ee

We parametrize them into $2\times 2$ matrices
\be
\ell=\mat{cc}{\lambda & 0 \\ z & \lambda^{-1}}\,,\quad
\lt=\mat{cc}{\tlambda & 0 \\ \tz & \tlambda^{-1}}\,,\quad
u=\mat{cc}{\alpha & -\bbeta \\ \beta & \balpha}\,,\quad
\ut=\mat{cc}{\talpha & -\btbeta \\ \tbeta & \btalpha}\,,
\ee
where $\lambda,\tlambda\in \R^+$ and other parameters are complex. With this parametrization, the Poisson brackets in \eqref{eq:poisson_all} and \eqref{eq:poisson_dagger} are explicitly 
\be\ba{llll}
\{\lambda,z\}=\f{i\ka}{2}\lambda z\,, &
\{\lambda,\zb \}=-\f{i\ka}{2}\lambda \zb\,, &
\{z,\zb \}=i\ka (\lambda^2-\lambda^{-2} )\,. &\\[0.15cm]
\{\alpha,\beta \}= -\f{i\kappa}{2}\alpha\beta\,, &
\{\alpha,\bbeta \}=-\f{i\kappa}{2}\alpha\bbeta\,, & 
\{\alpha,\balpha \}=i\ka \beta\bbeta\,,\\[0.15cm]
\{\balpha,\beta \}=\f{i\ka}{2}\balpha\beta\,, &
\{\balpha,\bbeta \}=\f{i\ka}{2}\balpha\bbeta\,, &
\{\beta,\bbeta \}=0\,,\\[0.15cm]
\{\lambda,\alpha\}=-\f{i\ka}{4}\lambda \alpha,, &
\{\lambda,\balpha \}=\f{i\ka}{4}\lambda\balpha\,, &
\{\lambda,\beta \}=\f{i\ka}{4}\lambda\beta\,, &
\{\lambda,\bbeta \}=-\f{i\ka}{4}\lambda\bbeta\,, \\ [0.15cm]
\{z,\alpha \}=-\f{i\ka}{4}\lb z\alpha+4\lambda^{-1}\beta \rb\,, &
\{z,\balpha \}=\f{i\ka}{4}z\balpha\,, &
\{z,\beta \}=\f{i\ka}{4}z\beta\,, &
\{z,\bbeta \}=-\f{i\ka}{4}\lb z\bbeta-4\lambda^{-1}\balpha \rb\,,\\[0.5cm]
\{\tlambda,\tz \}=-\f{i\ka}{2}\tlambda \tz\,,&
\{\tlambda,\btz \}=\f{i\ka}{2}\tlambda \btz\,,&
\{\tz,\btz \}=-i\ka(\tlambda^2-\tlambda^{-2})&\\[0.15cm]
\{\talpha,\tbeta \}= \f{i\kappa}{2}\talpha\tbeta\,, &
\{\talpha,\btbeta \}=\f{i\kappa}{2}\talpha\btbeta\,, & 
\{\talpha,\btalpha \}=-i\ka \tbeta\btbeta\,,\\[0.15cm]
\{\btalpha,\tbeta \}=-\f{i\ka}{2}\btalpha\tbeta\,, &
\{\btalpha,\btbeta \}=-\f{i\ka}{2}\btalpha\btbeta\,, &
\{\tbeta,\btbeta \}=0\,,\\[0.15cm]
\{\tlambda,\talpha\} = -\f{i\ka}{4}\tlambda\talpha\,, & 
\{\tlambda,\btalpha\} = \f{i\ka}{4}\tlambda\btalpha\,,&
\{\tlambda,\tbeta\}=-\f{i\ka}{4}\tlambda\tbeta\,,&
\{\tlambda,\btbeta\}=\f{i\ka}{4}\tlambda\btbeta\,, \\[0.15cm]
\{\tz,\talpha\}=\f{i\ka}{4}\tz\talpha\,,&
\{\tz,\btalpha\} =-\f{i\ka}{4}(\tz\btalpha+4\tlambda\tbeta )\,,&
\{\tz,\tbeta\}=\f{i\ka}{4}\tz\tbeta\,,&
\{\tz,\btbeta\}=-\f{i\ka}{4}(\tz\btbeta -4\tlambda\talpha)\\[0.5cm]
\{\lambda,\talpha \}=-\f{i\ka}{4}\lambda \talpha\,, &
\{ \lambda,\btalpha \}= \f{i\ka}{4}\lambda \btalpha\,, &
\{ \lambda,\tbeta \} = \f{i\ka}{4}\lambda\tbeta\,, &
\{ \lambda,\btbeta \}=-\f{i\ka}{4}\lambda\btbeta\,, \\[0.15cm]
\{z,\talpha \}=\f{i\ka}{4}(z\talpha -4 \lambda \tbeta) \,, &
\{z,\btalpha \}=-\f{i\ka}{4}z\btalpha\,, &
\{z,\tbeta \}=-\f{i\ka}{4}z\tbeta\,,&
\{z,\btbeta \}=\f{i\ka}{4}(z\btbeta + 4 \lambda \btalpha)\,, \\[0.15cm]
\{\zb,\talpha \}= \f{i\ka}{4}\zb \talpha\,, & 
\{ \zb,\btalpha \}=-\f{i\ka}{4}(\zb\btalpha-4 \lambda\btbeta)\,,&
\{\zb,\tbeta \}=-\f{i\ka}{4}(\zb\tbeta +4 \lambda \talpha)\,, &
\{\zb,\btbeta \}=\f{i\ka}{4}\zb\btbeta\,, \\[0.5cm]
\{\tlambda,\alpha \}=-\f{i\ka}{4}\tlambda \alpha\,, &
\{\tlambda,\balpha \}=\f{i\ka}{4}\tlambda \balpha\,, &
\{\tlambda,\beta \}-\f{i\ka}{4}\tlambda\beta\,, &
\{\tlambda,\bbeta \}=\f{i\ka}{4}\tlambda \bbeta\,,\\[0.15cm]
\{\tz,\alpha \}=-\f{i\ka}{4}\tz\alpha\,, &
\{\tz,\balpha \}=\f{i\ka}{4}(\tz\balpha -4 \tlambda^{-1}\beta)\,, &
\{\tz,\beta \}=-\f{i\ka}{4}\tz\beta\,,&
\{\tz,\bbeta \}=\f{i\ka}{4}(\tz\bbeta+4\tlambda^{-1}\alpha)\,,\\[0.15cm]
\{\btz,\alpha \}=-\f{i\ka}{4}(\btz\alpha - 4\tlambda^{-1}\bbeta)\,, &
\{\btz,\balpha \}=\f{i\ka}{4}\btz \balpha\,, &
\{\btz,\beta \}=-\f{i\ka}{4}(\btz\beta+4 \tlambda^{-1}\balpha)\,, &
\{\btz,\bbeta \}=\f{i\ka}{4}\btz\bbeta\,,
\ea
\label{eq:poisson_matrix_elements}
\ee
and others vanish.
These explicit Poisson brackets are used to check the validity of the spinor parametrization in Section \ref{sec:spinor_phase_space}. 

\section{$\UQ$ and $\SU_q(2)$}
\label{sec:UQ_and_SUq2}

We work with a real deformation parameter $q=e^{\ka\hbar}$ which includes the quantum parameter $\hbar$ and the cosmological constant information encoded in $\ka$.
The key point is to realize that the classical dual pair $(\SU(2)^*\equiv\AN(2),\SU(2))$, whose Lie algebra structures are encoded by the classical $r$-matrix, can be $q$-deformed into a pair of quasitriangular Hopf algebras, $(\UQ,\cR)$ and its dual $(\SU_q(2),\cR)$. Let us first recall their definitions. 

\begin{definition}[$(\UQ,\cR)$]
The quasitriangular Hopf algebra $(\UQ,\cR)$ is generated by the identity and $J_\pm,K=q^{\f{J_z}{2}}$ with the relations
\be
KJ_\pm K\mone = q^{\pm \demi} J_\pm\,,\quad
[J_+,J_-]=[2J_z]\,,\quad \text{ with }
[n]\equiv\f{q^{\f{n}{2}}-q^{-\f{n}{2}}}{q^{\f12}-q^{-\f12}}\,.
\label{eq:UQ_generators_2}
\ee
It forms a Hopf algebra with the following coproduct and antipode
\be\ba{llllll}
\cop (J_\pm):= J_\pm \otimes K + K^{-1}\otimes J_\pm \,,\quad &\quad 
\cop (K) := K\otimes K\,, \quad & \quad 
S(J_\pm):=-q^{\pm \f12}J_\pm\,,\quad & \quad
S(K):=K^{-1}\,.
\ea
\label{eq:cop_S_counit}
\ee
while the counit $\epsilon$ is defined by $\epsilon K=1$ and $\epsilon J_\pm =0$. It is furthermore quasitriangular, with the $\cR$-matrix $\cR\in \UQ\otimes \UQ$:
\be
\cR = q^{J_z \otimes J_z} \sum_{n=0}^{\infty} 
\f{(1-q^{-1})^n}{[n]!} q^{\f{n(n-1)}{4}} \lb q^{\f{J_z}{2}}J_+ \rb^n \otimes \lb q^{-\f{J_z}{2}}J_- \rb^n\,.
\label{eq:cR}
\ee
\end{definition}
The $\cR$-matrix is the quantum version of the classical $r$-matrix and it satisfies the {\it quantum Yang-Baxter equation} (QYBE) 
\be
\cR_{12}\cR_{13}\cR_{23} = \cR_{23}\cR_{13}\cR_{12}\,,
\ee
where we have used the standard notation $\cR_{12}=\sum \cR_{(1)}\otimes \cR_{(2)}\otimes\id\,,\cR_{23}=\sum\id\otimes \cR_{(1)}\otimes \cR_{(2)}\,,\cR_{13}=\sum\cR_{(1)}\otimes \id \otimes \cR_{(2)}$. 
In the fundamental representation ($j=1/2$), the generators are represented as $2\times 2$ matrices
\be
\rho(J_-)=\mat{cc}{0 & 0 \\ 1 & 0}\,,\quad
\rho(J_+)=\mat{cc}{0 & 1 \\ 0 & 0}\,,\quad
\rho(K)=\mat{cc}{q^{\f14} & 0 \\ 0 & q^{-\f14}}\,.
\ee
Thus the $\cR$-matrix \eqref{eq:cR} takes the form
\be
R=\mat{cccc}{q^{\f14} & 0 & 0 & 0 \\
 0 & q^{-\f14} &  q^{-\f14}(q^{\f12}-q^{-\f12}) & 0 \\ 
0 &0 & q^{-\f14} & 0 \\
 0 & 0 & 0 & q^{\f14}}\,.
\label{eq:R_4X4_2}
\ee
Clearly,  the classical $r$-matrix \eqref{eq:r_4X4} in the fundamental representation is recovered at the first order,
\be
R=\id\otimes \id +i\hbar r+O(\hbar^2)\,.
\ee
We are particularly interested in the $\UQ$ elements written as $2\times 2$ matrix operators. These elements, denoted as $Q^\pm=\{(q^\pm)^i_{\phantom{i}j}\in \UQ (i,j=\pm)\}$, are \cite{Majid:2000fo}
\be
Q^+=\mat{cc}{K & 0 \\ q^{-\f14}(q^{\f12}-q^{-\f12})J_+ & K^{-1}}\,,\quad
Q^-=\mat{cc}{K^{-1} & -q^{\f14}(q^{\f12}-q^{-\f12})J_-\\ 0 & K}\,.
\ee 
The coproduct and counit of $Q^\pm$ are given by
\be
\cop(Q^\pm)=Q^\pm\otimes Q^\pm\,,\quad
\epsilon(Q^\pm)=\id\,,\quad
\ie\,\, 
\cop((q^\pm)^i_{\phantom{i}j})=\sum_k (q^\pm)^i_{\phantom{i}k}\otimes (q^\pm)^k_{\phantom{k}j}\,,\quad
\epsilon((q^\pm))^i_{\phantom{i}j}=\delta^i_j\,. 
\ee
They satisfy 
\be
Q_1^{\pm}Q_2^{\pm} R=RQ_2^{\pm}Q_1^{\pm}\,,\quad Q_1^-Q_2^+ R= RQ_2^+Q_1^-\,.
\label{eq:commutation_Q}
\ee

\medskip

$\UQI$ is generated by the same generators as $\UQ$ with the same commutation relations \eqref{eq:UQ_generators_2} but possessing a different coproduct and antipode, denoted as $\com$ and $\bS$. 
They act on the generators as
\be
 \com (J_\pm):= J_\pm \otimes K^{-1} + K\otimes J_\pm \,,\quad 
 \com (K) := K\otimes K\,, \quad 
\bS(J_\pm):=-q^{\mp \f12}J_\pm\,,\quad 
 \bS(K):=K^{-1}\,.
 \label{eq:com_Sb_counit}
 \ee
The two coproducts and two antipodes are related by 
 \be
\com=\sigma\circ\cop \,,\quad 
 \bS =S^{-1}\,,
 \nn\ee
where $\sigma$ is the permutation operator acting on the tensor space as $\sigma(a\otimes b)=b\otimes a$.

\medskip 

$\UQI$ can in fact be represented on the representation spaces of $\UQ$. Indeed, since $q$-numbers are invariant under the exchange $q\leftrightarrow q^{-1}$, as  algebras $\UQ$ and $\UQI$ are isomorphic. The isomorphism between generators is given by 
\begin{align}
J_\pm = \tilde J_\pm,  \quad K^{\mp 1} = \tilde K^{\pm 1},    
\end{align}
where the tilde is used for $\UQ$.

\begin{definition}[$(\SU_q(2),\cR)$]
\label{def:SUq2}
The dual quasitriangular Hopf algebra $(\SU_q(2),\cR)$ is generated by the identity and the coordinate functions $T=\mat{cc}{t_{--} & t_{-+} \\ t_{+-} & t_{++}}$ on the space of $2\times 2 $ matrices satisfying 
\be
R T_1 T_2 = T_2 T_1 R\,,
\label{eq:RTT_1}
\ee
where $R$ is defined in \eqref{eq:R_4X4_2}, and quotient with the $q$-determinant $\det_q T\equiv t_{--}t_{++}-q^{-\f12}t_{-+}t_{+-}=\id$. 
The antipode, coproduct and counit are given by
\be
S(T)=\mat{cc}{t_{++} & -q^{\f12}t_{-+} \\ -q^{-\f12}t_{+-} & t_{--}}
\quad
\cop(T)=T\otimes T\,,\quad
\epsilon(T)=\id \,,\quad
\ie \,\,  
\cop(t^i_{\phantom{i}j}) = \sum_{k=\pm}  t^i_{\phantom{i}k}\otimes t^k_{\phantom{k}j}\,,\quad
\epsilon(t^i_{\phantom{i}j}) =\delta^i_j\,,\quad
i,j=\pm\,.
\label{eq:coproduct_counit_T_1}
\ee
This Hopf algebra is dual quasitriangular with the $\cR$-matrix defined in \eqref{eq:cR} which is viewed as a map $\cR:\SU_q(2)\otimes \SU_q(2)\rightarrow \bC$.  
\end{definition}
The commutation relation \eqref{eq:RTT_1} is equivalent to the following relations.
\be\begin{split}
&t_{--}t_{-+}=q^{-\f12}t_{-+}t_{--}\,,\quad
t_{--}t_{+-}=q^{-\f12}t_{+-}t_{--}\,,\quad
t_{-+}t_{++}=q^{-\f12}t_{++}t_{-+}\,,\\
&t_{+-}t_{++}=q^{-\f12}t_{++}t_{+-}\,,\quad
t_{-+}t_{+-}=t_{+-}t_{-+}\,,\quad
[t_{--},t_{++}]=-(q^{\f12}-q^{-\f12})t_{-+}t_{+-}\,.
\end{split}
\ee

The duality between $\UQ$ and $\SU_q(2)$ can be represented by the bilinear map between the operator matrices $Q^{\pm}$ and $T$ \cite{Majid:2000fo} (See $\eg$\cite{Chari1:995guide} for a more detailed proof of the duality relation): 
\be
\la T_1, Q_2^+ \ra =R\,,\quad
\la T_1, Q_2^- \ra =\RT^{-1}\,,\quad
\ie\,\,
\la t^i_{\phantom{i}j}, (q^+)^k_{\phantom{k}l} \ra = R^{i\phantom{j}k}_{\phantom{i}j\phantom{k}l}\,,\quad
\la t^i_{\phantom{i}j}, (q^-)^k_{\phantom{k}l} \ra = (R^{-1})^{i\phantom{j}k}_{\phantom{i}j\phantom{k}l}\,,
\label{eq:L_U_duality}
\ee
where $R_{21}=\sigma\circ R=\sum R_{(2)}\otimes R_{(1)}$.

\bibliographystyle{bib-style} 
\bibliography{QK}

\end{document}